\documentclass[a4paper,11pt]{article}
\pdfoutput=1 

\usepackage{jheppub} 

\usepackage[T1]{fontenc} 

\usepackage{lscape}
\usepackage{array}
\usepackage{hyperref}
\usepackage{amsfonts}
\usepackage{amssymb}
\usepackage{bbm} 
\usepackage{graphicx}
\usepackage{caption}

\usepackage{imakeidx}
\usepackage[utf8]{inputenc}
\usepackage[T1]{fontenc}
\makeindex

\newcommand{\tf}{\mathfrak{t}}

\renewcommand{\=}{\,= \,}

\newcommand\G{\Gamma}

\renewcommand{\i}{{\rm i}}

\newcommand{\z}{\zeta}

\newcommand\bi{\begin{itemize}}
\newcommand\ei{\end{itemize}}

\newcommand\bspl{\begin{split}}
\newcommand\espl{\end{split}}

\newcommand{\susyL}[1]{\underset{\text{SUSY Locus}}{\longrightarrow}}
\newcommand{\bea}{\begin{eqnarray}}
\newcommand{\eea}{\end{eqnarray}}

\renewcommand{\=}{\,= \,}

\newcommand{\Qq}{\widetilde{Q}^\prime}
\newcommand{\DetExp}{\frac{\prod_{1\leq j<i\leq n} \left(z_i-z_j\right)
   \left(y_j-y_i\right)}{\prod _{i\neq j=1}^n
   \left(z_i-y_j\right)}}
\newcommand{\simExp}{\underset{\Lambda\to\infty}{\sim_{\text{exp}}}}

\begin{document}

\title{Large black hole entropy from the giant brane expansion}

\author{Matteo Beccaria\,$^a$ and Alejandro Cabo-Bizet\,$^a$}

\affiliation[a]{Università del Salento, Dipartimento di Matematica e Fisica Ennio De Giorgi, and I.N.F.N. - sezione di Lecce, Via Arnesano, I-73100 Lecce, Italy}

\emailAdd{matteo.beccaria@le.infn.it,\,acbizet@gmail.com}

\abstract{ We show that the Bekenstein-Hawking entropy of large supersymmetric black holes in AdS$_5\times S^5$ emerges from remarkable
cancellations in the \emph{giant graviton} expansions recently proposed by Imamura, and Gaiotto and Lee, independently. A similar cancellation mechanism is shown to happen in the exact expansion in terms of free fermions recently put-forward by Murthy. These two representations can be understood as sums over independent systems of giant D3-branes and free fermions, respectively. At large charges, the free energy of each independent system \emph{localizes} to its asymptotic expansion near the leading singularity. The sum over the independent systems maps their localized free energy to the localized free energy of the superconformal index of~$U(N)$~$\mathcal{N}=4$ SYM. This result 
constitutes 
a non-perturbative test of the giant graviton expansion valid at any value of~$N$. Moreover, in the holographic scaling limit $N\to\infty$ at fixed ratio~$\frac{\text{Entropy}}{N^2}\,$, it recovers the 1/16 BPS black hole entropy by a saddle-point approximation of the giant graviton expansion.

}



\maketitle


\section{Introduction}
\label{sec1}

Recently, the counting of small~$\frac{1}{16}$-BPS states in~$4d$ $U(N)$ $\mathcal{N}=4$ SYM on~$\mathbb{R}\times S^3$~\cite{Kinney:2005ej,Romelsberger:2005eg} has been nicely related to the problem of counting gravitons and giant brane BPS excitations in~$AdS_5\times S^5\,$~\cite{Imamura:2021ytr,Arai:2019xmp}~\cite{Gaiotto:2021xce,Lee:2022vig}.

It is known that upon truncation at powers of~$q$ of order~$N$ or smaller, the $\frac{1}{16}$-BPS index $\mathcal{I}(q)\,$ matches the~$N$-independent index~$\mathcal{I}_{KK}(q)$ counting BPS multigravitons in~$AdS_5\times S^5\,$~\cite{Kinney:2005ej}. The coefficients in the $q$-series of~$\mathcal{I}(q)\,$ that depend on~$N$ appear only at powers of~$q$ of order~$N\,$ or larger~\cite{Murthy:2020rbd,Agarwal:2020zwm}. This is because the traces of products of more than~$N$ BPS gauge covariant letters  can be always written as a linear combination of multiple single trace gauge invariant states. The dependence on $N$ implied by these constraints is essential to obtain a growth of states as~$N$ grows. Should there be no such dependence, there would be no chance to match the order~$N^2$ growth predicted by the dual black hole entropy (which we know is not the case). 

The $q$-monomials with~$N$-dependent coefficients can be reorganized in linear combinations of subsums with an overall pre-factor~$q^{n N}$,~\footnote{Sometimes it will be more convenient to work with chemical potentials e.g.,~$\tf$, dual to the rapidities e.g.,~$q=e^{2\pi\i \tf}\,$.}  where~$n$ is a positive integer. Such reorganization is obviously non-unique.~\footnote{For instance, assume~$N=8$ then a monomial~$q^{16}$ in the total index can be divided in many ways into contributions coming from the subsums labelled by~$n=1$ and~$n=2\,$.} As recognized in ~\cite{Imamura:2021ytr,Arai:2019xmp}~\cite{Gaiotto:2021xce,Lee:2022vig}, there is at least one such reorganization for which the~$q^{n N}$-weighted subsums count~$n$ D3 brane excitations wrapping supersymmetric and contractible~$S^3$-cycles in~$S^5\,$.

Whether this organizational pattern continues to hold for the complete~$q$-series~$\mathcal{I}(q)$ remains an open question.~\footnote{For the Schur limit of the $\frac{1}{16}$-BPS index the correspondence applies to the complete~$q$-series~\cite{Gaiotto:2021xce,Lee:2022vig}.} For example, it is possible that new stringy excitations in~$AdS_5\times S^5$ are required at large enough~$N$ and charges of order~$N^2$ in order to keep the correspondence going.  This puzzle is important to elucidate because for such charges the number (counted with signs) of $\frac{1}{16}$-BPS gauge-invariant states in the gauge theory, the coefficients of the~$q$-monomials, grows with~$N$ as the exponential of the dual black hole entropy~\cite{Cabo-Bizet:2018ehj,Choi:2018hmj,Benini:2018ywd}. Thus, in a sense, it is a priori unclear whether such an entropy growth can be understood by working solely within the D3 brane systems prescribed by the proposal of~\cite{Imamura:2021ytr}. 

Another giant graviton-like reorganization of the index, an exact one by construction, has been recently put forward in~\cite{Murthy:2022ien}.~\footnote{This study covers a family a matrix integrals that include the superconformal index as a particular case.} This reorganization is not quite the same as the proposal of~\cite{Imamura:2021ytr} -- as explained in~\cite{Liu:2022olj} -- but it seems to be closely related to it as argued in~\cite{Murthy:2022ien} and~\cite{Eniceicu:2023uvd} . Being an exact expansion, it would be useful to understand the physics behind it and how close it is to the physics of the proposals of~\cite{Imamura:2021ytr} and~\cite{Gaiotto:2021xce}.~\footnote{It would very interesting to understand whether there is a systematic way to identify holographic dualities of this kind starting from the partition function of free gauge theories. The approach put forward in~\cite{Gaiotto:2021xce,Lee:2022vig} seems natural to start thinking about this problem. The approach of~\cite{Murthy:2022ien} gives a first step in such a direction as well. The next step though, which would be to understand how to translate the averages over free-fermion systems to partition functions of brane systems in~$AdS_5\times S^5\,$, seems more involved. Perhaps some of the ideas in~\cite{Berenstein:2022srd,Lin:2022gbu} may be useful, at least to study~$\frac{1}{4}$~\cite{Beccaria:2023zjw} and~$\frac{1}{8}$-BPS indices, and to understand what stringy/brane excitations the individual free-fermion contributions are counting.    }

The main goal of this paper is to study the giant graviton representations of~\cite{Imamura:2021ytr,Gaiotto:2021xce,Lee:2022vig} and~\cite{Murthy:2022ien} at large charges and to compare the results with the ones obtained with the canonical matrix integral representation~\cite{Choi:2018hmj,Honda:2019cio,ArabiArdehali:2019tdm,Kim:2019yrz,Cabo-Bizet:2019osg,Cassani:2021fyv}. Moreover, we will also aim at understanding whether at large-$N$ the entropy of dual 1/16 BPS black holes can be recovered from the perspective of giant-brane expansions. We advance that the answer to both these previous quests turns out to be positive. 

Using the representation of~\cite{Imamura:2021ytr} and working in the macrocanonical ensemble, at large charges and for all~$N$ we will show that an exponentially large number of cancellations occurs when summing over the giant brane number~$n$.
Such cancellations can be explained in terms of an extremization mechanism for the giant graviton number~$n\,$. At~$N\,\gg\, 1$ this mechanism explains how the dual black-hole entropy is recovered within the giant graviton expansion, and its derivation provides, in particular, a first-principle explanation of the large-$N$ extremization mechanism proposed in~\cite{Choi:2022ovw}. More generally, the mechanism here identified implies that the latter cancellations continue to happen at large charges for any value of~$N\,$, not just in the large-$N$ expansion.  
It will be also shown that a similar extremization mechanism holds for the exact giant graviton-like representation of~\cite{Murthy:2022ien} and checked -- against numerics -- how such mechanism exactly accounts for the exponentially large cancellations happening after summing over individual giant graviton-like subseries (in appendix~\ref{sec:ExactRep}, see plot~\ref{fig:GGIndexComparison}). 

In the representation of~\cite{Imamura:2021ytr,Gaiotto:2021xce,Lee:2022vig}, this extremization mechanism will tell us that the black hole entropy~\cite{Gutowski:2004ez,Gutowski:2004yv,Cvetic:2004ny} comes from the superposition of a pair of complex conjugated saddle points whose semiclassical contributions evaluate the sum over giant graviton brane number~$n$. The canonical matrix integral representation of the index~\cite{Kinney:2005ej} is known to be dominated by a pair of complex conjugated eigenvalue configurations too~\cite{Cabo-Bizet:2020ewf,Benini:2018ywd,Cabo-Bizet:2019eaf,Cabo-Bizet:2020nkr,Aharony:2021zkr}\cite{Agarwal:2020zwm}. The latter and the former pairs are related: they provide two different interpretations of the very same contributions to the index at large charges of order~$N^2$.~\footnote{ It would be interesting to understand what is the physical meaning in the microcanonical ensemble of the~$\mathbb{Z}_2$ operation that exchanges the two leading saddles. What are the two groups of 1/16 BPS states that carry charge~$\pm1$ under this operation?} It remains for the future to understand the physics of the excitations accounting for subleading corrections in both, the canonical matrix model and giant graviton(-like) expansions, and for both small and large black holes.~\footnote{In the context of the canonical matrix integral representation of the index, this problem has been partially analyzed in~\cite{Aharony:2021zkr,Mamroud:2022msu}.} ~\footnote{It would be also interesting to study how the defects recently studied in~\cite{Chen:2023lzq} deform the giant-graviton expansions. }

The paper is organized as follows. After a summary of results, in section~\ref{sec:LargeChargeCount} we explain how the large-charge approximation simplifies the counting of states, and introduce tools that will be useful later on. In section~\ref{sec:TheIndexIntro} we introduce conventions, and the two representations of the superconformal index that we will study. In section~\ref{sec:IndexLargeChargeExpansion}, and as warm-up for the analysis of the giant graviton indices, we compute the large charge asymptotics of the superconformal index using a novel approach that turns out to be convenient for our scope. In section~\ref{sec:GiantGravitons} we apply the previously mentioned asymptotic tools to understand how the large-charge growth of the index is matched by the large-charge counting of giant gravitons for all~$N\,$ not just at~$N\gg 1\,$.~ In appendix~\ref{sec:AppB} we explain the role played by the choice of contour of integration~\cite{Lee:2022vig}\cite{Beccaria:2023zjw} in the large charge expansion. In appendix~\ref{sec:ExactRep} we move on to study the exact representation of the superconformal index put-forward in~\cite{Murthy:2022ien} and conclude explaining how exponentially large cancellations among individual giant graviton-like contributions are understood in the macrocanonical ensemble.

\subsection{Summary of main results}

Let us briefly introduce and summarize our main results. Detailed expositions will be presented in the main body of the paper.

~As mentioned in the introduction, the authors of~\cite{Imamura:2021ytr,Gaiotto:2021xce,Lee:2022vig} proposed that the superconformal index~$\mathcal{I}$ of four-dimensional~$U(N)$~$\mathcal{N}=4$ SYM on~$S^3$ can be expanded in a sum over indices~$\mathcal{I}_{\underline{n}}:=\mathcal{I}_{KK}\,\mathcal{I}_{n_1,n_2,n_3}$ of stacks of~$n_1$,~$n_2$,~$n_3$ giant graviton D3-branes wrapping three contractible~$S^3$-cycles in~$S^5\,$.~\footnote{The fact the cycles are contractible implies the existence of tachyons: the low energy spectrum of this D3-branes is rather different from that of~$\mathcal{N}=4$ SYM.} The details of this proposal will be given in subsection~\ref{sec:GGProp}. Schematically, it looks as follows
\begin{equation}\label{eq:MacroGGProposal}
\mathcal{I}(\tf)\, \underset{?}{=} 
\,\mathcal{I}_{gg}(\tf)\,:=\,\sum_{n_1,n_2,n_3} \mathcal{I}_{\underline{n}}(\tf)\,.
\end{equation}
In this relation~$\tf$ denotes the set of chemical potentials~$\{\varphi_1,\varphi_2,\varphi_3,\tau\}$~\footnote{Later on we will use the convention~$\Delta_{I}\,=\,-\,2\pi \text{i}\varphi_{I}\,$, $I=1,2,3\,$, and~$\omega_1\,=\,-\,2 \pi \text{i}\tau$, after fixing~$\omega_2 \to \pm 2\pi \text{i}\,+\, \Delta_1+\Delta_2+\Delta_3\,-\,\omega_1 \,$.} dual to the charge operators
\begin{equation}\label{eq:Charges}
\mathfrak{Q}\,=\,\{\mathfrak{Q}_1,\mathfrak{Q}_2,\mathfrak{Q}_3,\mathfrak{J}\}\,.
\end{equation}
~\footnote{This is the same set of charges defined in~\eqref{eq:ChargesB} and that will be denoted as~$\underline{\widetilde{Q}^\prime}$ in section~\ref{sec:GiantGravitons}.}~The three spin-twisted~$R$-charges~$\mathfrak{Q}_{1,2,3}$ and the right spin that we denote here as~$\mathfrak{J}\,$, respectively.~\footnote{ By $R$-charges and spin we refer to the charges that have such an interpretation from the perspective of the $4d$~$\mathcal{N}=4$ SYM leaving in the boundary of~$AdS_5$. From the perspective of the giant branes the meaning of $R$-charge and spin is exchanged.} From now on we call them R-charges and spin, respectively.

Let the lattice of eigenvalues of the operators~\eqref{eq:Charges} over a basis of eigenstates spanning the space-of-(BPS)-states in~$\mathcal{N}=4$ SYM be
\begin{equation}
\mathcal{S}_{\mathcal{N}=4}=\{\mathfrak{Q}\}
\end{equation}
~\footnote{Purposely denoted with the same letter as the operators. We hope this does not create much confusion in the reader.} Let the lattice of eigenvalues of~\eqref{eq:Charges} over a basis of eigenstates in the space-of-(BPS)states of the~$n$-brane theory be
\begin{equation}
\mathcal{S}^{(n)}_{gg}=\{\mathfrak{Q}^{(n)}_{gg}\}
\end{equation}
Let us denote the union of all possible BPS lattices of charges  of $n$-brane theories as
\begin{equation}
S_{gg} \,=\,\{\mathfrak{Q}_{gg}\}\,=\underset{n}{\cup} S^{(n)}_{gg}\,.
\end{equation}
~\footnote{All these three charge lattices can be projected in~$\mathbb{R}^4\,$(with degeneracies). They are, closely related to weight lattices of~$SO(6)\times SO(4)\,$.} With these definitions in mind,  we put forward the following proposal to test giant graviton identities like~\eqref{eq:MacroGGProposal}. The indices of $\mathcal{N}=4$ SYM and of the proposed~$n$-giant brane theories can be encoded in formal Fourier expansions~\footnote{Their truncations are finite Fourier series. } at the domain~$\tf =\widetilde{\tf} \,\in\, \mathbb{R}^4$ (to avoid issues with convergence in the discussion below one can freely replace the index by its truncation to the finite sum of terms necessary to count states at certain level of charges )
\begin{equation}
\mathcal{I}(\tf) \,{=}\, \sum_{\mathfrak{Q}} \,a(\mathfrak{Q}) e^{2\pi\i \tf \mathfrak{Q}}\,,\qquad  \mathcal{I}_{\underline{n}}(\widetilde{\tf}) = \sum_{Q_{gg}^{(n)}} \,a_{\underline{n}}(\mathfrak{Q}^{(n)}_{gg}) e^{2\pi\i \widetilde{\tf} \mathfrak{Q}^{(n)}_{gg}}\,.
\end{equation}
The Fourier expansion coefficients of~$\mathcal{I}(\tf)$ can be computed by computing the Laurent expansion of its matrix integral (plethystic) representation around~$\mathfrak{q}\,=\,e^{2\pi\text{i}\tf}\,=\,0$ and they are bound to be integer numbers which can be either positive or negative. The Fourier expansion coefficients of~$\mathcal{I}_{\underline{n}}(\tf)$ can be computed from Laurent expansions of their plethystic representation (and truncations of it) by \underline{carefully expanding} its plethystic representation around~$\widetilde{\mathfrak{q}}=e^{2\pi\text{i}\widetilde{t}}=0\,$. By carefully expanding, we mean that we only take Laurent expansions in~$\widetilde{\mathfrak{q}}$'s around the origin when they appear as second or third arguments in the elliptic functions~$\Gamma_e$ and~$\theta_0$ that define the integrand of $\mathcal{I}_{\underline{n}}(\widetilde{\tf})$ (and that will be reported in~\eqref{eq:4dAdjointContribution} and\eqref{eq:2dBiadjoint index}).~\footnote{These are the elliptic (modular-like) parameters appearing after the semicolons.} Then, after integrating over gauge rapidities one obtains the~$a_{\underline{n}}$'s, which are also bound to be integer numbers that can be either positive or negative.~\footnote{In particular the Fourier series computed in this way will not start with~$1$ as for the usual index. That is because of the presence of tachyons. These contributions cancel out after suming over~$n$, provided one has correctly integrated out gauge rapidities.}

The total sum of giant brane indices~\eqref{eq:MacroGGProposal} can be also written as a Fourier series
\begin{equation}
\mathcal{I}_{gg}(\tf)\,=\,\sum_{\mathfrak{Q}_{gg}}a_{gg}(\mathfrak{Q}_{gg}) e^{2\pi\i \tf \mathfrak{Q}_{gg}}\,,
\end{equation}
where
\begin{equation}
a_{gg}(\mathfrak{Q}_{gg})\,:=\,\sum_{\underline{n}}a_{\underline{n}}(\mathfrak{Q}_{gg})\,.
\end{equation}
~{ For later convenience it should be said that the sum over~$\underline{n}$ in~\eqref{eq:MicroGGProposal} is not a series because~$a_{\underline{n}}(\mathfrak{Q}_{gg})$ vanishes for large enough values~$\underline{n}\,$, at a fixed~$\mathfrak{Q}_{gg}\,$.~\footnote{\label{fn:Truncation} This is because by definition the generating function of the integer number~$a_{\underline{n}}(\mathfrak{Q})$ is a~$\mathfrak{q}$-series that starts at a power larger than~$\mathfrak{q}^{n N}$. Thus, schematically speaking, at any~$\mathfrak{Q}$ the integrals~\eqref{eq:ALoc} that define the microcanonical indices~$a_{\underline{n}}(\mathfrak{Q})$ are forced to vanish for every~$\underline{n}$  larger enough than~$\mathfrak{Q}/N\,$.}}

It is clear that~$\mathcal{S}_{gg}$ is much larger than~$\mathcal{S}_{\mathcal{N}=4}\,$, and also that a necessary condition for the equality~\eqref{eq:MacroGGProposal} to hold is~$\mathcal{S}_{\mathcal{N}=4}\subset\mathcal{S}_{gg}\,$. Our discussion above implies that a microcanonical version of the proposal~\eqref{eq:MacroGGProposal} is
\begin{equation}\label{eq:MicroGGProposal}
a_{gg}(\mathfrak{Q}_{gg})\,:=\,\sum_{\underline{n}}a_{\underline{n}}(\mathfrak{Q}_{gg})\,\underset{?}{=}\,\begin{cases} a(\mathfrak{Q}) \,, \quad \text{if} \quad \mathfrak{Q}_{gg}\,=\,\mathfrak{Q}\,\in\,\mathcal{S}_{\mathcal{N}=4} \\\,0 \,,\qquad \,\quad \text{otherwise}
\end{cases}\,.
\end{equation}
Our proposal~\eqref{eq:MicroGGProposal} to test~\eqref{eq:MacroGGProposal} says that the sum over giant graviton numbers~$\underline{n}$ must project the BPS giant graviton spectrum~$\mathcal{S}_{gg}$ to the much smaller gauge-theory spectrum~$\mathcal{S}_{\mathcal{N}=4}\,$ of BPS states. In forthcoming work we will study~\eqref{eq:MicroGGProposal} at finite values of charges~$\mathfrak{Q}\,$.

~As said before, the proposal~\eqref{eq:MacroGGProposal} has been checked for small enough values of~$\mathfrak{Q}\sim N$~\cite{Imamura:2021ytr}\cite{Lee:2022vig}. Our goal in this paper is to show that at large charges~$\mathfrak{Q}\to\infty$ (and for all~$N$) a precise and more general version of the following asymptotic relation holds~\footnote{The precise definition of the symbol~$\sim$ will be explained below.}
\begin{equation}\label{eq:GGSummaryAssympt}
|\sum_{\underline{n}}\,a_{\underline{n}}(\mathfrak{Q})|\,\sim\, |a(\mathfrak{Q})|\, \sim\,  e^{\left(\sqrt{3}\right) 3^{1/3}\pi  \,c\,\,{\mathfrak{J}^{2/3}}{ N^{2/3}}}\,.
\end{equation}
This is, that at large charges and for all~$N$ the sum over the giant graviton microcanonical indices~$a_{\underline{n}}$ evaluated at charges~$\mathfrak{Q}\in \mathcal{S}_{\mathcal{N}=4}$ matches the exponential growth of $\frac{1}{16}$-BPS states at charges~$\mathfrak{Q}\,$.

In this relation the quantity~$c$ is an order~$1$ real contribution that depends on how fast the spin~$\mathfrak{J}$ grows in relation to the~$R$-charges, we will come back to comment on it below (e.g. a particularly simple case where~$c$ is simply a c-number will be reported in~\eqref{eq:AsymptoticsLargeSpin} but our results cover more general cases).

To illustrate, let us briefly explain how the particular result~\eqref{eq:GGSummaryAssympt} is obtained. In subsection~\ref{sec:LCLocalization} we will introduce a large-charge localization Lemma that will help us to compute~\emph{localized} contributions~$a^{loc}_{\pm,\underline{n}}$ to the giant graviton index. The microcanonical index of giant gravitons is defined as follows
\begin{equation}\label{eq:ALoc}
a_{\underline{n}}(\mathfrak{Q}):=\int_{\Gamma} d\tf\,\mathcal{I}_{\underline{n}}(\tf) \,e^{-2\pi \i \tf \mathfrak{Q}}\,.
\end{equation}
In this equation~$\Gamma$ is a period (of the integrand) in the four-dimensional moduli space of chemical potentials, denoted as~$\tf\,$. By saying that~$\Gamma$ is a period we mean that it is a cycle of periodicity of the integrand (following from quantization and periodicity of the dual flavour charge lattice~$\mathcal{S}^{(n)}_{gg}$). It is important to say that~$\Gamma$ is independent of the giant graviton number~$n\,$.~\footnote{That there is a common cycle for all the charge lattices~$\mathcal{S}^{(n)}_{g}$ can be seen from the definitions of $\mathcal{I}_{\underline{n}}(\tf)$ (given in~\eqref{eq:IntegralGGindex}). Namely, such integral is invariant under changes of rapidities~$\tf\to \tf +1$ for all~$n\,$.} In this equation the gauge-rapidities have been already integrated out using saddle-point approximation.~\footnote{The contour of integration over gauge rapidities could depend on~$n\,$, but in the large-charge approximation it is enough to localize its integral to its leading saddle point~$u_{ab}^\star=0$ which is independent on~$n\,$. Thus, effectively, if there is such dependence it disappears at large charges. The details on our conclusions regarding the integration over gauge rapidities are presented in appendix~\ref{sec:AppB}.}

The~$a^{loc}_{\pm,\underline{n}}$ are two equally-dominating contributions to~\eqref{eq:ALoc} in its asymptotic expansion at \underline{large~$R$-charges},~\underline{assuming generic growth for the spin}~$\mathfrak{J}\,$,  and any~$\underline{n}\,$
\begin{equation}
a_{\underline{n}}(\mathfrak{Q})\,\sim\, a^{loc}_{+,\underline{n}}(\mathfrak{Q})+a^{loc}_{-,\underline{n}}(\mathfrak{Q})\,.
\end{equation}
These two contributions~$\pm$ are complex conjugated to each other
\begin{equation}
a^{loc}_{\pm,\underline{n}}(\mathfrak{Q})=\int_{\Gamma_\pm} d\tf\,\mathcal{I}^{(\pm)}_{\underline{n}}(\tf) \,e^{-2\pi \i \tf \mathfrak{Q}}\,.
\end{equation}
The large-charge localization Lemma of~\ref{sec:LCLocalization} will tell us that the contours~$\Gamma_\pm$ can be understood as small subpieces of the contour~$\Gamma\,$, centered at the leading essential singularities of the integrand~$\mathcal{I}_{\underline{n}}(\tf)\,$ in the moduli space of chemical potentials~$\tf\,$. These singularities are located at specific values of the chemical potentials~$\underline{\varphi}=\{\varphi_1,\varphi_2,\varphi_3\}$ dual to $R$-charges. In the cases of interest to us, there are two types of such divergences that we label by the two choices of signs~$\pm\,$. The localized form of the integrands,~$\mathcal{I}^{(\pm)}_{\underline{n}}(\tf)$, are the leading asymptotic expansions of~$\mathcal{I}_{\underline{n}}(\tf)$ around the essential singularities~$\pm\,$.

After commuting the sum over~$n$ with the integrals over~$\tau$ in~\eqref{eq:ALoc} one obtains
\begin{equation}
\sum_{\underline{n}} a^{loc}_{\pm,\underline{n}}(\mathfrak{Q}) =\int_{\Gamma_\pm} d\tf \,\Bigl(\sum_{\underline{n}}\mathcal{I}^{\pm}_{\underline{n}}(\tf)\Bigr)\,e^{-2\pi \i \tf \mathfrak{Q}}\,.
\end{equation}
As it will be explained in the main body of the paper, the sum over~$\underline{n}\,$ can be replaced by an integral  over a compact domain whose asymptotic behaviour around the singularities~$\pm$ (and at large $R$-charges) can be obtained by the saddle point method
\begin{equation}
\sum_{\underline{n}} \mathcal{I}^{(\pm)}_{\underline{n}}(\tf) \,\sim\, \mathcal{I}^{(\pm)}_{\underline{n}^\star}(\tf)\,.
\end{equation}
The saddle point condition ends up taking a simple linear form that fixes~$n=\underline{n}^\star:=\underline{n}^\star(\tf)\,$ as a function of~$\tf$. The function~$n^\star(\tf)$ is defined by a linear relation of the schematic form
\begin{equation}\label{eq:SPGG}
\underline{\varphi} \,\cdot\,\underline{n}^\star_{\pm} \,=\frac{N}{\tau^2}f_{\pm}(\tf)\,,
\end{equation}
where $f_{\pm}(\tf)$ are cubic polynomials in~$\tf$ such that 
$|f_{\pm}(\tf)|$ is finite and non zero as $\tau\to0\,$. The explicit form of this equation will be specified in the main body of the paper.~\footnote{e.g. the simplest possible example comes from equation~\eqref{eq:LinearRelationsN}+\eqref{eq:FinalT} after constraining~$\Delta_3\, =\, -\Delta_{1}-\Delta_{2}-2\omega_1  \,\mp\,2\pi\i\,$, and $\Delta_2 =\Delta_1\,$, and then identifying~$\omega_1 \,\to\, -2\pi\i \tau$ and~$\Delta_1\,\to\,-2\pi\i \varphi\,$.}

To compute the asymptotic behaviour of~
$\int_{\Gamma_{\pm}} d\tf\, \mathcal{I}^{(\pm)}_{\underline{n}^\star_{\pm}}(\tf) e^{-2\pi\i \tf Q}$
at large~$\mathfrak{Q}$, not just at large $R$-charges as before, but also at large spin~$\mathfrak{J}\,$, we use again a saddle point evaluation
\begin{equation}\label{eq:Aloc}
\int_{\Gamma_{\pm}} d\tf\, \mathcal{I}^{(\pm)}_{\underline{n}^\star_{\pm}}(\tf) e^{-2\pi\i \tf \mathfrak{Q}}\,\sim\, \mathcal{I}^{(\pm)}_{\underline{n}^\star_{\pm}}(\tf) e^{-2\pi\i \tf \mathfrak{Q}}\,=:\,a^{loc}_{\pm,\underline{n}^\star_\pm}\,.
\end{equation}
This time the saddle-point condition fixes the chemical potentials~$\tf$, and in particular the one dual to spin~$\mathfrak{J}$,~$\tau$, to a function of charges~$\mathfrak{Q}$ (led by the spin~$\mathfrak{J}$)
\begin{equation}\label{eq:TauQ}
\tau\,=\,\tau^\star_{\pm}(\mathfrak{Q})\,= c_{\tau_{\pm}} \frac{N^{2/3}}{\mathfrak{J}^{1/3}}\,,
\end{equation}
with~$c_{\tau_{\pm}}$ being order~$1$ contributions that depend on how fast the spin~$\mathfrak{J}$ grows in relation to the $R$-charges. At this point we simply collect results and obtain
\begin{equation}
\sum_{\underline{n}}\,a_{\underline{n}}(\mathfrak{Q})\, \sim\, a^{loc}_{+,\underline{n}^\star_+}+a^{loc}_{-,\underline{n}^\star_-}
\end{equation}
which after trivial algebraic manipulations leads to the announced asymptotic relations~\eqref{eq:GGSummaryAssympt}. 

By composing~\eqref{eq:SPGG} with~\eqref{eq:TauQ} we obtain the scaling properties of the complex saddle point configuration that dominates the sum over giant gravitons
\begin{equation}\label{eq:CondensateValue0}
\underline{c}_{1,\pm}\cdot \underline{n}_{\pm}^\star\,\sim\,c_{2,\pm}\, \frac{\mathfrak{J}^{2/3}}{N^{1/3}}\,.
\end{equation}
In this equation~$\underline{c}_{1,\pm}$ and~$c_{2,\pm}\,$, again, represent order~$1$~\footnote{If one fixes the angular momentum~$\mathfrak{J}$ to be small and instead considers large $R$-charges then the conclusions are different (See the discussion in the last paragraph of subsection~\ref{sub:GeneralCharges}). In this paper we will not study in detail this other domain of the spectrum of charges. } contributions that depend on how fast the spin~$\mathfrak{J}$ grows in comparison with the $R$-charges. In particular, we note that~$c_{2,\pm}$ are complex quantities.~\footnote{They are related to the constant~$c$ in~\eqref{eq:GGSummaryAssympt}.} 

In summary, the asymptotic relations~\eqref{eq:GGSummaryAssympt} show that the giant graviton proposals of~\cite{Imamura:2021ytr}\cite{Gaiotto:2021xce,Lee:2022vig} capture the large charge (for all~$N$) asymptotic growth of the microcanonical superconformal index. Moreover, using the large-charge localization lemma we show that in the \emph{holographic} scaling limit
\begin{equation}\label{eq:SugraScaling}
\mathfrak{Q}\,\to\, \infty\,, \, \qquad \frac{\mathfrak{Q}}{N^2} \,=\,\text{fixed}\,,
\end{equation}
the giant-graviton representations exactly recover the entropy of $\frac{1}{16}$-BPS black holes in~$AdS_5\,$, at generic values of the ratio $\frac{\text{Entropy}}{N^2}\,$, where in our conventions~$G_5=\frac{2N^2}{\pi}\,$.~\footnote{Note that for the \emph{black hole scaling}~\eqref{eq:SugraScaling} the absolute value of the complex saddle points~\eqref{eq:CondensateValue0} becomes of order~$N$ as expected.} Namely,
\begin{equation}
|a_{gg}(\mathfrak{Q})|\,\to\,d_{BH}(\mathfrak{Q})\,:=\, e^{S_{BH}(\mathfrak{Q})}
\end{equation}
upon imposition of the non-linear constraint among charges that in the bulk corresponds to avoiding CTC's. More on this, will be said below.

\section{State-counting at large charges}\label{sec:LargeChargeCount}

The large charge approximation has been a useful tool in varied contexts as, for example, the computation of anomalous dimensions, correlation functions, partition functions, the conformal bootstrap, cf. ~\cite{Berenstein:2002jq,Alday:2007mf,Basso:2006nk,Komargodski:2012ek,Fitzpatrick:2012yx}~\cite{Alvarez_Gaume_2021}. Let us explain briefly how this tool applies to the counting of operators in quantum statistical system.~\footnote{In the context of superconformal and topologically twisted indices a particular case of one such large-charge approximation known as the Cardy-like approximation has been thoroughly studied in the last few years ~\cite{Choi:2018hmj,Honda:2019cio,ArabiArdehali:2019tdm,Kim:2019yrz,Cabo-Bizet:2019osg} ~\cite{GonzalezLezcano:2020yeb,Goldstein:2020yvj,Amariti:2020jyx,Amariti:2021ubd,Cassani:2021fyv,ArabiArdehali:2021nsx,Jejjala:2021hlt,Ardehali:2021irq,Cabo-Bizet:2021plf,Cabo-Bizet:2021jar,Jejjala:2022lrm,Amariti:2023rci}\cite{Choi:2019zpz, Nian:2019pxj,GonzalezLezcano:2022hcf,BenettiGenolini:2023rkq}\cite{Amariti:2023ygn}. Perturbative corrections to the leading asymptotic behaviour of four-dimensional~$\mathcal{N}=1$ superconformal indices in the large charge expansion have been exactly matched against higher-derivative corrections to the leading semiclassical onshell action of~$AdS_5$ black holes in the relevant dual supergravities~\cite{Bobev:2022bjm,Cassani:2022lrk,Cassani:2023vsa}. It would be very interesting to study the large charge expansion of the partition function at non-vanishing coupling\,, of say $\mathcal{N}=4$ SYM\,, at least in near-BPS sectors~\cite{Berkooz:2006wc,Berkooz:2008gc,Chang:2023zqk}\cite{Budzik:2023vtr}\cite{Caetano:2023zwe}. The goal being to try to extract universal lessons that could be compared against recent holographic expectations e.g.~\cite{Boruch:2022tno,Turiaci:2023jfa}. }

Consider a $2\pi$-periodic complex function~$f=f(x)=f(x+2\pi)$ with a set of singularities at~$x=x_{a,sing}\in \mathbb{R}$\,, $a=1,2,\ldots$, such that
\begin{equation}\label{eq:AssumptionFunctions}
f\left(x_{a,sing}+\frac{\delta x}{\Lambda }\right)\underset{\Lambda\to\infty}{\sim} \Lambda ^{n} \widetilde{f}_{a}\left(\delta
   x\right)\,, \qquad n\,>\,0\,,
\end{equation}
where the definition of the symbol~$\underset{\Lambda\to 0}{\sim}\,$, which denotes an asymptotic relation, is given in appendix~\ref{app:AsymptRelators}.

Let us consider the \emph{average}
\begin{equation}
d(Q):=\int_\Gamma dx\ e^{f(x)\,-\, \i x Q},\,\qquad Q\in \mathbb{Z}\,,
\end{equation} 
over a cycle~$\Gamma$ that can be decomposed in an integral combination of Lefschetz thimbles~$\Gamma_{x^\star}$ ending at saddle points~$x=x^\star$ of the exponent~$f(x)+\i x Q\,$. 

Under these assumptions, the leading asymptotic behaviour of~$d(Q)$ in the large charge approximation 
\begin{equation}\label{eq:ScalingLimitIntroduction}
Q\,=\,q \Lambda ^{n+1}\,,\, \Lambda\,\gg\, 1\,,\,q\,=\,\text{finite},
\end{equation}
is determined by the asymptotic form of the saddle points~$x^*\,$, which in the large charge regime become infinitelly close to the singularities~$x_{a,sing}$,
\begin{equation}
x^* = x_{a,sing} + \frac{\delta x^\star}{\Lambda},
\end{equation}
with
\begin{equation}\label{eq:SaddleFluctuations}
\delta x^* : \widetilde{f}_a^\prime(\delta x^*)\,-\,\i q\,=\,0\,.
\end{equation}
Then, under the previous assumptions and in the large charge approximation, we have
\begin{equation}
d(Q)\,\simExp\, e^{\Lambda^n \bigl(\widetilde{f}_{a^\star}(\delta x^\star_{a^\star})\,-\,\i(x_{a^\star,sing} \Lambda+\delta x^\star_{a^\star}) q\bigr)}
\end{equation} 
where~$a^\star$, $\delta x_{a^\star}^\star$ label the singularity $a=a^*$ and the solution~$\delta x^\star=\delta x_{a^\star}^\star$ of~\eqref{eq:SaddleFluctuations}, respectively, that maximize the real part of the exponent~$\Lambda^n \bigl(\widetilde{f}_{a}(\delta x^\star)-\i(x_{a,sing} \Lambda+\delta x^\star) q\bigr)\,$. The definition of the symbol~$\simExp\,$, which denotes an asymptotic relation, is given in appendix~\ref{app:AsymptRelators}.

\subsection{An illustrative  example}

As an example, we briefly discuss a simple toy model. 
Let us assume~$Q=q\Lambda^3$ to be a positive integer,
\begin{equation}
f(x)\,:=\,-\pi \i \csc ^2\bigg(\frac{x}{2}\bigg)\,.
\end{equation}
In this case, we have $n=2\,$, and~
\begin{equation}
\widetilde{f}_{a}(x)\,:=\,-\frac{4\pi\i}{x^2}\,.
\end{equation}
Let us fix the integration cycle as follows
\begin{equation}
\Gamma\,:=\,\Bigl\{y\,\in\, \mathbb{C}\,|\, y\,=\,x\,+\,\frac{\left(-1+\i \sqrt{3}\right) \sqrt[3]{\pi }}{\sqrt[3]{Q}}\,,\,x\,\in\, [0,2\pi) \Bigr\}\,.
\end{equation}
Obviously~$d(Q)$ is convergent, because~$\Gamma$ is compact and it does not intersect the set of singularities 
\begin{equation}
x_{a,sing}\,=\, 0\,+\, 2\pi\i (a-1)\,,\qquad a\,=\,1\,,\,2\,,\ldots\,.
\end{equation}
There are three saddle points around each singularity~$x_{a,sing}$. At large charge, they take the form
\begin{equation}\label{eq:SaddlesToyExample}
\delta x^\star\,:=\,\left\{\frac{\left(-1-i \sqrt{3}\right) \sqrt[3]{\pi }}{\sqrt[3]{Q}}\,,\,\frac{2
   \sqrt[3]{\pi }}{\sqrt[3]{Q}}\,,\,\frac{\left(-1+i \sqrt{3}\right) \sqrt[3]{\pi
   }}{\sqrt[3]{Q}}\right\}\, .
\end{equation}
Notice that we have engineered the integration cycle~$\Gamma$ to intersect the last saddle. This guarranties~$|d(Q)|\nearrow\infty$ for~$Q\nearrow \infty\,$. Indeed, one can check numerically for~$Q\sim100$ and larger, that the integral~$d(Q)$ localizes to the integrals over the infinitesimal vicinity of the contour~$\Gamma$ that becomes infinitely close to the singularities. More precisely, at large charges,~$d(Q)$ localizes to its saddle-point approximation which is, at leading order,~\footnote{Having into consideration the contributions from the two saddles whose thimbles are intersected by the contour of integration~$[0,2\pi]$, labelled by~$a=1,2,$ and one-loop logarithmic corrections about each one of them, one obtains an improvement of~\eqref{eq:SaddlePoints}. Comparing absolute values for simplicity, as we will eventually do, one obtains 
\begin{equation}
|d(Q)|\, \underset{\Lambda\to \infty}{\to}\,2\,\times\,\frac{\pi ^{2/3}}{\sqrt{3} Q^{2/3}} \,\times\, e^{\frac{\pi ^{2/3} e^{\frac{3}{2} \sqrt{3} \sqrt[3]{\pi } \sqrt[3]{Q^2}}}{\sqrt{3} Q^{2/3}}}\,.
\end{equation} 
Now the quotient between the left and right-hand sides is~$1$ at~$\Lambda\to \infty$.}
\begin{equation}\label{eq:SaddlePoints}
d(Q)\,\simExp\,\exp \left(\frac{6 \i \sqrt[3]{\pi } Q^{2/3}}{\left(-\i+\sqrt{3}\right)^2}\right)\,,  \qquad\text{for}\ Q\,\text{a positive integer}\,.
\end{equation}
The prediction coming from the saddle point intersected by~$\Gamma$ for~$Q\,\ll\,0\,$ is~$d(Q)\,=\,0\,$, which happens to be the correct answer as well, i.e., the answer we computed from the direct numerical evaluation of the integral~$d(Q)$ at~$Q\,\ll\,0\,$. This happens because the cycle~$\Gamma$ has zero intersection number with the Lefschetz thimble ending at the saddle point that produces exponential growth of the quantity~$e^{f(x^\star)-\i x^\star Q}$ at~$Q\ll 0\,$, which is the first one in~\eqref{eq:SaddlesToyExample}.

\subsection{Application to the  superconformal index}

In the case of the superconformal index, 
we are interested in computing integrals over multidimensional cycles of the form
\begin{equation}\label{eq:MicrocanonicalIndexIntro}
d(\underline{Q}^\prime)\,=\, \int_\Gamma \, d\underline{x} \,\int_{\Gamma_{gauge}}d\underline{u}\,e^{-S_{\text{eff}}(\underline{x};\underline{u}) -\i\underline{x}\cdot \underline{Q}^\prime}\,,
\end{equation}
at \underline{large and positive integer} charges~$ \underline{Q}^\prime\,$. Here, $\underline{x}$ denotes the set of four chemical potentials dual to four global charges~$\underline{Q}^\prime$.~\footnote{In terms of the usual notation for the chemical potentials of $\mathcal{N}=4$ SYM $\{{\Delta_1},{\Delta_2},{\Delta_3},{\omega_1},{\omega_2}\}$\,, with~$\Delta_3 =-\Delta_1-\Delta_2+\omega_1+\omega_2\,$, we define~$-\i x_{1,2,3}\,=\,\Delta_{1,2,3}$, and $-\i x_{4,5}\,=\,\omega_{1,2}\,.$}~$\Gamma$ and~$\Gamma_{gauge}$ are integration cycles that we assume can be decomposed in integral combinations of Lefschetz thimbles of $S_{\text{eff}}(\underline{x};\underline{u})\,$. The effective action~$-S_{\text{eff}}(\underline{x};\underline{u})$ is the logarithm of the integrand of the superconformal index~$\mathcal{I}(\underline{x})\,:=\int_{\Gamma_{gauge}} d\underline{u}\,e^{-S_{\text{eff}}(\underline{x};\underline{u})}\,$. As it will be shown below,~$S_{\text{eff}}(\underline{x};\underline{u})$ has leading singularities located at
\begin{equation}\label{eq:Singularities}
x_{4,sing}\,,\,x_{5,sing} \,=\, 2\pi\i (a_{4,5}-1)\,.
\end{equation}
The free energy takes the form
\begin{equation}\begin{split}
S_{\text{eff}}\Bigl(x_{1},x_2,{x}_{4,sing}+\frac{\delta x_4}{\Lambda},{x}_{5,sing}+\frac{\delta x_5}{\Lambda};\underline{u}\Bigr)&\,\underset{\Lambda\to \infty}{\sim}\, \widetilde{s}( x_1, x_2, \frac{\delta x_4}{\Lambda},\frac{\delta x_5}{\Lambda};\underline{u})\, ,
\end{split}
\end{equation}
where
\begin{equation}
\widetilde{s}( x_1, x_2, \frac{\delta x_4}{\Lambda},\frac{\delta x_5}{\Lambda};\underline{u})= \Lambda^{2}\, \widetilde{s}_\Lambda( x_1, x_2, \delta x_4,\delta x_5;\underline{u})\,,
\end{equation}
and most importantly, these two functions are~\emph{asymptotically-equal} 
\begin{equation}\label{eq:AsymptoticConditionHomogen}
\frac{\widetilde{s}_\Lambda( x_1, x_2, \delta x_4,\delta x_5;\underline{u})}{\widetilde{s}( x_1, x_2, \delta x_4,\delta x_5;\underline{u})} \underset{\Lambda\to\infty}{\sim} 1\,.
\end{equation}
Notice that the singularities~\eqref{eq:Singularities} are not points but a 2-cycle~$\Gamma_{1,2}$ spanned by the variables~$x_{1,2}$\,(times the integration cycle~$\Gamma_{gauge}$ over the gauge potentials). Then, following analogous reasoning as before and using~\eqref{eq:AsymptoticConditionHomogen}, it follows that at large charges
\begin{equation}\label{eq:ScalingLimitIntroduction}
Q^\prime_{1,2}\,=\,q^\prime_{1,2}\Lambda^2\,, \,Q^\prime_{4,5}\,=\,q^\prime_{4,5}\Lambda^3\, ,
\end{equation}
one has
\begin{equation}\label{eq:DegeneracySingularLocus}
d(\underline{Q}) \,\simExp\,\exp\Bigl(\Lambda^2\Bigl(-\widetilde{s}( x_1^\star, x_2^\star, \delta x_3^\star, \delta x_4^\star;\underline{u}^\star)-\i {x}^\star_1 q^\prime_1-\i {x}^\star_2 q^\prime_2-\i \delta {x}^\star_4 q^\prime_4-\i \delta{x}^\star_5 q^\prime_5\Bigr)\Bigr)\, ,
\end{equation}
where the~$\star$ denotes one of the saddle points that maximize the real part of the exponent in~\eqref{eq:DegeneracySingularLocus} among those intersected by the Lefschetz thimbles that compose the original cycle~$\Gamma\otimes\Gamma_{c}$. As in the simplest toy example before, such saddle points will be asymptotically close to the singular locus~$\Gamma_{1,2}\times\Gamma_{gauge}\,$ of~$S_{\text{eff}}(\underline{x};\underline{u})\,$ in the scaling limit~\eqref{eq:ScalingLimitIntroduction}.
Note that the first perturbative corrections in the~$\frac{1}{\Lambda}$-expansion are also captured by~\eqref{eq:DegeneracySingularLocus}. They are encoded in the Laurent expansion of $\widetilde{s}(\underline{x})$ around~$x_{4}=x_5=0$.

\subsection{Large-charge limit as a localization mechanism}\label{sec:LCLocalization}

Let us come back to a generic function~$f=f(x)$ 
with a regular singularity~$x_{sing}$
\begin{equation}
\i x_{sing} Q= 2\pi\i n\,, \qquad n\in\mathbb{Z}\,,
\end{equation}
such that
\begin{equation}\label{eq:AssumptionFunctions2}
f\left(x_{sing}+\frac{\delta x}{\Lambda }\right)\underset{\Lambda\to\infty}{\sim} \Lambda ^{n} \widetilde{f}_{\Lambda}\left(\delta
   x\right)\,, \qquad n\,>\,0\, ,
\end{equation}
with a single dominating saddle~$x^\star=x_{sing}+\frac{\delta x^\star}{\Lambda}$.
Let us further assume that given the equality
\begin{equation}\label{eq:HomogeneityProperty3}
\Lambda^n \widetilde{f}_\Lambda(x)= \widetilde{f}\left(\frac{x}{\Lambda}\right)\,,
\end{equation}
the functions~$\widetilde{f}_{\Lambda}(x)$ and~$\widetilde{f}(x)$ are asymptotically-equal~\eqref{eq:AsymptoticConditionHomogen}
\begin{equation}
\frac{\widetilde{f}_\Lambda(x)}{\widetilde{f}( x)} \,\underset{\Lambda\to\infty}{\sim}\, 1\,.
\end{equation}
Then, as we explained before, in the large-charge scaling limit
\begin{equation}
Q\,=\,q\Lambda^{n+1}\,,
\end{equation}
it follows that
\begin{equation}\label{eq:IntegralScaling0}
\int_{\Gamma} dx e^{f(x)-\i x Q}\,\simExp\, \int_{\Gamma_{\delta x^\star}} d(\delta x) e^{\Lambda^n (\widetilde{f}_\Lambda(\delta x)-\i \delta x q)}\,,
\end{equation}
where~$\Gamma_{\delta x^\star}$ is the Lefschetz thimble of~$\widetilde{f}_{\Lambda}(\delta x)-\i \delta x q$ intersecting the dominating saddle point~$\delta x^\star$.

After scaling the variable~$\delta x\to y \Lambda\,$, equations~\eqref{eq:IntegralScaling0} and~\eqref{eq:HomogeneityProperty3} imply
\begin{equation}
\int_{\Gamma} dx e^{f(x)-\i x Q}\,\simExp \,\int_{\Gamma_{y^\star}} dy e^{ \widetilde{f}(y)-\i y Q}\,, \qquad y^\star:=\frac{\delta x^\star}{\Lambda}\,,
\end{equation}
where~$\Gamma_{y^\star}$ is a Lefschetz thimble of~$\widetilde{f}(y)-\i y Q$ that ends up at the dominating saddle point~$y^\star\,$.~\footnote{Note that we have dropped out a factor of~$\Lambda$ which is subleading with respect to the~$e^{\Lambda^2}$-growth that comes from the exponential in the integrand.}

Thus, to compute the asymptotic behaviour of~$d(Q)$ at large values of charges
\begin{equation}
Q\sim \Lambda^{n+1}\, ,
\end{equation}
we only need to plug the asymptotic expansion of $\widetilde{f}(x)$ around~$x=0$
\begin{equation}
\widetilde{f}(x)= \frac{\widetilde{f}^{(-n)}}{x^n}+\frac{\widetilde{f}^{(-n+1)}}{x^{n-1}}\,+\,\text{O}(x^{2-n})\, ,
\end{equation}
into the integral
\begin{equation}
\int_{\Gamma_{y^\star}} dy e^{ \widetilde{f}(y)-\i y Q}\,.
\end{equation}
This integral will be called the \emph{large-charge-localization} or \emph{large-charge coarse grain} of the original integral~$\int_\Gamma e^{f(x)-\i x Q}\,$, and it is much simpler to study.
Roughly speaking, this localization mechanism tells us that at large charges the function~$f(x)$, which could be rather complicated, can be substituted by its asymptotic expansion~$\widetilde{f}(y)$ around the singularity~$y=0\,$, i.e., the singularity that attracts the leading saddle point~$y=y^\star\,$ at large charges. It should be also noted that the integration cycle needs also to be modified as indicated before. The subleading and perturbative terms in the asymptotic expansion of~$\widetilde{f}(y)$ give exact perturbative corrections to the leading prediction for the asymptotic growth of~$d(Q)$. 

The generalization of this localization mechanism to the case where~$f(\underline{x})$ depends on more than one variable (when the singularities can be not only points, but also cycles), is straightforward. 
For example, the microcanonical index~\eqref{eq:MicrocanonicalIndexIntro}, is such that the localized action~$\widetilde{s}$ (essentially the series expansion of the complete effective action about the leading singularity) 
\begin{equation}
\widetilde{s}\Bigl(x_1,x_2,\frac{x_4}{\Lambda},\frac{x_5}{\Lambda};\underline{u}\Bigr)
\,\underset{\Lambda\to \infty}{\sim}\,\Lambda^2 \widetilde{s}_{\Lambda}(\underline{x};\underline{u})\,,
\end{equation}
is (weakly) equal to~$\widetilde{s}_\Lambda$
\begin{equation}
\frac{\widetilde{s}_{\Lambda}(\underline{x};\underline{u})}{\widetilde{s}(\underline{x};\underline{u})}\, \underset{\Lambda\to \infty}{\sim}\,1\,.
\end{equation}
Then as a consequence of~\eqref{eq:DegeneracySingularLocus}
it follows the \emph{large-charge localization formula} or~\emph{lemma}:
\begin{equation}\label{eq:LocalizationFormula}
d(\underline{Q}) \,\simExp\,\int_{\Gamma_{\underline{y}^{\star},\underline{u}^\star}} d\underline{y} d\underline{u} \,e^{-\widetilde{s}(\underline{y};\underline{u})-\i \underline{y}\cdot \underline{Q}}\,.
\end{equation}
In this equation~$\Gamma_{\underline{y}^{\star},\underline{u}^\star}$ is a~$4+N$-dimensional integration contour. It is also a combination of Lefschetz thimbles of~$-\widetilde{s}(\underline{y};\underline{u})-\i \underline{y}\cdot \underline{Q}$ and it intersects the leading saddle point(s)~$\underline{y}^{\star}\,$,~$\underline{u}^\star\,$
\begin{equation}
\partial_{\underline{u}}\widetilde{s}(\underline{y};\underline{u})\Big|_{y=\underline{y}^{\star},\underline{u}=\underline{u}^\star}=0\,,\qquad \partial_{\underline{y}}\widetilde{s}(\underline{y};\underline{u})\Big|_{y=\underline{y}^{\star},\underline{u}=\underline{u}^\star}\,-\,\i \underline{Q}=0\, , 
\end{equation}
with intersection numbers defined by the decomposition of the original integration contour~$\Gamma\times \Gamma_{gauge}\,$ in terms of the Lefschetz thimbles associated to the original exponent~$S(\underline{y};\underline{u})\,-\,\i \underline{y}\cdot \underline{Q}$.

In conclusion, to compute the asymptotic behaviour of~$d(Q)$ at large values of charges we need, \underline{first}, to compute the asymptotic expansion of~$\widetilde{s}(y_1,y_2,y_4,y_5)$ around~$y_4=y_5=0$
\begin{equation}\label{eq:SeriesExpS}
\begin{split}
\widetilde{s}(\underline{y};\underline{u})= \frac{\widetilde{s}^{(1,1)}(y_1,y_2;\underline{u})}{y_4 y_5}+\frac{\widetilde{s}^{(1,0)}(y_1,y_2,y_4,y_5;\underline{u})}{y_4}\,+\,\frac{\widetilde{s}^{(0,1)}(y_1,y_2, y_4,y_5;\underline{u})}{y_5}\,+\,\text{subleading}\,
\end{split}
\end{equation}
which, by construction, is the same as the asymptotic expansion of the complete effective action~$S(\underline{y},\underline{u})$ around~$y_4=y_5=0$. \underline{Second}, we must compute the leading saddle point values~$\underline{y}^\star$ and~$\underline{u}^\star$ of the desired truncation of~\eqref{eq:SeriesExpS}. Then, at last, we obtain the following asymptotic formula
\begin{equation}\label{eq:LargeChargeLocalizationFormula}
d(\underline{Q})\,\simExp\,e^{-\widetilde{s}(\underline{y}^\star;\underline{u}^\star)-\i \underline{y}^\star\cdot \underline{Q}}\,.
\end{equation}

In the following sections we will use this recipe, and particularly its integral version, the large-charge localization formula~\eqref{eq:LocalizationFormula}, to compute asymptotic behaviours. 

\section{The $\frac{1}{16}$-BPS index at large charges}
\label{sec:TheIndexIntro}
The superconformal index of 4d $\mathcal{N}=4$ SYM on $\mathbb{R}\times S^3$ is defined as~\cite{Kinney:2005ej}
\begin{equation}\label{IndexInitial}
\mathcal{I}\,=\,\text{Tr}_{\mathcal{H}}\left[(-1)^F p_1^{J} p_2^{\overline{J}} w^{Q_1}_1 w^{Q_2}_2 w^{Q_3}_3\right]
\end{equation}
with the constraint
\begin{equation}\label{eq:ConstraintRapidities}
 \frac{w_1 w_2 w_3}{p_1 p_2}\,=\,1\,.
\end{equation}
Substituting it in~\eqref{IndexInitial} fixes the four-dimensional lattice of charges within the five-dimensional lattice spanned by~$\left\{J,\overline{J},Q_1,Q_2,Q_3\right\}$ that commutes with the two super (conformal) charges that define the index~$\mathcal{I}\,$.  

The commuting charges in~\eqref{IndexInitial} are defined as follows
\begin{equation}
\begin{split}
J&\,=\,E -J_L+J_R\,,\quad \bar{J}\,=\,E  
   -J_L-J_R\,, \\ Q_1&\,=\,-q_2-q_3\,,\quad Q_2\,=\,-q_1-q_3\,,\quad Q_3\,=\,-q_1-q_2\,,
   \end{split}
\end{equation}
in terms of the dilation operator~$E$, the left and right angular momenta~$J_{L,R}$ in the Cartan of the~$SO(4)=SU(2)\times SU(2)$ isometries of~$S^3$, and~$q_1$,~$q_2$ and~$q_3$ are the Cartan elements of the~$SO(6)$ R-symmetry.~\footnote{We use the conventions and values of charges of fundamental letters of e.g.~\cite{Chang:2013fba}.} The following definitions of rapidities and chemical potentials will be useful later on
\begin{equation}
\begin{split}
w_1&=e^{- \Delta _1}\,,\, w_2=e^{-  \Delta _2}\,,\, w_3\,=e^{-  \Delta _3}\\
   p_1&=e^{- \omega _1}\,,\, p_2=e^{-  \omega _2}\,.
\end{split}
\end{equation}
For gauge group~$U(N)$ the index can be written in the form~\cite{Kinney:2005ej}
\begin{equation}\label{eq:Index}
\mathcal{I}:=\oint \text{d}\mu \,\prod^{N}_{a=1} \prod^{N}_{b=1} \text{Pexp}\left(i(w;p_1,p_2) \,{U _{ab}}\right)\, ,
\end{equation}
where~$U_a$ is the a-th diagonal component of a diagonal unitary matrix, and~${U _{ab}}=U _{a}/{U _{b}}$. The measure in~\eqref{eq:Index} is defined as
\begin{equation}
\text{d}\mu\, :=\,\frac{1}{N!}\prod^{N}_{a=1} \,\frac{\text{d$U$}_{a}}{2\pi\text{i}U_a}\,\cdot\,\prod_{a\neq b=1}^N(1-{U_{ab}})\,,
\end{equation}
and
\begin{equation}
i(w;p_1,p_2)=i\left(w_1,w_2,w_2;p_1,p_2\right)\,:=\,1-\frac{\left(1-w_1\right) \left(1-w_2\right) \left(1-w_3\right)}{(1-p_1) (1-p_2)}\,.
\end{equation}
The plethystic exponential is defined as usual
\begin{equation}
\text{Pexp}\left(R\left(x_1,{\ldots},x_d\right)\right):=e^{\sum^{\infty}_{l=1} \frac{R\left(x_1^l,{\ldots},x_d^l\right)}{l}}\, ,
\end{equation}
for any rational function~$R$ of~$d$ rapidities~$x_{1},\ldots, x_d$. In particular, 
\begin{equation}
\text{Pexp}\left({U _{ab}}\right)=\frac{1}{1-{U _{ab}}}\,,\qquad a\,\neq\, b\,.
\end{equation}
Summarizing different representation for the index that can be found in various references~\cite{Kinney:2005ej}\cite{Romelsberger:2005eg}\cite{Dolan:2008qi} (see also, for instance~\cite{Benini:2018mlo}) we recall that
\begin{equation}\label{eq:DefPlethIndex}
\begin{split}
\mathcal{I}&=\mathcal{N} \oint\prod^{N}_{a=1} \,\frac{\text{d$U$}_{a}}{2\pi\text{i}U_a}\cdot\prod^{N}_{\text{a\,$\neq$\,b}\,=\,1} \exp \left(-\sum^{\infty}_{l=1} \frac{\left(1-w_1^l\right) \left(1-w_2^l\right) \left(1-w_3^l\right)}{l \left(1-p_1^l\right) \left(1-p_2^l\right)}\,{U^{l}_{ab}}\right) \\&=\mathcal{N} \oint\prod^{N}_{a=1} \,\frac{\text{d$U$}_{a}}{2\pi\text{i}U_a}\cdot\prod^{n}_{\text{a\,$\neq$\,b}\,=\,1}\,\frac{\prod^3_{I=1} \Gamma _e\left(w_I U_{ab};p_1,p_2\right)}{\Gamma _e\left(U_{ab};p_1,p_2\right)}\, , 
\end{split}
\end{equation}
where the normalization (or zero modes) factor is defined as
\begin{equation}
\mathcal{N}\,:=\,\mathcal{N}(N,w_1,w_2,w_3;p_1,p_2)\,=\,\frac{\left(\text{Pexp}\left(i(w;p_1,p_2)\right)\right)^N}{N!}\, ,
\end{equation}
and
\begin{equation}
\text{Pexp}\left(i(w;p_1,p_2)\right)\,:=\,\left(p_1;p_1\right) \left(p_2;p_2\right) \prod^{3}_{I=1} \Gamma _e\left(w_I;p_1,p_2\right)\,.
\end{equation}
\subsection*{Two ways of implementing the constraint among rapidities}\label{sec:ChoicesConstraint}
The constraint
\begin{equation}
 \frac{w_1 w_2 w_3}{p_1 p_2}\,=\,1\, ,
\end{equation}
can be implemented in various ways.

\paragraph{Expansion A)} The implementation (A)
\begin{equation}\label{eq:FixingBalancingCondition1}
w_3\,:=\, \frac{p_1 p_2}{w_1 w_2}\,,
\end{equation}
(and analogously for the case obtained by the permutation of the indices~$1,2,3$ of~$w$'s) defines the following series expansion
\begin{equation}\label{eq:RepA}
\mathcal{I}=\sum_{J^\prime,\,\overline{J}^\prime,Q_1^\prime,Q_2^\prime}d\left(J^\prime,\overline{J}^\prime,Q_1^\prime,Q_2^\prime\right) p_1^{J^\prime} p_2^{\bar{J}^{\prime}} w_1^{Q_{1}^{\prime}} w_2^{Q_{2}^{\prime}} \, , 
\end{equation}
in terms of the four charges
\begin{equation}\label{eq:ChargesA}
J^\prime:=J+Q_3\,,\,\bar{J}^{\prime}:=\bar{J}+Q_3\,,\,Q_{1,2}^{\prime}\,:=\,Q_{1,2}\,-\,Q_3\,.
\end{equation}

For~\eqref{eq:RepA} to be a well-defined expansion, i.e. for it to follow from the original representation~\eqref{eq:Index}, requires imposing the following condition (A)
\begin{equation}\label{eq:ConvergenceDomain1}
|p_1 p_2|< |w_1 w_2|\, , 
\end{equation}
which together with
\begin{equation}\label{eq:ConvergenceDomain0}
|p_a|\,,\,|w_{1,2,3}|\,<\,1\,,
\end{equation}
guarantees absolute convergence of the series in the exponent of the plethystic exponential defining the index~\eqref{eq:DefPlethIndex}. 

\paragraph{Scaling limit A)} For later purposes, we note that the condition~\eqref{eq:ConvergenceDomain1} implies that in a scaling limit to the boundary of the convergence region of representation A)
\begin{equation}\label{eq:ScalingLimit1}
p_{1,2} \,\to\,1-\epsilon\,\,\to\,1^-\,,
\end{equation}
necessarily
\begin{equation}
\text{Re}(\Delta_I)\to\,0^+\,.
\end{equation}
where~$I=1,2,3\,$.
Thus, we are free to assume that in such a scaling limit
\begin{equation}
\Delta_{I}\to \text{Im}(\Delta_{I})\,\i\, , 
\end{equation}
where~$\text{Im}(\Delta_{I})$ is a generic real number (which eventually we will require to be different from~$2\pi n$, with~$n$ integer).

\paragraph{Expansion B)} The implementation (B)
\begin{equation}\label{eq:FixingBalancingCondition2}
p_{2}\,:=\, \frac{w_1 w_2 w_3}{p_1}\,,\qquad p_2\,\neq\, p_1\,,
\end{equation}
(and analogously for the case obtained by the permutation of the indices~$1,2$ of~$p$'s) defines the following series expansion
\begin{equation}\label{eq:RepB}
\mathcal{I}=\sum_{\widetilde{J}^\prime,\widetilde{Q}^\prime_1,\widetilde{Q}^\prime_2,\widetilde{Q}^\prime_3}\,\widetilde{d}\left(\widetilde{J}^\prime,\widetilde{Q}^\prime_1,\widetilde{Q}^\prime_2,\widetilde{Q}^\prime_3\right)\, p_1^{\widetilde{J}^\prime} w_1^{\widetilde{Q}^\prime_1} w_2^{\widetilde{Q}^\prime_2} w_3^{\widetilde{Q}^\prime_3}\, ,
\end{equation}
that counts degeneracies as a function of the four charges
\begin{equation}\label{eq:ChargesB}
\widetilde{J}^\prime := J -\overline{J}\,,\qquad\widetilde{Q}^\prime_{1,2,3}:= Q_{1,2,3}+\overline{J}\,.
\end{equation}
These charges relate to~\eqref{eq:ChargesA} as follows
\begin{equation}\label{eq:ChangeChargesAB}
\widetilde{J}^\prime \,=\,J^\prime-\overline{J}^\prime\,,\, \widetilde{Q}_{1,2}^\prime\,=\, Q_{1,2}^\prime\,+\,\overline{J}^{\prime}\,,\, \widetilde{Q}^\prime_3\,=\, \overline{J}^\prime\,.
\end{equation}
Obviously, the two degeneracies $d$ and $\widetilde{d}$ are related by the composition conditions~\eqref{eq:ChangeChargesAB}. 

For~\eqref{eq:RepB} to be a well-defined expansion of the index~$\mathcal{I}$, i.e. for it to follow from the original representation~\eqref{eq:Index}, requires imposing the following condition (B)
\begin{equation}\label{eq:ConvergenceDomain2}
|w_1 w_2 w_3|< |p_1|\,.
\end{equation}

\paragraph{Scaling limit B)}\label{sec:ScalingB} For later purposes, we note that the condition~\eqref{eq:ConvergenceDomain2} implies that in a scaling limit to the boundary of the convergence region of representation B)
\begin{equation}\label{eq:ScalingLimit2}
 w_{1,2,3}\,\to\, 1-\epsilon_w\,\to\,1^-\,,
\end{equation}
necessarily
\begin{equation}
\text{Re}(\omega_1)\,\to\,0^-\,.
\end{equation}
Hence, we are free to assume
\begin{equation}
\omega_{1}\to \text{Im}(\omega_{1})\,\i\, ,
\end{equation}
where~$\text{Im}(\omega_{1})$ is generic real number (which eventually we will require to be different from~$2\pi n$, where~$n$ is an arbitrary integer number).

We will use the expansion B) for the study of the giant graviton representation. As mentioned before, the domain of convergence of the giant graviton Hamiltonian traces is different from the one of the $\frac{1}{16}$-BPS index of~$\mathcal{N}=4$ SYM. In such an analysis, extensive use of analytic continuation will be required.

\subsection{The giant graviton proposal}\label{sec:GGProp}
The giant graviton expansion proposed in~\cite{Imamura:2021ytr} is
\begin{equation}\label{eq:GiantGravImamura}
\mathcal{I}\,\underset{?}{=}\,\mathcal{I}_{KK}\, \mathcal{I}_{GG}\, , 
\end{equation}
where~$\mathcal{I}_{KK}$ is the generating function of~$\frac{1}{16}$-BPS multi-graviton excitations at~$N=\infty$ (closed strings contributions)
\begin{equation}
\begin{split}
\mathcal{I}_{KK}&=\exp\Biggl(\sum_{l=1}^\infty \frac{1}{l}\Bigl(\frac{w_1^l}{1-w_1^l}+\frac{w_2^l}{1-w_2^l}
   +\frac{w_3^l}{1-w_3^l}-\frac{p_1^l}{1-p_1^l}-\frac{p_2^l}{1-p_2^l}\Bigr)\Biggr) 
\\&=\prod^{\infty}_{l=1} \frac{\left(1-w^l{}_1\right) \left(1-w^l{}_2\right)
   \left(1-w^l{}_3\right)}{\left(1-p_1^l\right) \left(1-p_2^l\right)}\,,
   \end{split}
   \end{equation}
and~$I_{GG}$ is the giant graviton index
\begin{equation}
\mathcal{I}_{GG}\,=\,\sum
   _{n_3=0}^{\infty }  w_1^{N n_1}w_2^{N n_2}w_3^{N n_3} \,\mathcal{I}_{n_1,n_2,n_3}\,.
\end{equation}
Here, $\mathcal{I}_{n_1,n_2,n_3}$ is the index of~$n_{1}$, $n_2$ and~$n_3$ stacks of D3 branes wrapping three different~$S^3$ cycles within the internal space~$S^5$ ($i=1,2,3$), times the index of open strings ending on pairs of stacks~\cite{Imamura:2021ytr}.

Concretely,
\begin{equation}\label{eq:IntegralGGindex}
\mathcal{I}_{\underline{n}}(\tf)\,\equiv\,\mathcal{I}_{n_1,n_2,n_3}=\oint_{\Gamma_{gauge}} \text{d}\mu_1 \text{d}\mu_2 \text{d}\mu_3
 \, \mathcal{I}^{4 d}_{n_1,n_2,n_3} \mathcal{I}^{2 d}_{n_1,n_2,n_3}\,,
\end{equation}
with measure
\begin{equation}\label{eq:MeasureGGint}
\text{d}\mu_{I}\, :=\,\frac{1}{n_I!}\prod^{n_I}_{a=1} \,\frac{\text{d$U$}^{(I)}_{a}}{2\pi\text{i}U^{(I)}_a}\,\cdot\,\prod_{a\neq b=1}^{n_I}(1-\frac{U^{(I)}_a}{U^{(I)}_b})\,.
\end{equation}
The closed contour~$\Gamma_{gauge}$, which is not the trivial unit-circle, has been proposed and tested at small values of~$N$ and charges in~\cite{Lee:2022vig}~\cite{Imamura:2021ytr}. Another seemingly valid definition has been given in~\cite{Lee:2022vig}.~\footnote{We have recently reported on this for the Schur index~\cite{Beccaria:2023zjw}.} For reasons that will be explained in Appendix~\ref{sec:AppB} the explicit form of the closed contour~$\Gamma_{gauge}$ plays (almost) no role in the large-charge expansion. To understand this one must rely on results that will be derived in subsection~\ref{sec:LargeBlackHole}. So, from now on we postpone any discussion on~$\Gamma_{gauge}$ until appendix~\ref{sec:AppB}.   

The objects:
\begin{equation}\label{DefinitionPlethystic4d}
\mathcal{I}^{4 d}_{n_1,n_2,n_3}:=\prod_{I=1}^{3}\mathcal{I}^{4 d}_{I} \, ,\qquad \mathcal{I}^{2 d}_{n_1,n_2,n_3}:=\prod_{I=1}^{3}\mathcal{I}^{2 d}_{I,I+1}\,,\, \quad I+3\,\sim\, I,
\end{equation}
are the contributions of 4d $\mathcal{N}=4$ vector multiplets corresponding to worldvolume massless excitations of a stack of~$n_I$ D3-branes wrapping the 3-sphere~$I$, and 2d $U(n_I)\times U(n_{I+1})$ bi-adjoint hypermultiplets corresponding to massless open strings excitations stretching between the stacks of D3 branes~$I$ and~$I+1$, respectively. By definition~$\mathcal{I}_{\{0,0,0\}}=1$.

The 4d adjoint contributions are
\begin{equation}\label{eq:4dAdjointContribution}
\begin{split}
\text{d}\mu_I\mathcal{I}^{4 d}_{I}&:= \text{d}\mu_I \,\prod^{n_I}_{a=1} \prod^{n_I}_{b=1} \text{Pexp}\Bigl(i(w_I^{-1},p_1,p_2;w_J,w_K) \,{U^{(I)}_{ab}}\Bigr) \\ & \,=\, \mathcal{N}^{4d}_I\,\prod^{n_I}_{a=1} \,\frac{\text{d$U^{(I)}$}_{a}}{2\pi\text{i}U^{(I)}_a}\cdot\prod^{n_I}_{\text{a\,$\neq$\,b}\,=\,1} \exp \left(-\sum^{\infty}_{l=1} \frac{\left(1-w_I^{-l}\right) \left(1-p_1^l\right) \left(1-p_2^l\right)}{l \left(1-w_J^l\right) \left(1-w_K^l\right)}\,{U^{(I)l}_{ab}}\right) \\
&\,=\,\mathcal{N}^{4d}_I\,\prod^{n_I}_{a=1} \,\frac{\text{d$U^{(I)}$}_{a}}{2\pi\text{i}U^{(I)}_a}\cdot\prod^{n_I}_{\text{a\,$\neq$\,b}\,=\,1} \,\frac{ \Gamma _e\left(\frac{1}{w_I} U^{(I)}_{ab};w_J,w_K\right)\Gamma _e\left(p_1 U^{(I)}_{ab};w_J,w_K\right)\Gamma _e\left(p_2 U^{(I)}_{ab};w_J,w_K\right)}{\Gamma _e\left(U^{(I)}_{ab};w_J,w_K\right)}\, ,
\end{split}
\end{equation}
for~$I\neq J\neq K=1,2,3\,$, and the zero-mode contributions are defined as
\begin{equation}
\mathcal{N}^{4d}_I\,:=\,\mathcal{N}(n_I,w^{-1}_I,p_1,p_2;w_J,w_K)\,.
\end{equation}
We define the~$a$-th component of the diagonal unitary matrices as~$U^{(I)}_{{a}}$ and their quotient~$U^{(I)}_{ab}:=\frac{U^{(I)}_{a}}{U^{(I)}_{b}}$.

The contributions to the index coming from a 2d~$U(n_1)\times U(n_2)$ bi-fundamental field are
\begin{equation}\label{eq:2dBiadjoint index}
\begin{split}
\mathcal{I}^{2 d}_{I,I+1}&:= \prod^{n_I}_{a=1} \prod^{n_{I+1}}_{b=1} \text{Pexp}\left(i_h(p_1,p_2;w_J) \,\left(U^{(I,I+1)}_{ab}+\frac{1}{U^{(I,I+1)}_{ab}}\right)\right) \\
&=\prod^{n_I}_{a=1} \prod^{n_{I+1}}_{b=1}\frac{\theta _0\left(\frac{1}{U^{(I,I+1)}_{ab}}\sqrt{\frac{p_1 w_J}{p_2}};w_J\right) \theta _0\left(U^{(I,I+1)}_{ab} \sqrt{\frac{p_1
   w_J}{p_2}};w_J\right)}{\theta _0\left(\frac{\sqrt{\frac{w_J}{p_1 p_2}}}{U^{(I,I+1)}_{ab}};w_J\right) \theta _0\left(U^{(I,I+1)}_{ab} \sqrt{\frac{w_J}{p_1
   p_2}};w_J\right)}\, ,
\end{split}
\end{equation}
where
\begin{equation}
i_h(p_1,p_2;w):=\sqrt{\frac{w}{p_1 p_2}}\frac{(1-p_1) (1-p_2) }{1-w}\,.
\end{equation}
In this expression~$J\neq I, I+1\,\text{mod}\,3\,$. We define the quotient of diagonal components of different unitary matrices as~$U^{(I,I+1)}_{ab}:=\frac{U^{(I)}_a}{U^{(I+1)}_b}\,$.~\footnote{Following the conventions of the original proposal of~\cite{Imamura:2021ytr} here we have assumed~$a_{loop}\,=\,a_{12}a_{23}a_{31}\,=\,1\,$. In that case, without loss of generality we can assume~$a^{(I,I+1)}\,=\,1$ (See equation (11) in~\cite{Imamura:2021ytr}). More generally, the analysis in section~\eqref{sec:LargeBlackHole} can be straightforwardly reproduced for any other choice of~$a^{(I,I+1)}\,$, however, the only for~$a_{loop}\,=\,1\,$ we obtain consitent results.}

 \subsection{The free fermion representation of the index}
 
An exact expansion of the index as an average over an ensemble of free fermion systems was put-forward in~\cite{Murthy:2022ien}. As we explained in the introduction, it takes again the form
of a giant-graviton expansion, different from the physically motivated D-brane expansion. Still, it is a mathematical exact rearrangement of the index and it will be interesting to 
consider its properties. In particular,  we will discuss 
in Appendix D the detailed way it  reproduces the large black hole entropy. In this representation, the index reads
 \begin{equation}\label{eq:GGFormula}
 \mathcal{I}=\mathcal{I}_{KK} \left(\sum^{\infty}_{n=0} \mathcal{J}_{n}(N)\right)\, ,
 \end{equation}
 where
 \begin{equation}
 \begin{split}
 \mathcal{J}_n(N)=\frac{(-1)^n}{n!} \oint\prod^{n} _{i=1}& \frac{\text{d}y_i \text{d}z_i}{\left(2 \pi  i y_i\right) \left(2 \pi 
    i z_i\right)}\frac{
    \left({y_i}/{z_i}\right)^{N+1}}{ \left(1-{y_i}/{z_i}\right)} \cdot \,\det{\Bigl( \frac{1}{1-\frac{y_j}{z_i}}}
   \Bigr) _{i,j=1} \\ \qquad &\exp \left(\sum^{\infty}_{l=1} \frac{j_{n}\left(p_1^l,p_2^l;w^l\right) \sum^{n}_{i,j=1}
   \left(z^l_i-y^l_i\right)
    \left(z^{-l}_j-y^{-l}_j\right)}{l}\right)\, , 
    \end{split}
 \end{equation}
 and
 \begin{equation}
 j_n(p_1,p_2;w):=1-\frac{\left(1-p_1\right) \left(1-p_2\right)}{\left(1-w_1\right)
    \left(1-w_2\right) \left(1-w_3\right)}\,,
 \end{equation}
 \begin{equation}
 \text{with}\ \frac{w_1 w_2 w_3}{p_1 p_2}\,=\,1\,.
 \end{equation}
The object~$\mathcal{J}_n(N)$ is a Hubbard-Stratonovich transformation of a determinant of two-point functions in an auxiliary theory of free fermions~\cite{Murthy:2022ien}. 
 
Using the identity
\begin{equation}
\det \left(\frac{1}{1-\frac{y_j}{z_i}}\right) 
\,=\, \prod_{i=1}^n z_i\,\cdot\,\frac{\prod_{1\leq j<i\leq n} \left(z_i-z_j\right)
   \left(y_j-y_i\right)}{\prod _{i,j=1}^n
   \left(z_i-y_j\right)}
   =\prod_{i=1}^{n}\frac{1}{1-\frac{y_i}{z_i}}\,\cdot\,\frac{\prod_{1\leq j<i\leq n} \left(z_i-z_j\right)
   \left(y_j-y_i\right)}{\prod _{i\neq j=1}^n
   \left(z_i-y_j\right)}
\end{equation}
together with the change of variables
\begin{equation}
(z_i\,,\, y_i)\,\to\, \Bigl(z^\prime_i\,=\,z_i\,,\,\zeta_i\,=\,\frac{y_i}{z_i} \Bigr)\,,   
\end{equation}
 (and ignoring the~$\prime$ in the~$z^\prime_i$'s
from now on) one reaches the form that we will work with
   \begin{equation}\label{eq:JN}
 \begin{split}
 \mathcal{J}_n(N)=\frac{(-1)^n}{n!} \oint\prod^{n} _{i=1}& \frac{\text{d}\zeta_i \text{d}z_i}{\left(2 \pi  i \zeta_i\right) \left(2 \pi 
    i z_i\right)}\frac{
    \left(\zeta_i\right)^{N+1}}{ \left(1-{\zeta_i}\right)^2} \, \cdot\text{Det}(\underline{z},\underline{\zeta})\cdot\\ \qquad &\exp \left(\sum^{\infty}_{l=1} \frac{j_{n}\left(p_1^l,p_2^l;w^l\right) \sum^{n}_{i,j=1}\frac{z^l_i}{z^l_j}
   \left(1-\zeta^l_i\right)
    \left(1-\zeta^{-l}_j\right)}{l}\right)\, ,
    \end{split}
\end{equation}
where
\begin{equation}
\text{Det}=\text{Det}(\underline{z},\underline{\zeta})\,:=\,\DetExp\,.
\end{equation}

\subsection{The index at large charges}
\label{sec:IndexLargeChargeExpansion}
Let us fix the constraint~\eqref{eq:FixingBalancingCondition1} and study the large charge asymptotic behaviour of the microcanonical index
\begin{equation}\label{eq:IntegralFourierCoefficients}
d\left(\underline{Q}^\prime\right)=\int_{0}^{2\pi\i}\frac{ \text{d$\Delta $}_1 \text{d$\Delta
   $}_2 }{(2 \pi \i)^2}\,\int_{\omega_1^\star}^{4\pi\i+\omega_1^\star}\frac{ \text{d$\omega $} _1 }{(4 \pi \i)}\,\int_{\omega_2^\star}^{4\pi\i+\omega_2^\star}\frac{ \text{d$\omega $}_2 }{(4 \pi \i)}\,\int^{1}_{0} \frac{d\underline{u}}{N!}\,e^{-S_{\text{eff}}({\underline{x};\underline{u}})-\i \underline{x}\cdot \underline{Q}^\prime}\, ,
\end{equation}
where
\begin{equation}\label{eq:ChemPotsCharges}
\begin{split}
-\i x_{1,2}&\,=\,\Delta_{1,2},\qquad -\i x_{4,5}\,=\,\omega_{1,2}\,, \qquad Q^\prime_{1,2}=Q^\prime_{1,2}\,,\,\qquad \,\,Q^\prime_{4,5}\,=\,J^\prime\,,\,\overline{J}^\prime\,.
\end{split}
\end{equation}
The $4\pi\i$ is because the charges~$Q^\prime_{4,5}$ are quantized in units of~$1/2$. The two saddle point positions $\omega_{1,2}^\star$  (which are not pure imaginary)  will be determined below.

The effective action
\begin{equation}\label{eq:TruncatedEffAction}
\begin{split}
-S_{\text{eff}}({\underline{x};\underline{u}})\,:=&\,-\,\sum_{\text{a$\neq$b}=1}^{N}\sum _{l=1}^{\infty} \frac{\left(1-w^l_1\right) \left(1-w^l_2\right) \left(1-\left(\frac{p_1
   p_2}{w_1 w_2}\right){}^l\right)}{l \left(1-p_1^l\right) \left(1-p_2^l\right)}\,\cos{(2\pi l u_{ab})}\\&\,-\,N \,\sum _{l=1}^{\infty}\frac{1}{l}\Biggl( \frac{\left(1-w^l_1\right) \left(1-w^l_2\right) \left(1-\left(\frac{p_1
   p_2}{w_1 w_2}\right)^l\right)}{ \left(1-p_1^l\right) \left(1-p_2^l\right)}-1\Biggr)\,,
   \end{split}
\end{equation}
has singularities located at
\begin{equation}
x_4=x_{4,sing}=0\,,\qquad x_5=x_{5,sing}=0\,,\qquad \text{and}\  \text{periodic images}\,.
\end{equation}
Around these singularities:
\begin{equation}\label{eq:RelSeffsmallS}
S_{\text{eff}}\Bigl(x_{1},x_2,{x}_{4,sing}+\frac{\delta x_4}{\Lambda},{x}_{5,sing}+\frac{\delta x_5}{\Lambda};\underline{u}\Bigr)\,\underset{\Lambda\to \infty}{\sim}\, \Lambda^{2}\, \widetilde{s}( x_1, x_2, \delta x_4,\delta x_5;\underline{u})\,.
\end{equation}
Using the formal Taylor expansion~\cite{Narukawa}
\begin{equation}\label{eq:AsymptoticsEpsilon}
\frac{1}{\left(e^{\frac{\delta x_4}{\Lambda} l \epsilon }-1\right) \left(e^{\frac{\delta x_5}{\Lambda} l
   }-1\right)}=\sum _{k=0}^{\infty } \frac{B_{2,k}(\delta x_4,\delta x_5 )}{k!}\, \Bigl( \frac{l}{\Lambda}\Bigr)^{k-2}\, ,
\end{equation}
on the denominator in the right-hand side of~\eqref{eq:TruncatedEffAction} one computes the small-$1/\Lambda$ expansion of the effective action~$S_{\text{eff}}$
\begin{equation}\label{eq:functionStilde}
\begin{split}
\widetilde{s}(\underline{y};\underline{u})&= \frac{\widetilde{s}^{(1,1)}(y_1,y_2;\underline{u})}{y_4 y_5}+\frac{\widetilde{s}^{(1,0)}(y_1,y_2, y_4, y_5;\underline{u})}{y_4}\,+\,\frac{\widetilde{s}^{(0,1)}(y_1,y_2, y_4, y_5;\underline{u})}{y_5}\\ &\,+\,c_4\,N \log y_{4}\,+\, c_5\,N \log y_{5}\,+\,c_6\,N \log (y_4+y_{5})\,+\,\ldots\,,
\end{split}
\end{equation}
where 
\begin{equation}\label{eq:SubleadingSs}
\widetilde{s}^{(1,0)}(y_1,y_2, y_4, y_5;\underline{u})\,,\, \qquad \widetilde{s}^{(1,0)}(y_1,y_2, y_4, y_5;\underline{u})\, ,
\end{equation}
are linear functions of~$y_4$ and~$y_5\,$,~\footnote{... which can be straightforwardly extracted from~\eqref{eq:TruncatedEffAction}. We do not write in here these expressions because their explicit form will not be relevant for our goals. For our present goals it will be enough to start from~\eqref{eq:TruncatedEffAction} to recover the contribution that these terms give to the effective action, as it will be explained below.} and dots denote contributions that vanish in the infinitely large scale transformation~$y_{4,5}\to \frac{y_{4,5}}{\Lambda}\,$ at $\Lambda\to\infty\,$. 

For the moment let us focus on the leading contribution
\begin{equation}\label{eq:stilde}
-\widetilde{s}^{(1,1)}(y_1,y_2;\underline{u}):=- \sum_{a, b=1}^{N}\sum _{l=1}^{\infty} \frac{\left(1-w^l_1\right) \left(1-w^l_2\right) \left(1-\left(\frac{1}{w_1 w_2}\right)^l\right)}{l^3}\,e^{2\pi\i l u_{ab}}\,.
\end{equation}
Below we will show how to compute the subleading contributions. Recalling the expansion
\begin{equation}
\text{Li}_n(z):=\sum^{\infty}_{l=1} \frac{z^l}{l^n}\, ,
\end{equation}
\eqref{eq:stilde} can be rewritten as:
\begin{equation}
-\widetilde{s}^{(1,1)}(y_1,y_2;\underline{u}):=+ \frac{4 \pi ^3\i}{3}\,\sum_{a, b=1}^{N}\,{\sum_{I=1}^{3} \overline{B}_3\left[u_{ab}+\frac{\Delta_I}{2\pi\i}\right]}\,,
\end{equation}
with
\begin{equation}\label{eq:ConstraintLargeCharges}
\Delta_3 \to-\Delta_1-\Delta_2\,.
\end{equation}
In this equation
\begin{equation}\label{eq:TruncatedBernoulliPolynomials}
\overline{B}_n[\Delta]:=-\frac{n!}{(2 i \pi )^{n} } \Bigl(\text{Li}_n\left(e^{2 i \pi  \Delta }\right)+(-1)^n \text{Li}_n\left(e^{-2 i \pi  \Delta
   }\right)\Bigr)\, ,
\end{equation}
is the periodic Bernoulli polynomial of order~$n\,$. For example, for~$n=3$ one gets
\begin{equation}
\overline{B}_3(x)=B_3(x-\lfloor x\rfloor )\quad,\quad B_3(x)\,:=\,x^3-\frac{3 x^2}{2}+\frac{x}{2}\,.
\end{equation}
The contributions~$\widetilde{s}^{(1,0)}$ and $\widetilde{s}^{(0,1)}$, can be computed analogously.

Remarkably the large-charge degeneracy of states up-next-to-leading order in the large-$\Lambda$ expansion (up to order $\mathcal{O}(\Lambda)$) is computed as follows 
\begin{equation}
\begin{split}
d(\underline{Q})&\simExp\,\sum_{\underline{x}^\star,\underline{u}^\star}\,e^{-\widetilde{s}(\underline{x}^\star;\underline{u}^\star)-\i \underline{x}^\star\cdot \underline{Q}}\, \\ &\simExp\,\sum_{\underline{x}^\star,\underline{u}^\star}\,e^{+\frac{4 \pi ^3\i}{3}\,\sum_{a, b=1}^{N}\,\frac{\sum_{I=1}^{3} \overline{B}_3\left[u^\star_{ab}+\frac{\Delta^\star_I}{2\pi\i}\right]}{\omega^\star_1\omega^\star_2}+\omega^\star
   _1 J^\prime+\omega^\star _2 \bar{J}^\prime +\Delta^\star_1 Q^\prime_1+\Delta^\star _2 Q^\prime_2}\,,
   \end{split}
\end{equation}
where the variables~$(\underline{x}^\star;\underline{u}^\star)=(\i\omega^\star_{1,2},\i\Delta^\star_{1,2};\underline{u}^\star)$ denote the leading saddle points of:
\begin{equation}\label{eq:SaddlePTilded0}
+\frac{4 \pi ^3\i}{3}\,\sum_{a, b=1}^{N}\,\frac{\sum_{I=1}^{3} \overline{B}_3\left[u_{ab}+\frac{\Delta_I}{2\pi\i}\right]}{\omega_1\omega_2}+\omega
   _1 J^\prime +\omega _2 \bar{J}^\prime+\Delta_1 Q^\prime_1+\Delta _2 Q^\prime_2\, ,
\end{equation}
i.e. the saddle points of~\eqref{eq:SaddlePTilded0} with respect to~$(\omega_{1,2},\Delta_{1,2}, u_{ab})\,$. Those that maximize the real part of~\eqref{eq:SaddlePTilded0} but this time with the constraint
\begin{equation}\label{eq:ConstraintLargeChargesSubleading}
\Delta_3 \,\to\, -\Delta_1-\Delta_2+\omega_1+\omega_2\,.
\end{equation}
instead of~\eqref{eq:ConstraintLargeCharges}. 

It is easy to prove that the complete answer at next-to-leading order~$\mathcal{O}(\Lambda)$ is recovered by simply substituting the rule~\eqref{eq:ConstraintLargeCharges} by~\eqref{eq:ConstraintLargeChargesSubleading}, and only considering the asymptotic expansion of the gauge saddle-point solution at leading order at large-$\Lambda\,$. From now on we denote the latter asymptotic value as~$u^\star$ (the saddle point of~$s^{(1,1)}$). This is because any next-to-leading correction to the effective action coming from~$\frac{1}{\Lambda}$ deformations to~$u^*=\mathcal{O}({\Lambda}^0)$ would vanish when evaluated at~$u^\star\,$. This is because by definition such correction to the effective action is proportional to the saddle-point condition that~$u^\star$ satisfies by definition. Thus, to evaluate the contribution at next-to-leading order to the effective action we just need to evaluate the original form of the latter~\eqref{eq:TruncatedEffAction} at~$u^\star =\mathcal{O}(\Lambda^0)\,$ and expand the result up to order~$\mathcal{O}(\Lambda^1)\,$. Following this procedure we obtain
\begin{equation}\label{eq:StildeDD}
\widetilde{s}(\underline{x};\underline{u}^\star)\,=\,-\,\frac{4 \pi ^3\i}{3}\,\sum_{a, b=1}^{N}\,\frac{\sum_{I=1}^{3} \overline{B}_3\left[u^\star_{ab}+\frac{\Delta_I}{2\pi\i}\right]}{\omega_1\omega_2}\,+\,\mathcal{O}(\Lambda^0)\,
\end{equation}
with the relation~$\Delta_3 \,\to\, -\Delta_1-\Delta_2+\omega_1+\omega_2$\,.
The terms linear in~$\omega_1$ and~$\omega_2$ in~\eqref{eq:ConstraintLargeChargesSubleading} come from the powers of~$p_1$ and~$p_2$ in the numerator of the sumands in the first line of~\eqref{eq:TruncatedEffAction}.

\subsection*{The contribution of zero modes: computing $c_4\,$, $c_5\,$, and~$c_6$}

The contribution of zero modes in the second line of~\eqref{eq:TruncatedEffAction} determines the coefficients of the logarithmic divergencies~$\log y_{4,5}\,$ and~$\log(y_4+y_5)$. The easiest way to compute these contributions is to write
\begin{equation}
\begin{split}
-\,N& \,\sum _{l=1}^{\infty}\frac{1}{l}\Biggl( \frac{\left(1-w^l_1\right) \left(1-w^l_2\right) \left(1-\left(\frac{p_1
   p_2}{w_1 w_2}\right)^l\right)}{ \left(1-p_1^l\right) \left(1-p_2^l\right)}-1\Biggr)\\&\,=\,N \,\sum _{l=1}^{\infty}\frac{1}{l} \frac{ \left(1-p_1^l\right) \left(1-p_2^l\right)\,-\,\left(1-w^l_1\right) \left(1-w^l_2\right) \left(1-\left(\frac{p_1
   p_2}{w_1 w_2}\right)^l\right)}{ \left(1-p_1^l\right) \left(1-p_2^l\right)}\, ,
   \end{split}
\end{equation}
and Taylor-expand the denominator, keeping as many terms as necessary. Then, we sum (over~$l$) the coefficients of each monomial in the Taylor expansion. The result is a linear combination of polylogarithms. Many of such polylogarithms contribute to the terms~\eqref{eq:SubleadingSs}.
The remaining ones take the form
\begin{equation}\label{eq:LogSCI}
\,+\,c_4\,N \log y_{4}\,+\, c_5\,N \log y_{5}\,+\,c_6\,N \log (y_4+y_{5})\,+\,\ldots\, ,
\end{equation}
where for~$0<\omega_{1,2}=-\i y_{4,5}<1$
\begin{equation}
\begin{split}
c_4\,=\,c_5\,=\,\frac{1}{12} \left(\frac{y_4}{y_5}+\frac{y_5}{y_4}-3\right)\,,\,\quad c_6\,=\,\frac{1}{6} \left(-\frac{y_4}{y_5}-\frac{y_5}{y_4}-3\right)\,,
\end{split}
\end{equation}
the dots in~\eqref{eq:LogSCI} denote terms that vanish after rescaling~$y_{4,5}\to y_{4,5}/\Lambda\,$ and taking~$\Lambda\to\infty\,$. Logarithmic contributions with similar origins as~\eqref{eq:LogSCI} will appear in the study of the giant graviton expansions. They are subleading contributions (of type-$F$) that will not affect the leading asymptotics we are looking for, but for future developments it may be useful to explain how to compute them. 

\subsection*{Evaluating the saddle points}\label{ref:Interference}
The saddle-point condition~
\begin{equation}
\partial_{\underline{u}}\widetilde{s}(\underline{x};\underline{u})=0\, ,
\end{equation}
has a leading solution (independent of other chemical potentials)~\cite{Cabo-Bizet:2019osg,Cabo-Bizet:2019eaf},
\begin{equation}
\underline{u}^\star_{ab}=0\,.
\end{equation}
The remaining saddle point conditions 
\begin{equation}\label{eq:SaddleConditions}
\partial_{\underline{x}}\widetilde{s}(\underline{x};\underline{u}^\star)\,=\,-\i\underline{Q} \,,
\end{equation}
are piecewise polynomial conditions and can be solved straightforwardly. In this subsection we focus on counting operators with charges
\begin{equation}\label{eq:SectionCharges}
J^\prime=\overline{J}^\prime \sim \Lambda^3 \neq 0\,,\qquad Q^\prime_1 =Q^\prime_2= 0 \,. 
\end{equation}
The solvability of conditions~\eqref{eq:SaddleConditions} requires
\begin{equation}
\omega_{1}=\omega_{2}\,,\,\qquad \Delta_1=\Delta_2\,.
\end{equation}
Then solutions of~\eqref{eq:SaddleConditions} at leading order in the large~$\Lambda$ expansion are
\begin{equation}
\frac{\Delta_{1\pm}^\star}{2\pi\text{i}}=\pm\frac{1}{3}\text{mod}\, 1\, ,
\end{equation}
the extrema of~$\overline{B}_3\left[\frac{\Delta_{1}}{2\pi\i}\right]\,$, and
\begin{equation}\label{eq:ChemicalPotentialtoCharges}
\omega_{1\pm}^\star\,=\,\mp  {\text{i}}^{1/3}\,\frac{2 \pi  {N}^{2/3}}{3^{2/3}}\,{\frac{1}{|{J}^{\prime}|^{\frac{1}{3}}}}\sim\frac{N^\frac{2}{3}}{\Lambda}\,.
\end{equation}
These two saddle points~$\underline{x}^\star_{\pm}=\{\i\omega^\star_{\pm},\i\Delta^\star_{\pm}\}$ contribute as follows
\begin{equation}\label{eq:Oscillations}
e^{-\widetilde{s}(\underline{x}^\star_{\pm};\underline{u}^\star)-\i \underline{x}^\star_\pm\cdot \underline{Q}}\,=\, \exp{\Bigl(\left(\sqrt{3}\mp\text{i}\right) 3^{1/3}\pi  \,{J^{\prime 2/3}}{N^{2/3}}\Bigr)}\, ,
\end{equation}
to the asymptotic growth of the microcanonical index \underline{along the region of charges~\eqref{eq:SectionCharges}} and at very leading order in the large-$\Lambda$ expansion
\begin{equation}
\begin{split}
|d(\underline{Q})|&\simExp\,|e^{-\widetilde{s}(\underline{x}^\star_{+};\underline{u}^\star)-\i \underline{x}^\star_{+}\cdot \underline{Q}}+e^{-\widetilde{s}(\underline{x}_{-}^\star;\underline{u}^\star)-\i \underline{x}_-^\star\cdot \underline{Q}}|\, \\
&\simExp \exp{\Bigl(\left(\sqrt{3}\right) 3^{1/3}\pi  \,{j^{\prime 2/3}}{(\Lambda^3 N)^{2/3}}\Bigr)}\,|2\cos{\Bigl(3^{1/3}\pi  \,{j^{\prime 2/3}}{(\Lambda^3 N)^{2/3}}\Bigr)}|
\\
\label{eq:AsymptoticsLargeSpin}
 &\,\simExp\, \exp{\Bigl(\left(\sqrt{3}\right) 3^{1/3}\pi  \,{J^{\prime2/3}}{ N^{2/3}}\Bigr)}\,.
\end{split}
\end{equation}
We note that this result is valid at any finite~$N$. It is, however, only valid at leading order in the large-$\Lambda$ expansion (i.e. in the large charge expansion). Namely, this particular form is only the leading asymptotic expansion of~$|d(\underline{Q})|\,$. Note also that in order to have order~$N^2$ growth for~$N\gg 1$ we have to demand~$J^\prime = N^2 \Lambda^3\mathcal{O}(1)$ which means that \underline{at very leading order} in the large charge expansion, the asymptotic expression~\eqref{eq:AsymptoticsLargeSpin} only captures the growth of states with spin~$\frac{J^\prime}{N^2}=\mathcal{O}(\Lambda^3)\to\infty$ and ~$\frac{\text{Entropy}}{N^2}=\mathcal{O}(\Lambda^2)\to \infty\,$.

The complete black hole entropy at any finite ratio~$\frac{\text{Entropy}}{N^2}$ is recovered by using the localized form~$\widetilde{s}$ up to next-to-leading order, concretely
\begin{equation}\label{eq:StildeDD2}
\widetilde{s}(\underline{x};\underline{u}^\star=0)\,=\,-\,\frac{4 \pi ^3\i N^2}{3}\,\frac{\sum_{I=1}^{3} \overline{B}_3\left[\frac{\Delta_I}{2\pi\i}\right]}{\omega_1\omega_2}\,+\,\mathcal{O}(\Lambda^0)\,
\end{equation}
with the substitution rule
\begin{equation*}
\Delta_3 \,\to\, -\Delta_1-\Delta_2+\omega_1+\omega_2\,.
\end{equation*}

The numerator of this localized form of the effective action~$\sum_{I=1}^{3} \overline{B}_3\left[\frac{\Delta_I}{2\pi\i}\right]$ is a piece-wise cubic  polynomial. Its profile along the real locus~$\frac{\Delta_{I}}{2\pi\text{i}}\,\in\,\mathbb{R}$ is reproduced by translation of two cubic polynomial profiles leaving in two independent fundamental domains~$\frac{\Delta_{I}}{2\pi\text{i}}$ that we will denote from now on by appending the symbol~$\pm\,$ on flavour chemical potentials. In such domains we find
\begin{equation}\label{eq:SubleadingEntropyFunctionIndex}
\begin{split}
\widetilde{s}(\underline{x}_{\pm};\underline{u}^\star)\,&+\,\mathcal{O}(\Lambda^0)\, = \,\mathcal{F}_{BH}\,{=}\,\frac{N^2}{2}\,\frac{\Delta_1 \Delta_2 \Delta_3}{\omega_1\omega_2}\,,  \\\Delta_{1}&\,+\,\Delta_{2}\,+\,\Delta_{3}-\omega_1-\omega_2 = \pm 2\pi\text{i}\,,
\end{split}
\end{equation}
where~$\mathcal{F}_{BH}$ is the effecitve action that reproduces the~$\frac{1}{16}$-BPS black hole entropy at any ratio~$\frac{\text{Entropy}}{N^2}\,$, as first observed in~\cite{Hosseini:2017mds}.

From now on when we refer to the~$d(Q)$ of the superconformal index we will mean not just its leading asymptotic form~\eqref{eq:AsymptoticsLargeSpin} in the region of charges~\eqref{eq:SectionCharges} but more generally the finite-$N$ degeneracy~$d(Q)$ computed by plugging~\eqref{eq:SubleadingEntropyFunctionIndex} into the localization formula~\eqref{eq:LargeChargeLocalizationFormula}; and which particularized to the large-$N$ expansion~\eqref{eq:SugraScaling} is known to match the exponential of the 1/16 BPS black hole entropy at any region of charges and for entropies such as the ratio~$\frac{\text{Entropy}}{N^2}$ remains finite and arbitrary~\cite{Hosseini:2017mds}.~\footnote{The explicit form for this~$|d(Q)|$ upon the imposition of a non-linear constraint among the four charges~$Q\,$, can be found in the original reference. An alternative way of deriving it can be found in Appendix B of~\cite{Cabo-Bizet:2018ehj}. Using this way one obtain the complete answer without imposing the non-linear constraint among charges. Here we avoid the reproduction of those results, and instead refer the reader looking for such level details to those references.} 
 
\paragraph{Some comments on the more general region of charges}\label{sub:GeneralCharges}
\label{eq:CoefficientLCGrowth}
Let us assume
\begin{equation}\label{eq:SectionCharges2}
J^\prime=\overline{J}^\prime \sim \Lambda^3 \neq 0\,,\qquad Q^\prime_1 =Q^\prime_2 \sim \Lambda^2\neq 0 \,. 
\end{equation}
Working with the analytic continuation to complex~$\chi:=\Delta_{1,2}/(2\pi\i)$ of the function~\eqref{eq:SaddlePTilded}, which was originally defined for~$\chi \in\mathbb{R}$, the extremization conditions take the form
\begin{equation}\label{eq:SaddleValues}
\begin{split} 
J^\prime=\overline{J}^\prime&=\frac{24 \text{i} \pi^3 N^2 (\chi-1)^2 (2 \chi-1)}{\omega_1^3} \\
Q^\prime_{1,2}&=-\frac{12 \pi ^2 N^2
   (\chi-1) (3 \chi-2)}{\omega_1^2}\,.
\end{split}
\end{equation}
Plugging~$\chi=\frac{2}{3}+\frac{1}{3}\alpha$ in~\eqref{eq:SaddleValues} we solve for
\begin{equation}
\omega_1=\omega^\star _1
:=\pm \frac{2 \pi  N \sqrt{\alpha^\star(1-\alpha^\star)  }}{\sqrt{Q^{\prime}_{1,2}}}\,\sim\,\Lambda^{-1}\,,
\end{equation}
where the complex saddle value~$\alpha=\alpha^\star$ is defined by the cubic equation
\begin{equation}\label{eq:CubicAlphaEquation}
1\,+\,\alpha^\star\,-\,2\alpha^{\star2}\, +\, \text{r}\,\alpha^{\star 3} \,=\,0\,,\qquad \text{r}:= 81\,N^2\,\frac{ J^{\prime2} }{Q^{\prime 3}_{1,2}}\,\sim\,\Lambda^0\,.
\end{equation}
The asymptotic growth of degeneracies comes from the root
$\alpha^\star$ with positive and maximal imaginary part of
\begin{equation}\label{eq:RefIntroduction}
\frac{2 \i \pi  (5 \alpha +1) Q^{\prime}_{1,2}}{3 \alpha }\longrightarrow \frac{2\pi}{3}  \frac{\text{Im}(\alpha^\star)}{|\alpha^\star|^2} Q_{1,2}^\prime\,.
\end{equation}
We note that only if
\begin{equation}\label{eq:InequalityR}
\left| r+\frac{20}{27}\right| >\frac{14 \sqrt{7}}{27}\, ,
\end{equation}
equation~\eqref{eq:CubicAlphaEquation} has non-real roots. Consequently, only in the chamber of charges consistent with~\eqref{eq:InequalityR} the present saddle point approximation predicts an exponential growth of states. For example if~$|\text{r}|$ is large enough
\begin{equation}
\alpha^\star \approx \text{(-\,\text{r})}^{-\frac{1}{3}}\,,
\end{equation}
and one recovers the asymptotic growth computed in the previous case~\eqref{eq:AsymptoticsLargeSpin}. On the contrary if~$\text{r}\approx0$ (i.e. for small enough~$J^\prime$ at fixed~$Q^\prime_{1,2}$) none of the saddle points of~$-\widetilde{s}(\underline{x};\underline{u}^\star)-\i \underline{x}\cdot \underline{Q}$ carries exponential growth: the leading saddle value becomes a highly oscillating phase times a bounded function. This feature is not surprising because we expect many more operators at large spin and fixed $R$-charge, than the other way around.

\section{Large charge entropy from giant gravitons}
\label{sec:GiantGravitons}

Let us define the following particularization of chemical potentials~$\underline{x}$ and charges~$\underline{\widetilde{Q}^\prime}$
\begin{equation}\label{eq:ConventionsXGG}
\begin{split}
-\i\underline{x}&\,=\,-\i\{x_1,x_2,x_3,x_4\}=\{\Delta_1,\Delta_2,\Delta_3,\omega_1\}\,,\\
\underline{\widetilde{Q}^\prime}&\,=\,\{\Qq_1,\Qq_2,\Qq_3,\widetilde{J}^\prime\}\,.
\end{split}
\end{equation}

Then we move on to compute the asymptotic growth of the giant graviton index~\eqref{eq:GiantGravImamura} at large positive integer charges~$\underline{\widetilde{Q}}^\prime$ 
\begin{equation}\label{eq:GGMicroIndexDefinition1}
\widetilde{d}_{GG}(\underline{\widetilde{Q}}^\prime)\,=\, \int_{\Gamma} d\underline{x}\,\sum_{n_1=0}^{\lfloor \Qq_1/N\rfloor}\sum_{n_2=0}^{\lfloor \Qq_2/N\rfloor}\sum_{n_3=0}^{\lfloor \Qq_3/N\rfloor}\int_{\Gamma_{gauge}} \frac{d\underline{u}}{n_1!n_2!n_3!}\, e^{-S^{(n_1,n_2,n_3)}_{\text{eff}}(\underline{x};\underline{u})-\i \underline{x}\cdot \underline{\widetilde{Q}}^\prime}\,,
\end{equation}
or more precisely, in a large-charge expansion (around~$\Lambda\to \infty$) defined by the scaling properties
\begin{equation}\label{eq:LargeRCharge}
\Qq_{1,2,3} = \Lambda^2 \widetilde{q}_{1,2,3}^\prime\,,\qquad  \widetilde{q}_{1,2,3}^\prime\,=\,\text{finite}\,.
\end{equation}
Before re-summation over the giant-graviton numbers $n$ has been taken, nothing will be assumed about the scaling properties of~$\widetilde{J}^\prime\,$ which can be an arbitrary function of~$\Lambda\,$. Eventually, we will assume~$\widetilde{J}^\prime\,$ to grow as~$\mathcal{O}(\Lambda^3)$. However, initially, the scaling properties of~$\widetilde{J}^\prime\,$ play no role in localizing the single giant-graviton contributions nor the sum over wrapping numbers~$n\,$. This is because there are no essential singularities at~$\omega_1\to0$ in the effective action of single-giant brane contributions. The scaling of~$\widetilde{J}^\prime\,$ will be essential to recover the growth of the giant graviton representation only after re-summation over the index~$n$ has been performed. Namely, in the last step when (as it will be shown below) the localization becomes equivalent to the one previously studied for the superconformal index.

As we summarized before
\begin{equation}\label{eq:EffectiveActionGiantGravitons}
{e^{-S^{(n_1,n_2,n_3)}_{\text{eff}}(\underline{x};\underline{u})}}\,:=\,w_1^{n_1 N} w_2^{n_2 N} w_3^{n_3 N} \,I^{4d}_{n_{1},n_{2},n_{3}}\,I^{2d}_{n_{1},n_{2},n_{3}}\, ,
\end{equation}
where the functions~$I^{2d,4d}_{n_{1},n_{2},n_{3}}$, defined in~\eqref{DefinitionPlethystic4d}, depend on the~$n_1+n_2+n_3$ gauge potentials~
\begin{equation}
\underline{u}=\{u^{(1)}_a,u^{(2)}_a,u^{(3)}_a\}\,.
\end{equation}
These potentials exponentiate to the rapidities~$U^{(I)}_a:=e^{2\pi\i u^{(I)}_a}\,$. Note that we have truncated the sums over $n_{1,2,3}\,$. That is because the truncated terms do not contribute to the counting of degeneracies at charges smaller or equal than~$\Qq_1$,~$\Qq_2$, and~$\Qq_3\,$ (the explanation was given in footnote~\ref{fn:Truncation}). 

The procedure to follow is summarized in the following steps:
\begin{itemize}
\item[1.] Commute the integral over~$\underline{x}$ with the sums over~$\underline{n}$~\footnote{The integral over~$\underline{x}$ can be commuted with the truncated sum over~$\underline{n}\,$, which  is finite. } 
\begin{equation}\label{eq:AuxiliaryEqu}
\int_{\Gamma} d\underline{x}\int_{\Gamma_{gauge}} d\underline{u}\, e^{-S^{(n_1,n_2,n_3)}_{\text{eff}}(\underline{x};\underline{u})-\i \underline{x}\cdot \underline{\widetilde{Q}}^\prime}\, .
\end{equation}
\item[2.] At large~$R$-charge the integral over~$\underline{u}$ is evaluated at its saddle point~$\underline{u}^\star$, while the integral over~$\underline{x}$ is localized as follows
\begin{equation}\label{eq:AsymptFormExpl}
\int_{\Gamma} d\underline{x}\int_{\Gamma_{gauge}} d\underline{u}\, e^{-S^{(n_1,n_2,n_3)}_{\text{eff}}(\underline{x};\underline{u})-\i \underline{x}\cdot \underline{\widetilde{Q}}^\prime}\,\simExp\,\int_{\Gamma_{\underline{x}^\star}} d\underline{x}\, e^{-\widetilde{s}^{(n_1, n_2, n_3)}(\underline{x};\underline{u}^\star)-\i \underline{x}\cdot \underline{\widetilde{Q}}^\prime}\,\,.
\end{equation} 
\item[3.] Use~\eqref{eq:AsymptFormExpl} in~\eqref{eq:AuxiliaryEqu} and substitute the result in~\eqref{eq:GGMicroIndexDefinition1}. Then commute the integral over~$\underline{x}$ with the sums over~$\underline{n}$ to obtain~\footnote{These integrals can be commuted because the localized integrand does not have poles: the logarithmic divergencies in the exponential are either suppressed or can be absorbed in a redefinition of gauge variables~$u\,$.}
\begin{equation}\label{eq:GGMicroIndexAsympt}
\widetilde{d}_{GG}(\underline{\widetilde{Q}}^\prime)\,\simExp\, \int_{\Gamma_{\underline{x}^\star}} d\underline{x}\,\sum_{n_1=0}^{\lfloor \Qq_1/N\rfloor}\sum_{n_2=0}^{\lfloor \Qq_2/N\rfloor}\sum_{n_3=0}^{\lfloor \Qq_3/N\rfloor}\, e^{-\widetilde{s}^{(n_1,n_2,n_3)}(\underline{x};\underline{u}^\star)-\i \underline{x}\cdot \underline{\widetilde{Q}}^\prime}\,.
\end{equation}
\item[4.] Evaluate the asymptotic behaviour of the sum over~$\underline{n}$ (in the large charge regime~\eqref{eq:LargeRCharge} we can safely drop the floor's) 
\begin{equation}\label{eq:ActionEffGGDefinition}
e^{-\widetilde{s}_{GG}(\underline{x})}\,:=\,\sum_{n_1=0}^{ \Qq_1/N}\sum_{n_2=0}^{ \Qq_2/N}\sum_{n_3=0}^{\Qq_3/N}\,e^{-\widetilde{s}^{(n_1, n_2, n_3)}(\underline{x};\,\underline{u}^\star)}\,.
\end{equation}
\item[5.] Substitute the \emph{entropy function of the gas of giant gravitons} $\widetilde{s}_{GG}(\underline{x})$ into~\eqref{eq:GGMicroIndexAsympt}, and localize the remaining integral over~$\underline{x}$ to the leading saddle point~$\underline{x}^\star$ which is the one attracted by the leading singularity of~$\widetilde{s}_{GG}(\underline{x})\,$
\begin{equation}
\widetilde{d}_{GG}(\underline{\widetilde{Q}}^\prime)\,\simExp\, e^{-\widetilde{s}_{GG}(\underline{x})}\,.
\end{equation}

\item[6.] At last, compare
\begin{equation}
\widetilde{d}_{GG}(\underline{\Qq}) \quad\text{ and}\quad \widetilde{d}(\underline{\Qq})\,=\,d(\underline{Q})\,, \qquad \text{(at large~$\widetilde{Q}^\prime\equiv{Q}$ and any~$N$)\,.}
\end{equation}

\end{itemize}

\subsection{A first approximation capturing the entropy of small black holes }
Let us start with step 2. Following our large charge localization lemma, we look for the leading singularities of~$S_{\text{eff}}^{(n_1,n_2,n_3)}$ which happen to be located at
\begin{equation}
x_{1,2,3}=x_{1,2,3,sing}\,:=\,0\,\text{mod}\, 1,
\end{equation}
and in their vicinity (the details behind the derivation of this formula are postponed to the following subsection)
\begin{equation}\label{eq:PropertiesGG}
\begin{split}
S^{(n_1,n_2,n_3)}_{\text{eff}}&({x}_{1,sing}+\frac{\delta{x}_{1}}{\Lambda},x_{2,sing}+\frac{\delta{x}_2}{\Lambda},x_{3,sing}+\frac{\delta{x}_3}{\Lambda},x_{4};\underline{u})\\&\underset{\Lambda\to \infty}{\sim}\Lambda\,\widetilde{s}_\Lambda^{(n_1, n_2,n_3)}(\delta x_1,\delta x_2, \delta x_3,x_4;\underline{u}) \\ & \underset{\Lambda\to \infty}{\sim}\,\widetilde{s}^{( n_1, n_2, n_3)}(\frac{\delta{x}_{1}}{\Lambda},\frac{\delta{x}_{2}}{\Lambda}, \frac{\delta{x}_{3}}{\Lambda},x_4;\underline{u})\,.
\end{split}
\end{equation}
This expansion holds at any value of~$n_1$,~$n_2\,$, and~$n_3\,$. Moreover,~$\widetilde{s}_\Lambda^{(n_1,n_2,n_3)}$ and~$\widetilde{s}^{(n_1,n_2,n_3)}$ are asymptotically-equal
\begin{equation}\label{eq:WHomogeneityGG}
\frac{\widetilde{s}_\Lambda^{(n_1, n_2,n_3)}(\delta x_1,\delta x_2, \delta x_3,x_4;\underline{u})}{\widetilde{s}^{(n_1, n_2,n_3)}(\delta x_1,\delta x_2, \delta x_3,x_4;\underline{u})} \underset{\Lambda\,\to\,\infty}{\sim}\,1\,.
\end{equation}
Assuming (for the moment)
\begin{equation}
\underline{u}^\star\, =\,O(\Lambda^{-1}) \quad \text{as} \quad \Lambda\to \infty\,, 
\end{equation}
we obtain for all~$\underline{n}$ and for all~$ N$ 
\begin{equation}\label{eq:tildeGGEffAct2}
\begin{split}
\widetilde{s}^{(n_1, n_2, n_3)}(\underline{x};\underline{u}^\star) &=\, T(\underline{x}) \Bigl(\underline{n}\cdot \underline{x}\Bigr)^2\,-\,\i N \Bigl(\underline{n}\cdot \underline{x}\Bigr)\,,
\end{split}
\end{equation}
with
\begin{equation}\label{eq:TDefB2}
T(\underline{x})\,:=\,-\frac{\frac{\pi ^2}{3}- \text{Li}_2\left(\frac{1}{p_1}\right)- \text{Li}_2\left(p_1\right)}{
   \Delta _1 \Delta _2 \Delta _3}\,+\,\widetilde{r}(\underline{x})\,=\,-\frac{\pi ^2 \left(1-6 \overline{B}_2\left(\frac{\omega_1}{2 \pi 
   i}\right)\right)}{3 \Delta _1 \Delta _2 \Delta _3}\,+\,\widetilde{r}(\underline{x}),
\end{equation}
where, again, these equations will be derived from scratch in the following section. The~$\overline{B}_2(x)$ in equation~\eqref{eq:TDefB2} is the periodic Bernoulli polynomial of order~$2$
\begin{equation}
\overline{B}_2(x)=B_2(x-\lfloor x\rfloor )\quad,\quad B_2(x)\,:=\,x^2\,-\, x\,+\,\frac{1}{6}\,.
\end{equation}

In this equation,~$\widetilde{r}$ comes from a subleading contribution to~$\widetilde{s}^{(n_1, n_2, n_3)}(\underline{x};\underline{u}^\star)$ which is a scale-invariant combination of~$\Delta_{1}\,$,~$\Delta_2\,$, and~$\Delta_3\,$.
Naively, one would say that discarding this contribution would not change the leading asymptotic behaviour of the giant graviton index (in microcanonical ensemble) at large charges and spin. However, as we will show below such an assumption turns out to be incorrect. In particular, at large~$N\,$, discarding~$\widetilde{r}$ does not give a chance to recover the counting of microstates of large BPS black holes. Instead, it allows, at most, to recover the entropy of small black holes i.e. those with large values of charges~$Q$, such that~$N^2\,\gg\,|Q|\,\gg\,1$~\cite{Choi:2022ovw}.

The contribution~$\widetilde{r}(\underline{x})\,$ turns out to be such that
\begin{equation}
\begin{split}
\frac{\widetilde{r}(\frac{\delta{x}_{1}}{\Lambda},\frac{\delta{x}_{2}}{\Lambda}, \frac{\delta{x}_{3}}{\Lambda},x_4)}{T(\frac{\delta{x}_{1}}{\Lambda},\frac{\delta{x}_{2}}{\Lambda}, \frac{\delta{x}_{3}}{\Lambda},x_4)} &\underset{\Lambda\to \infty}{\sim}\,0\,, \qquad x_4 \,\notin\, \mathbb{Z}\, ,
\\
\frac{\widetilde{r}(\frac{\delta{x}_{1}}{\Lambda},\frac{\delta{x}_{2}}{\Lambda}, \frac{\delta{x}_{3}}{\Lambda},x_4)}{T(\frac{\delta{x}_{1}}{\Lambda},\frac{\delta{x}_{2}}{\Lambda}, \frac{\delta{x}_{3}}{\Lambda},x_4)} \,&\underset{\Lambda\to \infty}{\sim}\,1\,,\qquad x_4 \,\in\, \mathbb{Z}\,.
\end{split}
\end{equation}
Thus,~$\widetilde{r}$ is subleading if~$x_4$ is far enough from~$\mathbb{Z}\,$. On the other hand if~$x_4$ is at distance~$O(\frac{1}{\Lambda})$ to the integers~$\mathbb{Z}\,$,~$\widetilde{r}$ becomes leading in the expansion~\eqref{eq:LargeRCharge}. Thus,~$\widetilde{r}$ cannot be ignored without the risk of missing leading contributions at large-charge saddle points infinitely attracted to integer values of the chemical potential~$x_4$. 
 
Step 4. further clarifies the relevance of~$\widetilde{r}\,$. The entropy functional~$-\widetilde{s}_{GG}(\underline{x})$ of the gas of giant gravitons is defined from
\begin{equation}\label{eq:ActionEffGGResumed}
e^{-\widetilde{s}_{GG}(\underline{x})}\,:=\,\sum_{n_1=0}^{ \Qq_1/N}\sum_{n_2=0}^{ \Qq_2/N}\sum_{n_3=0}^{\Qq_3/N}\,e^{-\widetilde{s}^{(n_1, n_2, n_3)}(\underline{x};\,\underline{u}^\star)}\,.
\end{equation}
To compute~\eqref{eq:ActionEffGGResumed} at large~$\widetilde{Q}^\prime_{1,2,3}$ (as detailed in~\eqref{eq:LargeRCharge}) it is convenient to change variables:
\begin{equation}
n_{1,2,3} = \frac{\Lambda^2}{N} \,\delta n_{1,2,3}\,.
\end{equation}
In the new variables the sums over~$n_{1,2,3}$ become integrals
\begin{equation}\label{eq:RefEq0}
\sum_{n_{1,2,3}=1}^{ \Qq_{1,2,3}/N}\,\underset{\Lambda\to \infty}{\to}\,\frac{\Lambda^2}{N}\,\int^{q_{1,2,3}}_0 d[\delta n_{1,2,3}]\,.
 \end{equation}
Precisely,
\begin{equation}\label{eq:IntegrandFlatDirections}
e^{-\widetilde{s}_{GG}(\underline{x})}\,\simExp\,\Bigl(\prod_{a=1}^3\frac{\Lambda^2}{N}\Bigr)\,\int_0^{\widetilde{q}^\prime_{1}} d[\delta n_1]\int_0^{\widetilde{q}^\prime_{2}} d[\delta n_2]\int_0^{\widetilde{q}^\prime_{3}} d[\delta n_3]\,e^{-\widetilde{s}^{(n_1, n_2, n_3)}(\underline{x};\underline{u}^\star)}\,.
\end{equation}
From~\eqref{eq:tildeGGEffAct2} it follows that this integral is Gaussian. Assuming for the time being that the~$x_a$ are real and positive (the general result can be obtained by analytic continuation) then in the variables
\begin{equation}\label{eq:changevarCondensate}
X=\delta n_1 x_1+\delta n_2 x_2+\delta n_3
   x_3\,,\quad Y\,=\, \delta n_2 x_2\,,\quad Z\,=\,\delta n_3 x_3\,,
\end{equation}
the integral measure (which acts upon an integrand that depends only on~$X$) becomes
\begin{equation}
\,\int_0^{\widetilde{q}^\prime_1 } d[\delta n_1]\int_0^{\widetilde{q}^\prime_2} d[\delta n_2]\int_0^{\widetilde{q}^\prime_3} d[\delta n_3]\,\to\, \int_0^{\underline{\widetilde{q}^\prime}\cdot \underline{x}} \frac{{d X}}{x_1x_2x_3} \,\cdot\, \mathcal{A}^{2d}_{\Sigma(X)} \, ,
\end{equation}
where
\begin{equation}
\mathcal{A}_{\Sigma(X)}\,:=\,\int_{\Sigma(X)} {dY} {dZ}\, , 
\end{equation}
is the area of a two-dimensional region~$\Sigma(X)$ spanned by pairs~$(Y,Z)\in\mathbb{R}^2$ such that
\begin{equation}
0<X-Y-Z<\widetilde{q}^\prime_1x_1\,,\qquad 0<Y<\widetilde{q}^\prime_2 x_2\,, \qquad 0<Z<\widetilde{q}^\prime_3 x_3\,.
\end{equation}
As~$\mathcal{A}_{\Sigma(X)}$ is the area of a polygonal surface whose perimeter has length growing linearly with~$X$ and/or~$\widetilde{q}^\prime_a {x}_a$, then~$\mathcal{A}_{\Sigma(X)}$ is always bounded from above by a polynomial function of~$X$ and~$\widetilde{q}^\prime_a {x}_a\,$. This is all we need to know about~$\mathcal{A}_{\Sigma(X)}$. 

Implementing the change of variables~\eqref{eq:changevarCondensate} and evaluating the one-loop saddle point approximation at large~$\Lambda$ one obtains
\begin{equation}\label{eq:SaddlePointCond}
\begin{split}
\int_0^{\widetilde{q}^\prime_1} d[\delta n_1]\int_0^{\widetilde{q}^\prime_2} d[\delta n_2]\int_0^{\widetilde{q}^\prime_3} d[\delta n_3]\,e^{-\widetilde{s}^{(n_1, n_2, n_3)}(\underline{x};\underline{u}^\star)}&\,\simExp\,\frac{N}{{\Lambda^2}}\,\frac{\sqrt{\pi}\mathcal{O}(\underline{x},\underline{\widetilde{q}^\prime})}{2\sqrt{T(\underline{x})}} e^{-\frac{N^2}{4 T(\underline{x})}}\,,
\end{split}
\end{equation}
where~$\mathcal{O}(\underline{x},\underline{\widetilde{q}^\prime})$ is the value of~$\frac{\mathcal{A}_{\Sigma(X)}}{x_1 x_2 x_3}$ at the saddle point locus
\begin{equation}\label{eq:LinearRelationsN}
(\underline{n}^\star\cdot \underline{x}) \,=\,\frac{\i N}{2T(\underline{x})}\quad \implies \quad(X^\star) \,=\,\frac{\i}{2T(\underline{x})}\,\frac{N^2}{\Lambda^2}, \qquad \forall\, \underline{x}\,.
\end{equation}
Collecting results one obtains
\begin{equation}
\begin{split}
e^{-\widetilde{s}_{GG}(\underline{x})}&\,
\simExp\,\frac{\Lambda^4}{N^2}\,\,\frac{\sqrt{\pi}\mathcal{O}(\underline{x},\underline{\widetilde{q}^\prime})}{2\sqrt{T(\underline{x})}} e^{-\frac{N^2}{4 T(\underline{x})}}\,.
\end{split}
\end{equation}
where
\begin{equation}\label{eq:Divergencies}
\frac{\Lambda^4}{N^2}\,\,\frac{\sqrt{\pi}\mathcal{O}(\underline{x},\underline{\widetilde{q}^\prime})}{2\sqrt{T(\underline{x})}} \simExp 1\,,\qquad \forall \,\underline{x}\,.
\end{equation}
At~$\underline{x}$ far enough from the zeroes of~$T$ the left-hand side of~\eqref{eq:Divergencies} diverges as the area spanned by two flat directions that open up in the moduli space of giant gravitons~$(n_{1},n_{2},n_{3})$ in the expansion~\eqref{eq:LargeRCharge} of the integrand~\eqref{eq:IntegrandFlatDirections}. This is because at leading order in such an expansion the integrand of~\eqref{eq:IntegrandFlatDirections} depends on a single direction in the three-dimensional space of~$(n_1,n_2,n_3)\,$'s: the other two directions become flat, and thus, summing over giant gravitons configurations along such directions produces an overall factor proportional to~$\frac{\Lambda^4}{N^2}\,$.

If and only if~$x$ is close enough to the zeroes of~$T(\underline{x})$, i.e. at distances of order~$\frac{1}{\Lambda}$ of them, then
\begin{equation}\label{eq:GGEffAction}
\widetilde{s}_{GG}(\underline{x})\, \underset{\Lambda\to \infty}{\sim} \,\frac{N^2}{4 T(\underline{x})}\,
\end{equation}
grows exponentially fast with~$\Lambda\,$.
Indeed, our large-charge localization lemma implies that the zeroes of~$T(\underline{x})$ which are the leading singularities of~$\widetilde{s}_{GG}\,$, determine the leading large-charge asymptotic behaviour of the integral
\begin{equation}\label{eq:GrowthGiantGravitons}
\begin{split}
\int_{\Gamma^\prime_{\underline{x}^\star}} d\underline{x}d\underline{u}\,e^{-\widetilde{s}_{GG}(\underline{x};\underline{u})-\i \underline{x}\cdot \underline{\widetilde{Q}}^\prime}&\,\simExp\,e^{-\frac{N^2}{4 T(\underline{x}^\star)}-\i \underline{x}^\star\cdot \underline{\widetilde{Q}}^\prime} \\ & \,\simExp\,\widetilde{d}_{GG}(\underline{\Qq})\,=\,a_{gg}(\mathfrak{Q})\,.
\end{split}
\end{equation}
where~$\underline{x}^\star$ is the leading saddle of~$\,-\frac{N^2}{4 T(\underline{x}^\star)}-\i \underline{x}^\star\cdot \underline{\widetilde{Q}}^\prime$ attracted by the zeroes of~$T(x)$. 

\paragraph{Control over subleading corrections in Step 2 is essential to recover large spin growth}
Step 6: Is the asymptotic growth in the index of giant gravitons~$\widetilde{d}_{GG}(\underline{\Qq})$ equal to the asymptotic growth of the superconformal index? i.e.
\begin{equation}\label{eq:ComparisonGrowth0}
\widetilde{d}_{GG}(\underline{\Qq}) \,\simExp\, \widetilde{d}(\underline{\Qq})\,=\,d(\underline{Q})\,, \qquad \text{(at any~$N$)\,}\,?
\end{equation}
In the chambers
\begin{equation}\label{eq:ChambersOmega1}
-1\,<\,\pm\text{Re}[\frac{\omega_1}{2\pi\i}]\,<\,0\,,
\end{equation}
the function~$T(\underline{x})$ is 
\begin{equation}\label{eq:Tnaive}
T(\underline{x})\,=\,\frac{ \omega _1 \left(\mp2 \pi\i\,-\,\omega _1 \right) }{2 \Delta _1 \Delta _2 \Delta _3}\,+\,\widetilde{r}(\underline{x})\,.
\end{equation}
If one naively substitutes~\eqref{eq:Tnaive} into the saddle point formula~\eqref{eq:GrowthGiantGravitons} assuming~$\widetilde{r}(\underline{x})\,\to\,0$, then one does not obtain the exponential growth at large spin of the superconformal index~\eqref{eq:AsymptoticsLargeSpin} (i.e. the degree of the singularity~$\omega_1\to 0$ or~$\mp\, 2\pi\i$ would be~$1<2$). 

Indeed, at large~$N$ and assuming~$\widetilde{r}(\underline{x})\,\to\,0$, the localized action~\eqref{eq:GGEffAction} can lead, at best, to the asymptotic growth of microstates of small black holes~\cite{Choi:2022ovw}\cite{Kinney:2005ej}. For example, if we assume~$\widetilde{r}=0$ and focus on the particular locus of charges~\cite{Choi:2022ovw}~\footnote{This is only for the moment, to recover the small-black hole entropy contributions, in the following sections we will comeback to the large-charge regime we are interested at~\eqref{eq:LargeRCharge} with $\widetilde{J}^\prime =\mathcal{O}(\Lambda^3)\,$.}
\begin{equation}
\qquad \widetilde{J}^\prime=:j=0\,,\qquad{\widetilde{Q}}_1^\prime={\widetilde{Q}}_1^\prime={\widetilde{Q}}_1^\prime=:-q
\end{equation}
then extremizing the entropy function
\begin{equation}
-\frac{N^2}{4 T(\underline{x})}-\i \underline{x}\cdot \underline{\widetilde{Q}}^\prime
\end{equation}
with respect to the chemical potentials~$\underline{x}$
\begin{equation}
\begin{split}
-\i\underline{x}&\,=\,-\i\{x_1,x_2,x_3,x_4\}=\{\Delta_1,\Delta_2,\Delta_3,\omega_1\}\,,\\
\underline{\widetilde{Q}^\prime}&\,=\,\{\Qq_1,\Qq_2,\Qq_3,\widetilde{J}^\prime\}\,,
\end{split}
\end{equation}
one obtains at the saddle point values
\begin{equation}\label{eq:saddlesSmallBHs}
\omega_1 = \mp \pi i\,, \qquad \Delta_1^2=\Delta_2^2=\Delta_3^2=\Delta^2 = \frac{2 \pi^2 q}{N^2}\,,
\end{equation}
and for~$q>0$ the following prediction for the entropy
\begin{equation}
\frac{2 \sqrt{2} \pi  q^{3/2}}{N}\,.
\end{equation}
\footnote{Note that for finite~$N$ and~$q\to 1$ the singularity of the localized action~$\frac{N^2}{4 T(\underline{x})}$ that attracts the saddle~\eqref{eq:saddlesSmallBHs} is not~$\omega_1=0$ but~$\Delta\,=\,\infty\,$.}
In the asymptotic regime~$N^2\,\gg\, q\,\gg\,1$ this is the leading term of the Bekenstein-Hawking entropy of small and supersymmetric black holes in~$AdS_5$ with equal left and right angular momenta~$j=0\,$~\cite{Choi:2022ovw}.~\footnote{Compare with the leading contribution in the first line of equation (2.26) of~\cite{Choi:2022ovw}.}

In order for~\eqref{eq:ComparisonGrowth0} to hold, namely in order to obtain the asymptotic growth of the most generic index at large charges which are not too small in comparison with~$N^2\,$, it is necessary that 
\begin{equation}\label{eq:FinalT}
T(\underline{x})\,=\,T_0(\underline{x})\,:=\,-\frac{\left(\omega _1\right) \left(\underline{-\Delta _1-\Delta _2-\Delta _3}+\omega _1\pm2  \pi\i
   \right) }{2 \Delta _1
   \Delta _2 \Delta _3}\, .
\end{equation}
In particular, this also means that if the underlined contribution does not match the microscopic prediction of~$\widetilde{r}(\underline{x})\,$ then the growth of the series of giant graviton indices can not account for the large charge growth of the complete superconformal index. In the following subsection we proceed to check whether $\widetilde{r}(\underline{x})$ equals
\begin{equation}\label{eq:DesiredCorrections}
\,+\,\frac{\omega _1\left(\underline{\Delta _1+\Delta _2+\Delta _3}
   \right) }{2 \Delta _1
   \Delta _2 \Delta _3}\,.
\end{equation}

\subsection{Refined calculation and large black hole entropy}\label{sec:LargeBlackHole}

In this subsection the localized form~$\widetilde{s}^{(n_1,n_2,n_3)}$ of the giant graviton effective action~$S^{(n_1,n_2,n_3)}_{\text{eff}}\,$ is computed. We follow the steps summarized below the equation~\eqref{eq:AuxiliaryEqu}. 

The first step is to compute the asymptotic expansion~$S^{(n_1,n_2,n_3)}_{\text{eff}}\,$ near its leading singularity(ies). 

Let us divide the effective action in three pieces (and omit the supra indices~$n_{1,2,3}$ for a moment)
\begin{equation}\label{eq:SeffGG}
S_{\text{eff}}^{(n_1,n_2,n_3)}(\underline{x};\underline{u})\,=\,\sum_{I=1} ^{3}\,S^{(I)}_{\text{ZM}}(\underline{x})\,+\,\sum_{I=1} ^{3}\,S^{(I)}_{\text{Vect}}(\underline{x};\underline{u})\,+\,\sum_{I=1} ^{3}\,S^{(I)}_{\text{Hypers}}(\underline{x};\underline{u})\,,
\end{equation}

\begin{equation}
S^{(I)}_{\text{ZM}}(\underline{x})\,=\,\sum_{l=1}^{\infty}\frac{n_{I}}{l}\frac{(1-w_{I}^{-l})(1-p_{1}^{l})(1-p_{2}^{l})-(1-w_{J}^{l})(1-w_{K}^{l})}{(1-w_{J}^{l})(1-w_{K}^{l})}\,+\,\log(n_I!)\,,
\end{equation}

\begin{equation}
S^{(I)}_{\text{Vect}}(\underline{x};\underline{u})= \sum_{l=1}^{\infty}\frac{\left(1-p_1^l\right) \left(1-p_2^l\right) \left(1-w_I^{-l}\right) }{l \left(1-w_J^l\right) \left(1-w_{K}^l\right)}\sum_{a\neq b=1}^{n_J}{U_{ab}^{(I)l}
   }\,,
\end{equation}
\begin{equation}
S^{(I)}_{\text{Hypers}}(\underline{x};\underline{u})= -\sum_{l=1}^{\infty}\frac{\left(1-p_1^l\right) \left(1-p_2^l\right) w_{J}^{-l/2} w_{K}^{-l/2} }{l \left(1-w_I^l\right)}\,\sum_{a=1}^{n_J}\sum_{b=1}^{n_K}\left({U_{ab}^{(J,K)}}+{U_{ba}^{(K,J)}}\right)\,,
\end{equation}

\begin{equation}
U_a^{(I)}\,:=\,\exp(2\pi\i u^{(I)}_a)\,,\, \qquad a\,=\,1\,,\ldots\,,n_I\,.
\end{equation}

We proceed to compute the expansion~$\Lambda\to\infty$ of
\begin{equation}
S_{\text{eff}}^{(n_1,n_2,n_3)}\Bigl(\frac{x_1}{\Lambda},\frac{x_2}{\Lambda},\frac{x_3}{\Lambda},x_4;\underline{u}\Bigr)
\end{equation}
or equivalently the expansion of each of the four contributions in~\eqref{eq:SeffGG} and extract its localized form~$\widetilde{s}^{(n_1,n_2,n_3)}$.

To compute this expansion we proceed as follows
\begin{itemize}
\item[ 1.] Substitute
\begin{equation}
w_{a}\rightarrow  e^{-\epsilon \Delta_{a}}\,, \,\qquad 
\end{equation}
in the \underline{denominators of the summands} of the zero-modes action~$S^{(I)}_{\text{ZM}}$ and in the \underline{numerators} and \underline{denominators} of the summands of the non-zero modes actions~$S^{(I)}_{\text{Vect}}$ and~$S^{(I)}_{\text{Hypers}}$ and expand about~$\epsilon \,=\,\frac{1}{\Lambda}\sim 0\,$.

\item[2.] Perform the sums $\sum_{l=1}^{\infty}$ in the result obtained after step 1.
\item[3.] Substitute 
\begin{equation}
p_2 \to \frac{w_1 w_2 w_3}{p_1}\,,\qquad w_{a} \to e^{-\epsilon \Delta_a}\,,
\end{equation}
in the result obtained after steps 1. and 2. and expand the answer around~$\epsilon=0$ up to order~$O(\epsilon^0)$ being careful about logarithmic singularities.
\item [4.] Lastly, truncate the series at order $O(\epsilon^0)$, and re-scale back the variables 
\begin{equation}
\Delta_a \to \frac{\Delta_a}{\epsilon}\,,
\end{equation}
to obtain an~$\epsilon$-independent effective action. Such an answer is the contribution of~$S_{\text{ZM},\,\text{Vect},\,\text{Hypers}}$, respectively, to the localized action~$\widetilde{s}^{n_1,n_2,n_3}(\underline{x};\underline{u})\,$.

\end{itemize}

Using these steps allows us to keep control over logarithmic corrections that appear in the expansion~$\Lambda\to \infty$ (coming from the action of vector zero-modes). Proceeding otherwise these non-analyticities would evidence themselves as infinite coefficients in the would-be-Laurent expansion around~$\Lambda = \infty\,$.

\emph{The large charge effective action of zero modes:} \label{paragraph:ZeroModes}
~Let us start computing the large charge effective action of zero modes following steps 1-4. To illustrate the procedure let us focus on a single zero mode contribution of the vector multiplet 1:
\begin{equation}
S^{(1)}_{\text{ZM}}(\underline{x})=\sum_{l=1}^{\infty}\frac{n_{1}}{l}\frac{(1-w_{1}^{-l})(1-p_{1}^{l})(1-p_{2}^{l})-(1-w_{2}^{l})(1-w_{3}^{l})}{(1-w_{2}^{l})(1-w_{3}^{l})}\,.
\end{equation}
After steps 1. and 2. we obtain for all~$n_{1}$, at order~$O(\epsilon^{-2})$
\begin{equation}\label{eq:ZMepsilonExp20}
\begin{split}
&-\frac{\mathrm{Li}_3(p_1) n_1}{\epsilon^2\Delta_2 \Delta_3} -\frac{\mathrm{Li}_3(p_2) n_1}{\epsilon^2\Delta_2 \Delta_3} +\frac{\mathrm{Li}_3(p_1 p_2) n_1}{\epsilon^2\Delta_2 \Delta_3} \\
&-\frac{\mathrm{Li}_3(1/w_1) n_1}{\epsilon^2\Delta_2 \Delta_3} +\frac{\mathrm{Li}_3(p_1/w_1) n_1}{\epsilon^2\Delta_2 \Delta_3} +\frac{\mathrm{Li}_3(p_2/w_1) n_1}{\epsilon^2\Delta_2 \Delta_3} \\
&-\frac{\mathrm{Li}_3(p_1 p_2/w_1) n_1}{\epsilon^2\Delta_2 \Delta_3} +\frac{\mathrm{Li}_3(w_2) n_1}{\epsilon^2\Delta_2 \Delta_3} +\frac{\mathrm{Li}_3(w_3) n_1}{\epsilon^2\Delta_2 \Delta_3} -\frac{\mathrm{Li}_3(w_2 w_3) n_1}{\epsilon^2\Delta_2 \Delta_3}\, ,
\end{split}
\end{equation}
and at order~$O(\epsilon^{-1})$
\begin{equation}\label{eq:ZMepsilonExp1}
\begin{split}
&-\frac{n_1 \text{Li}_2\left(p_1\right)}{2 \epsilon\Delta _2}-\frac{n_1
\text{Li}_2\left(p_2\right)}{2 \epsilon\Delta _2}+\frac{n_1 \text{Li}_2\left(p_1
p_2\right)}{2 \epsilon\Delta _2}-\frac{n_1 \text{Li}_2\left(p_1\right)}{2 \epsilon\Delta
_3}-\frac{n_1 \text{Li}_2\left(p_2\right)}{2 \epsilon\Delta _3}\\&+\frac{n_1
\text{Li}_2\left(p_1 p_2\right)}{2\epsilon \Delta _3}+\frac{n_1 \text{Li}_2\left(\frac{p_1}{w_1}\right)}{2 \epsilon\Delta _2}+\frac{n_1
\text{Li}_2\left(\frac{p_2}{w_1}\right)}{2 \epsilon\Delta _2}-\frac{n_1
\text{Li}_2\left(\frac{p_1 p_2}{w_1}\right)}{2 \epsilon\Delta _2}+\frac{n_1
\text{Li}_2\left(\frac{p_1}{w_1}\right)}{2 \epsilon\Delta _3}\\&+\frac{n_1
\text{Li}_2\left(\frac{p_2}{w_1}\right)}{2 \epsilon\Delta _3}-\frac{n_1
\text{Li}_2\left(\frac{p_1 p_2}{w_1}\right)}{2 \epsilon\Delta _3}-\frac{n_1 \text{Li}_2\left(\frac{1}{w_1}\right)}{2 \epsilon\Delta _2}+\frac{n_1
\text{Li}_2\left(w_2\right)}{2 \epsilon\Delta _2}+\frac{n_1 \text{Li}_2\left(w_3\right)}{2
\epsilon\Delta _2}\\ &-\frac{n_1 \text{Li}_2\left(w_2 w_3\right)}{2 \epsilon\Delta _2}-\frac{n_1
\text{Li}_2\left(\frac{1}{w_1}\right)}{2 \epsilon\Delta _3}+\frac{n_1
\text{Li}_2\left(w_2\right)}{2 \epsilon\Delta _3}+\frac{n_1 \text{Li}_2\left(w_3\right)}{2
\epsilon\Delta _3}-\frac{n_1 \text{Li}_2\left(w_2 w_3\right)}{2 \epsilon\Delta _3}\,,
\end{split}
\end{equation}
and at order~$O(\epsilon^0)$
\begin{equation}\label{eq:ZMepsilonExp2}
\begin{split}
&\frac{\left(\Delta_2^2+3\Delta_2\Delta_3+\Delta_3^2\right)n_1}{12\Delta_2\Delta_3} \\
&\quad\times\left.\Biggl(\log\left(1-p_1\right)+\log\left(1-p_2\right)-\log\left(1-p_1p_2\right)+\log\left(1-\frac{1}{w_1}\right) \right. \\
&\qquad\qquad\left.-\log\left(1-\frac{p_1}{w_1}\right)-\log\left(1-\frac{p_2}{w_1}\right)+\log\left(1-\frac{p_1p_2}{w_1}\right) \right. \\
&\qquad\qquad\left.-\log\left(1-w_2\right)-\log\left(1-w_3\right)+\log\left(1-w_2w_3\right)\right.\Biggr)\,.
\end{split}
\end{equation}
Then, after adding~\eqref{eq:ZMepsilonExp20},~\eqref{eq:ZMepsilonExp1} and~\eqref{eq:ZMepsilonExp2} and implementing step 3., we obtain at order~$\epsilon^{-1}$
\begin{equation}
-\frac{1}{\epsilon}\,\frac{\pi ^2 \Delta _1 n_1 \left(1-6 \overline{B}_2\left(-\frac{i \omega _1}{2 \pi
   }\right)\right)}{3 \Delta _2 \Delta _3}\, ,
\end{equation}
 and at order~$\epsilon^{0}$
 \begin{equation}
\begin{split}
&+\,\frac{\Delta_1 n_1\left(\Delta_1+\Delta_2+\Delta_3\right)\omega_1}{2 \Delta_2 \Delta_3}\,+\,\log\Xi_{1}(\epsilon \underline{x}) n_1\, ,
\end{split}
\end{equation}
where (assuming for the moment~$\epsilon\,>\,0$,~$\Delta_{I}\,>\,0$)
\begin{equation}\label{eq:Xi1}
\begin{split}
\log\Xi_{1}(\underline{x})
&=\frac{ (\Delta_1 \Delta_2+\Delta_1\Delta_3-\Delta_2\Delta_3)}{2 \Delta_2 \Delta_3}\\ & -\frac{\left(\Delta _2^2-3 \Delta _3 \Delta _2+\Delta _3^2\right) \log \left(-\Delta _2 \Delta _3\right)}{12 \Delta _2 \Delta _3}\\
&+\frac{\left(\Delta _2^2+3 \Delta _3 \Delta _2+\Delta _3^2\right) \log \left(\Delta _2+\Delta _3\right)}{6 \Delta _2 \Delta _3}
\\
&+\frac{\left(6 \Delta _1^2+6 \left(\Delta _2+\Delta _3\right)\Delta _1+\Delta _2^2+\Delta _3^2+3 \Delta _2 \Delta _3\right) \log \left(\frac{\Delta _1}{\Delta _1+\Delta _2+\Delta _3}\right)}{12 \Delta _2 \Delta _3}\,.
\end{split}
\end{equation}~\footnote{We note that the term in the first line can be absorbed in a redefinition of the argument of the second out of the three logarithms in the second line and the second and third out of the six logarithms in the fourth line. This implies that the~$\log \Xi_1(\underline{x}) n_1$ contribution is of type-$F$ and thus it will not contribute at the degree of accuracy we are looking for. We will keep track of these contributions though, as we may learn something for the future.}
At last, implementing step 4 we obtain the contribution of the zero mode 1 to the large charge action~$\widetilde{s}^{(n_1,n_2,n_3)}(\underline{x};\underline{u})$
\begin{equation}
\begin{split}&-\frac{\pi ^2 \Delta _1 n_1 \left(1-6 \overline{B}_2\left(-\frac{i \omega _1}{2 \pi
   }\right)\right)}{3 \Delta _2 \Delta _3}\,+\,\frac{\Delta_1 n_1\left(\Delta_1+\Delta_2+\Delta_3\right)\omega_1}{2 \Delta_2 \Delta_3} \\
&\,+\,\log\Xi_{1}(\underline{x}) n_1\, .
\end{split}
\end{equation}
The contribution coming from the zero modes 2 and 3 are computed analogously. The general result is
\begin{equation}\label{eq:IntermediateZM}
\begin{split}&-\frac{\pi ^2 \Delta_I n_I \left(1-6 \overline{B}_2\left(-\frac{i \omega _1}{2 \pi
   }\right)\right)}{3 \Delta_{J} \Delta _{K}}\,+\,\frac{\Delta_I n_1\left(\Delta_1+\Delta_2+\Delta_3\right)\omega_1}{2 \Delta_J \Delta_K}\\
&\,+\,\log\Xi_{I}(\underline{x}) n_I\, ,
\end{split}
\end{equation}
where the definition of~$\log\Xi_I$ can be recovered from~\eqref{eq:Xi1} by the obvious permutation of subscripts.
Assuming~\eqref{eq:ChambersOmega1}, equation~\eqref{eq:IntermediateZM} can be rewritten as
\begin{equation}\label{eq:ZM}
\frac{\Delta_I^2 n_I \,\omega _1 \left(\mp\,2 i \pi+\Delta _1+\Delta _2+\Delta _3-\omega _1\,
   \right)}{2 \Delta _1 \Delta _2 \Delta _3}\,+\,\log\Xi_{I}(\underline{x}) n_I\,.
\end{equation}
Note that the contributions~$\log\Xi_{I}(\underline{x}) n_I$ coming from zero-modes are of the type~$F$ defined around~\eqref{eq:FTypeContribution} and thus we can ignore them in the following. However, to gain insight into their meaning, we will keep track of them from now on.

\emph{The large charge effective action of vector non-zero modes}
To start let us focus on the contribution of the vector multiplet 1:
\begin{equation}
S^{(1)}_{\text{Vect}}(\underline{x})\,=\,\sum_{a\neq b=1}^{n_1}\sum_{l=1}^{\infty}\frac{1}{l}\frac{(1-w_{1}^{-l})(1-p_{1}^{l})(1-p_{2}^{l})}{(1-w_{2}^{l})(1-w_{3}^{l})}\,U^{(1) l}_{ab}\, .
\end{equation}
After steps 1.- 3. we obtain for all~$n_{1}$, at order~$\epsilon^{-1}$
\begin{equation}\label{eq:Vector1EpsilonM1}
\frac{1}{\epsilon}\sum_{a\neq b=1}^{n_1}\,\,\frac{\pi^2\Delta _1 \left(6
\overline{B}_2\left(-u^{(1)}_{ab}-\frac{i \omega _1}{2 \pi }\right)\,-\,\frac{6}{\pi^2}\text{Li}_2\left(U^{(1)}_{ab}\right)\right)}{3\Delta _2 \Delta _3 }\, ,
\end{equation}
and at order~$\epsilon^0$
\begin{equation}\label{eq:SubleadingVector1}
\sum_{a\neq b=1}^{n_1}\,\frac{\Delta _1 \left(\Delta _1+\Delta _2+\Delta _3\right) \left(\log
   \left(1-\frac{1}{p_1}U^{(1)}_{ab}\right)-\log \left(1-{p_1 U^{(1)}_{ab}}\right)\right)}{2
   \Delta _2 \Delta _3}\,.
\end{equation}
Using the relations
\begin{equation}
\begin{split}
\log \left(1-\frac{p_1 U^{(1)}_a}{U^{(1)}_b}\right) &\,=\, \log \left(-\frac{p_1 U^{(1)}_a}{U^{(1)}_b}\right)+\log
   \left(1-\frac{U^{(1)}_b}{p_1 U^{(1)}_a}\right)\\
   \log \left(-\frac{p_1 U^{(1)}_a}{U^{(1)}_b}\right)&\,=\, \log \left(-\frac{U^{(1)}_a}{U^{(1)}_b}\right)-\omega _1\,,\\
   \sum_{a\neq b=1}^{n_1}\log \left(-\frac{U^{(1)}_a}{U^{(1)}_b}\right) &\,=\,\log \left(\prod_{a\neq b=1}^{n_1}\left(-\frac{U^{(1)}_a}{U^{(1)}_b}\right)\right)=\log\left(1\right)\,=\,0\,,
   \end{split}
\end{equation}
\eqref{eq:SubleadingVector1} simplifies into
\begin{equation}\label{eq:Vector1Epsilon0}
\sum_{a\neq b=1}^{n_1}\,\frac{\Delta _1 \left(\Delta _1+\Delta _2+\Delta _3\right) \omega _1}{2 \Delta _2
   \Delta _3}\,.
\end{equation}
At last, adding~\eqref{eq:Vector1EpsilonM1} and~\eqref{eq:Vector1Epsilon0} and implementing step 4 we obtain the contribution of the vector modes 1 to the coarse grained action~$\widetilde{s}^{(n_1,n_2,n_3)}(\underline{x};\underline{u})$
\begin{equation}
\begin{split}
&\sum_{a\neq b=1}^{n_1}\,\, \frac{\pi^2\Delta _1 \left(6
\overline{B}_2\left(-u^{(1)}_{ab}-\frac{i \omega _1}{2 \pi }\right)\,-\,\frac{6}{\pi^2}\text{Li}_2\left(U^{(1)}_{ab}\right)\right)}{3\Delta _2 \Delta _3 }\\ &+n_1(n_1-1)\,\frac{\Delta _1 \left(\Delta _1+\Delta _2+\Delta _3\right) \omega _1}{2 \Delta _2
   \Delta _3}\,.
   \end{split}
\end{equation}
The contribution coming from the vector modes 2 and 3 are computed analogously. The general result is
\begin{equation}\label{eq:Vectors}
\begin{split}
&\sum_{a\neq b=1}^{n_I}\,\, \frac{\pi^2\Delta _I \left(6
\overline{B}_2\left(-u^{(I)}_{ab}-\frac{\i \omega _1}{2 \pi }\right)\,-\,\frac{6}{\pi^2}\text{Li}_2\left({U^{(I)}_{ab}}\right)\right)}{3\Delta _J \Delta _K }\\ &+n_I(n_I-1)\,\frac{\Delta _I \left(\Delta _1+\Delta _2+\Delta _3\right) \omega _1}{2 \Delta _J
   \Delta_K}\,.
   \end{split}
\end{equation}

\emph{The large charge effective action of hypermultiplets}
To start let us focus on the contribution of the hypermultiplet 3:
\begin{equation}
S^{(3)}_{\text{Hypers}}(\underline{x};\underline{u})= -\,\sum_{a=1}^{n_1}\sum_{b=1}^{n_2}\sum_{l=1}^{\infty}\frac{\left(1-p_1^l\right) \left(1-p_2^l\right) w_{1}^{-l/2} w_{2}^{-l/2} }{l \left(1-w_3^l\right)}\left(U^{(1,2)}_{ab}+U^{(2,1)}_{ba}\right)
\end{equation}
After steps 1.- 3. we obtain for all~$n_{1}$, at order~$\epsilon^{-1}$
\begin{equation}\label{eq:Hyper3EpsilonM1}
\frac{1}{\epsilon}\,\sum_{a=1}^{n_1}\sum_{b=1}^{n_2}\frac{2 \left(\pi ^2 B_2\left(u^{(1,2)}_{ab}-\frac{\i \omega _1}{2 \pi
   }\right)+\pi ^2 B_2\left(u^{(2,1)}_{ba}-\frac{\i \omega _1}{2 \pi
   }\right)-\text{Li}_2\left(U^{(1,2)}_{ab}\right)-\text{Li}_2\left(U^{(2,1)}_{ba}\right)\right)}{\Delta _3}
\end{equation}
and at order~$\epsilon^0$
\begin{equation}\label{eq:HyperEpsilon0First}
\frac{\left(\Delta _1+\Delta _2+\Delta _3\right) \left(\log \left(1-\frac{U^{(1,2)}_{ab}}{p_1}\right)-\log \left(1-{p_1 U^{(1,2)}_{ab}}\right)+\log \left(1-\frac{ U^{(2,1)}_{ba}}{p_1
  }\right)-\log \left(1-{p_1}U^{(2,1)}_{ba}\right)\right)}{2 \Delta _3}\,.
\end{equation}
Using the relations
\begin{equation}
\begin{split}
\log \left(1-\frac{p_1 U^{(1)}_a}{U^{(2)}_b}\right) &\,=\, \log \left(-\frac{p_1 U^{(1)}_a}{U^{(2)}_b}\right)+\log
   \left(1-\frac{U^{(1)}_a}{p_1 U^{(2)}_b}\right)\, ,\\
   \log \left(-\frac{p_1 U^{(1)}_a}{U^{(2)}_b}\right)&\,=\, \log \left(-\frac{U^{(1)}_a}{U^{(2)}_b}\right)-\omega _1\,,
   \end{split}
\end{equation}
together with the analogous one for $U^{(2,1)}_{ba}$ and
\begin{equation}
  \log \left(-\frac{U^{(1)}_a}{U^{(2)}_b}\right)+\log \left(-\frac{U^{(2)}_b}{U^{(1)}_a}\right) \,=\,\log\left(1\right)\,=\,0\,,
\end{equation}
\eqref{eq:HyperEpsilon0First} reduces to
\begin{equation}\label{eq:Hyper3Epsilon0}
\sum_{a=1}^{n_1}\sum_{b=1}^{n_2}
\,\frac{\left(\Delta _1+\Delta _2+\Delta _3\right) \omega _1}{\Delta _3}\,.
\end{equation}
At last, adding~\eqref{eq:Hyper3EpsilonM1} and~\eqref{eq:Hyper3Epsilon0} and implementing step 4 we obtain the contribution of the hypermultiplet 3 to the coarse grained action~$\widetilde{s}^{(n_1,n_2,n_3)}(\underline{x};\underline{u})$
\begin{equation}\label{eq:Hyper3}
\begin{split}
\sum_{a=1}^{n_1}\sum_{b=1}^{n_2}\,&\frac{2 \left(\pi ^2 B_2\left(u^{(1,2)}_{ab}-\frac{\i \omega _1}{2 \pi
   }\right)-\text{Li}_2\left(U^{(1,2)}_{ab}\right)\,+\,\text{symm}\,\right)}{\Delta _3}
    \\ &\,+\, n_1 n_2 \,\frac{\left(\Delta _1+\Delta _2+\Delta _3\right) \omega _1}{\Delta _3}\,.
    \end{split}
\end{equation}
The contribution coming from the hypermultiplets 1 and 2 are computed analogously. The general result is
\begin{equation}\label{eq:Hypers}
\begin{split}
\sum_{a=1}^{n_J}\sum_{b=1}^{n_K}\,&\frac{2 \left(\pi ^2 \overline{B}_2\left(u^{(J,K)}_{ab}-\frac{\i \omega _1}{2 \pi
   }\right)-\text{Li}_2\left(U^{(J,K)}_{ab}\right)\,+\,\text{symm}\,\right)}{\Delta _I}
    \\ &\,+\, n_J n_K \,\frac{\left(\Delta _1+\Delta _2+\Delta _3\right) \omega _1}{\Delta _I}\,.
    \end{split}
\end{equation}
\emph{The coarse grained action:}
Collecting~\eqref{eq:ZM},~\eqref{eq:Vectors}, and~\eqref{eq:Hypers} we obtain, at last,
\begin{equation}\label{eq:CoarseGrainedEffActionGiantG0}
\begin{split}
\widetilde{s}^{(n_1,n_2,n_3)}(\underline{x};\underline{u})\,:=&\,\sum_{I=1}^{3}\sum_{a,\, b=1}^{n_I}\,\, \frac{2\Delta _I \left(\pi^2
\overline{B}_2\left(-u^{(I)}_{ab}-\frac{\i \omega _1}{2 \pi }\right)\,-\,\text{Li}_2\left({U^{(I)}_{ab}}\right)\right)}{\Delta _J \Delta _K }\\ +&\sum_{I=1}^3\sum_{a=1}^{n_J}\sum_{b=1}^{n_K}\,\frac{2 \left(\pi ^2 \overline{B}_2\left(u^{(J,K)}_{ab}-\frac{\i \omega _1}{2 \pi
   }\right)-\text{Li}_2\left(U^{(J,K)}_{ab}\right)\,+\,(J,a) \leftrightarrow (K,b)\,\right)}{\Delta _I}\\ \,+\,&(n_1 \Delta_1+n_2\Delta_2+n_3\Delta_3)^2\frac{(\Delta_1+\Delta_2+\Delta_3)}{2\Delta_1\Delta_2\Delta_3}\,\omega_1\\\,+\,&n_1 \,\log\Xi_1(\underline{x})\,+\,n_2\,\log\Xi_2(\underline{x})\,+\,n_3\, \log\Xi_3(\underline{x})\,.
\end{split}
\end{equation}
We note that contributions coming from zero modes have been obtained in a certain choice of branch that has simplified computations for us. Other choices of branch would give us a different answer. However, as we will show next, these contributions can be absorbed in a redefinition of the gauge variables which does not affect the leading saddle point evaluation. This is, at least at large charges the ambiguities coming from zero modes are indistinguishable from gauge-choice ambiguities and thus they do not affect the indices of giant graviton branes. 
\paragraph{The gauge saddle point~$\underline{u}^\star$}
The next step is to find the leading saddle points~$\underline{u}^\star$~of
\begin{equation}\label{eq:integralInterm}
\,\frac{1}{n_1! n_2! n_3!}\,\int_{\Gamma_{u^\star}} d\underline{u}\, e^{-\widetilde{s}^{(n_1,n_2,n_3)}(\underline{x};\underline{u})}\,:=\,\frac{1}{n_1! n_2! n_3!}\,\int_{\Gamma_{u^\star}} \prod_{i=1}^{n_1} du^{(1)}_i\cdot\prod_{i=1}^{n_2} du^{(2)}_i\cdot\prod_{i=1}^{n_3} du^{(3)}_i \cdot e^{-\widetilde{s}^{(n_1,n_2,n_3)}(\underline{x};\underline{u})}
\end{equation}
in the small $\epsilon=\frac{1}{\Lambda}$ expansion at fixed~$n_{1,2,3}$, this is, assuming
\begin{equation}
\Delta_I\,=\,O({\Lambda}^{-1})\,, \qquad \Lambda\,\to\,\infty\,.
\end{equation}
After changing integration variables
\begin{equation}\label{eq:ChangeVariables}
u^{(I)}_a\,\to\, {u^{(I)}_a}{\Xi^{(I)}(\underline{x})}\,,
\end{equation}
equation~\eqref{eq:integralInterm} transforms into an integral over a new contour~$\Gamma^\Xi_{u^\star}$
\begin{equation}
\frac{1}{n_1! n_2! n_3!}\int_{\Gamma^{\Xi}_{u^\star}} \prod_{i=1}^{n_1} {du^{(1)}_i}\cdot\prod_{i=1}^{n_2} du^{(2)}_i\cdot\prod_{i=1}^{n_3} du^{(3)}_i \cdot e^{-\widetilde{s}^{\Xi(n_1,n_2,n_3)}(\underline{x};\underline{u})}
\end{equation}
with the new action taking the form 
\begin{equation}
\begin{split}
\widetilde{s}^{\Xi(n_1,n_2,n_3)}(\underline{x};\underline{u})&\,:=\,\widetilde{s}^{(n_1,n_2,n_3)}(\underline{x};{u}^{(1)}\Xi_1,{u}^{(2)}\Xi_2,{u}^{(3)}\Xi_3)\\ &-\underline{n}_1\,\log\underline{\Xi}_1\,-\,\underline{n}_2\,\log\underline{\Xi}_2\,-\,\underline{n}_3\,\log\underline{\Xi}_3\,.
\end{split}
\end{equation}
Following our large charge localization rules we scale the chemical potentials
\begin{equation}
\Delta_a\to \epsilon \Delta_a\,,
\end{equation}
and plug the ansatz
\begin{equation}
\underline{u}^{(I)\star}_a \,=\, \sum_{k=0}^{\infty}\,u^{(I)}_{a,k}(\underline{x})\,\epsilon^k\,+\,h^{(I)}_{a,0}(\underline{x})\,\epsilon^2\log{\epsilon}\,+\,\ldots\, , 
\end{equation}
\footnote{In this equation the~$\underline{u}$'s and~$h$'s depend on combinations of~$\Delta_{1,2,3}$ and the dots stand for possible higher order logarithmic terms, vanishing at~$\epsilon=0$. The factor of~$\epsilon^2$ is the minimal integer power of~$\epsilon$ that it does not produce singularities of the form~$\frac{\log\epsilon}{\epsilon}$ in the saddle-point conditions, i.e., it produces a singularity of the form~$\log\epsilon$ which can be used to cancel other such singular contributions to the saddle-point condition. Because we have used the change of variables~\eqref{eq:ChangeVariables} it follows from the saddle point conditions that~$h^{(I)}_{a,0}=0\,$.} into the saddle point equations following from the action
\begin{equation}\label{eq:CoarseGrainedEffAction}
\widetilde{s}^{\Xi(n_1,n_2,n_3)}(\frac{{x}_1}{\Lambda},\frac{{x}_2}{\Lambda},\frac{{x}_3}{\Lambda},x_4;\underline{u})\,.
\end{equation}
Then, we expand about~$\Lambda=\infty$ and extract recurrence relations among the~$u$'s and the~$h$'s.
One obvious saddle point solution to this recurrence relations is~\footnote{There are other saddle points~$\underline{u}^\star$ of~$\widetilde{s}^{(n_1,n_2,n_3)}(\underline{x};\underline{u})$ corresponding to~$n_{1,2,3}$\,-\,th roots of unity. Here we will focus on the leading ones~\eqref{eq:GaugeSaddle} (See analogous discussions in~\cite{Cabo-Bizet:2019osg,Cabo-Bizet:2019eaf,Cabo-Bizet:2020nkr}).}
\begin{equation}\label{eq:GaugeSaddle}
u^{(I)\star}_{a}\,=\, u^{(I)\star}_{a,0}\,=\,\frac{u_0\text{mod}\,1}{\Xi_I}\, \implies \frac{U^{(I)}_{a}}{U^{(J)}_{b}}\,=\,1\, ,
\end{equation}
where~$u_0$ is a zero mode that is integrated out trivially and we can set it to~$u_0=0$ without loss of generality (the integrand does not depend on this mode and thus, the corresponding integral gives~$1$).~\eqref{eq:GaugeSaddle} is a saddle point of the coarse grained action~\eqref{eq:CoarseGrainedEffAction}~\footnote{First, because~$\widetilde{s}^{\Xi(n_1,n_2,n_3)}$ is even in~$\underline{u}$; second, because~$\widetilde{s}^{\Xi(n_1,n_2,n_3)}$ depends only on differences of~$\underline{u}$'s, and third, because~$\widetilde{s}^{\Xi(n_1,n_2,n_3)}$ has continuous first derivatives on~$\underline{u}\,$. }, it is, on the other hand, a logarithmic singularity of the original effective action~$\widetilde{S}^{(n_1,n_2,n_3)}_{\text{eff}}\,$. 
The saddle-point of the original effective action must have a non vanishing $\epsilon-$subleading contribution which must be \emph{non-coincident}, i.e., such that
\begin{equation}
u^{(I)\star}_a \,\neq\, u^{(I)\star}_b\,, \qquad \text{if} \qquad a\,\neq\, b\,. 
\end{equation}
For our purposes knowing the explicit form of the small-$\epsilon$ correction to~\eqref{eq:GaugeSaddle} is not necessary. All that we need to know is of its existence, which as it was just explained, it has to be the case. The existence of one such non-coincident solution implies the existence of other~$\,\sim\, n_1! n_2! n_3!-1$ identical copies obtained by permutations of the gauge indices. Summing over these solutions cancels
the~$\frac{1}{n_1! n_2! n_3!}$ prefactor in~\eqref{eq:integralInterm}.

Thus, in the large $R$-charge expansion~\eqref{eq:LargeRCharge}
\begin{equation}
\begin{split}
\int_{\Gamma_{u^\star}} d\underline{u}\, e^{-\widetilde{s}^{(n_1,n_2,n_3)}(\underline{x};\underline{u})}&\,\simExp e^{-\widetilde{s}^{\Xi(n_1,n_2,n_3)}(\underline{x};\underline{u}^\star)}\,,
\end{split}
\end{equation}
where
\begin{equation}\label{eq:tildeGGEffActFinal}
\begin{split}
\widetilde{s}^{\Xi(n_1, n_2, n_3)}(\underline{x};\underline{u}^\star) &=\, T(\underline{x}) \Bigl(\underline{n}\cdot \underline{x}\Bigr)^2\,-\,\i N \Bigl(\underline{n}\cdot \underline{x}\Bigr)\,\\
T(\underline{x}) & \,=\,-\frac{\omega _1\left({-\Delta _1-\Delta _2-\Delta _3}+\omega _1\pm2  \pi\i
   \right) }{2 \Delta _1
   \Delta _2 \Delta _3}\,.
\end{split}
\end{equation}
At last, in Step~6., i.e. after using~\eqref{eq:tildeGGEffActFinal} together with~\eqref{eq:GrowthGiantGravitons}, we conclude that
\begin{equation}\label{eq:ComparisonGrowthFinal}
a_{gg}(\mathfrak{Q})\,:=\,\widetilde{d}_{GG}(\underline{\Qq}) \,\simExp\, \widetilde{d}(\underline{\Qq})\,=\,d(\underline{Q})\,=:\,a(\mathfrak{Q})\,, \qquad \text{(at any~$N$)\,.}
\end{equation}
Namely, that the large charge asymptotic growth of the superconformal index of $U(N)$ $\mathcal{N}=4$ SYM on $S^3$, at any~$N$, equals the large charge asymptotic growth of the giant graviton index. In virtue of the explanation given around equation~\eqref{eq:SubleadingEntropyFunctionIndex}, the asymptotic relation~\eqref{eq:ComparisonGrowthFinal} implies that, in the large-$N$ limit~\eqref{eq:SugraScaling}, the giant graviton index reproduces the asymptotic growth of states accounting for the Bekenstein-Hawking entropy of the dual~$\frac{1}{16}$ BPS states.

The large-charge analysis for the representation of~\cite{Murthy:2022ien} is summarized in appendix~\ref{sec:ExactRep}.

\section*{Acknowledgements}
~We thank J. Caetano, S. Kim,  J. H. Lee, and D. Orlando for useful conversations. We also acknowledge financial support from the INFN grant GSS (Gauge Theories,
Strings and Supergravity).  This work benefited from the participation of ACB in the workshops ``Precision Holography" and ``Exact Results and Holographic Correspondences" held at CERN and MITP, respectively. ACB would also like to thank the Isaac Newton Institute for Mathematical Sciences, Cambridge, for support and hospitality during the programme Black holes: bridges between number theory and holographic quantum information, where work on this paper was undertaken. This work was supported by EPSRC grant EP/R014604/1.

\appendix
 
\section{Conventions }
\label{app:AsymptRelators}

Let us assume two functions~$X_\Lambda=X_{\Lambda}(\underline{\mu})$ and~$Y=Y(\underline{\mu})$ of a set of variables~$\underline{\mu}$. Let us assume the explicit dependence of~$X_\Lambda$ on~$\Lambda$ to be such that its limit function in the~$\Lambda\to\infty$ is well-defined. Let us select a subset of variables~$\underline{\alpha}\,\subset\,\underline{\mu}$ and denote its complement as~$\underline{\gamma}$. Let us assume that~$\underline{\alpha}=\underline{\alpha}_0$ is a singularity of~$X$ and~$Y\,$. Thus, if we define 
\begin{equation}
\underline{\alpha} = \underline{\alpha_0} +\underline{\delta\alpha}\,,\qquad \underline{\delta \alpha}\,=\,\frac{\underline{\delta \alpha_{ren}}}{\Lambda}\, ,
\end{equation}
it follows that~$X_\Lambda, Y\to \infty$ in the limit $\Lambda\to \infty$ defined by keeping~$\underline{\delta \alpha_{ren}}$ fixed. Based on the previous definitions we will say that in such~$\Lambda\to\infty$ limit
\begin{equation}
X_\Lambda\underset{\Lambda\to\infty}{\sim} Y\, ,
\end{equation}
if and only if~\footnote{In this paper~$\underset{x\to x_0}{\to}$ means that the limit~$x\to x_0$ of the quotient among the left and right-hand sides of the symbol is~$1\,$. }
\begin{equation}\label{eq:FTypeContribution}
\frac{X_\Lambda}{Y}\, \underset{\Lambda\to \infty}{\to}\,A\log\Lambda\,+\,(\log F + c_{branch})\, .
\end{equation}
The~$A$, and~$F$ are functions of the~$\underline{\delta\alpha}$ (they can also depend on the~$\underline{\gamma}$) such that
\begin{equation}\label{eq:ContributionsAF}
A(\underline{\delta\alpha})\,=\,A(\underline{\delta\alpha_{ren}})\,,
\end{equation}
i.e. is invariant under homogeneous scaling of the~$\underline{\delta\alpha}$ and~$C$~{is a c-number} (independent of the~$\underline{\mu}$). Obviously, if~$\underline{\delta \alpha}$ is a single variable then~$A$ is a c-number as well. Also, if the explicit dependence of~$X_\Lambda$ on $\Lambda$ is trivial, we can safely assume~$A=0$. The function~$F$ does not need to be scale invariant, but it needs to have only power-like 
zeroes and singularities in such a way that~$\log F$ {has only logarithmic divergences}. The function~$c_{branch}$ is fixed in terms of~$\log F\,$ it is defined as a generic choice of branch cut of~$\log F$ and thus choosing the appropriate branch we can always assume, and we will do so from now on, that~$c_{branch}=0\,$.

Contributions to~$F$ can have perturbative~\footnote{Explicit examples of this kind of contributions are given in equations~\eqref{eq:LogSCI} and~\eqref{eq:Xi1}.} or non-perturbative~\footnote{An explicit example of this kind of contributions is given in equation~\eqref{eq:AsymptoticsLargeSpin}.} origin. Perturbatively, they could originate from subleading one-loop determinant contributions or subleading corrections to the effective action. Non-perturbatively, they could originate from the superposition of complex conjugated saddle-points (e.g., see subsection~\ref{ref:Interference}, second line of equation~\eqref{eq:Oscillations}).~\footnote{To properly fit the definitions before, one would need to invert the dependence on charge variable $J$ in terms of chemical potentials, as determined implicitly by equation~\eqref{eq:ChemicalPotentialtoCharges}.} In this paper we will not try to fix all these contributions (which are subleading with respect to the leading asymptotics we are looking after). Analogously, we will say that in the expansion~$\Lambda\to\infty$, as defined before,
\begin{equation}
X_\Lambda\,\underset{\Lambda\to\infty}{\sim_{\text{exp}}}\, Y\,,
\end{equation}
if and only if
\begin{equation}
\frac{X_\Lambda}{Y}\, \underset{\Lambda\to \infty}{\to}\,\Lambda^A\,F\,.
\end{equation}

\section{Elliptic functions}

The~$q$-Pochammer symbol~$(\zeta;q)\equiv (\zeta;q)_{\infty}$ has the following product representation
\begin{equation}
(\zeta;q)\= \prod_{j=0}^{\infty}(1\,-\,q^{j}\,\zeta)\,.
\end{equation}
The quasi-elliptic function has the following product representation
\begin{equation}\label{ThetaDef}
\theta_0(\z;q) \,=\,  (1-\z) \prod_{j=1}^\infty (1-q^j \z) \, (1-q^j \z^{-1}) \,.
\end{equation}
The elliptic Gamma functions has the following product representation
\begin{equation}\label{GammaeDef}
\G_{e}(\z;p,q) \=  \prod_{j,\,k=0}^{\infty}
\frac{1\,-\, \zeta^{-1} p^{j+1} q^{j+1}}{1\,-\,\zeta \,p^j\, q^k} \,.
\end{equation}

\section{On the contour of integration $\Gamma_{gauge}$}\label{sec:AppB}
In this appendix we explain how the details of the contour of integration~$\Gamma_{gauge}$, {\em cf.} (\ref{eq:IntegralGGindex}), are relevant  to compute the asymptotic growth of~$\mathcal{I}_{n_1,n_2,n_3}$ in the expansion~$\Delta_a\to0$ at fixed ratios among~$\Delta$'s. 

\subsection*{Resolving the physical poles} Let us comeback to the definition of giant-graviton indices~\eqref{eq:IntegralGGindex} 
\begin{equation}\label{eq:IntegralGGindexApp}
\mathcal{I}_{n_1,n_2,n_3}\,=\,\oint_{\Gamma_{gauge}} \text{d}\mu_1 \text{d}\mu_2 \text{d}\mu_3
 \, \mathcal{I}^{4 d}_{n_1,n_2,n_3} \mathcal{I}^{2 d}_{n_1,n_2,n_3}\,.
\end{equation}
It will be convenient to change integration variables from
\begin{equation}
{U}^{I}_{a}\,, \qquad a\,=\,1\,\ldots\,, n_{I} \,,
\end{equation}
to the \emph{affine variables}~\cite{Cabo-Bizet:2021plf}
\begin{equation}
\begin{split}
\widetilde{U}^{(I)}_{a,a+1}&\, :=\, \frac{U^{(I)}_{a}}{U^{(I)}_{a+1}}\,, \qquad a\,=\,1\,,\,\ldots\,,\, n_{I}-1\,, \\
\widetilde{U}^{(I)}_0 &\,:=\,\Biggl(\prod_{a=1}^{n_I} U^{(I)}_{a}\Biggr)^{1/n_I}\,.
\end{split}
\end{equation}
In terms of the new variables the original fundamental, adjoint, and bi-fundamental variables can be recovered as follows~\footnote{Here we assume the rules $(X Y)^z= X^z Y^z$ and~$(X/Y)^z= X^z /Y^z\,$.}
\begin{equation} \label{eq:InitialVarsTildedVars}
\begin{split}
U^{(I)}_{a} &\,= \,\Biggl(\prod_{j=1}^{n_I-1}\widetilde{U}^{(I)}_{a,a+j}\Biggr)^{1/n_I} \,\widetilde{U}^{(I)}_{0} \,,\\
U^{(I)}_{a b}&\,:=\,\frac{U^{(I)}_{a}}{U^{(I)}_{b}}\,=\,\widetilde{U}^{(I)}_{a, b}\,:=\,\prod_{k=a}^{b-1}\,\widetilde{U}^{(I)}_{k,k+1}\,. \\
U^{(I J)}_{a b}&\,:=\,\frac{U^{(I)}_{a}}{U^{(J)}_{b}} \,=\, \,\frac{\Biggl(\prod_{j=1}^{n_I-1}\widetilde{U}^{(I)}_{a,a+j}\Biggr)^{1/n_I}}{\Biggl(\prod_{j=1}^{n_J-1}\widetilde{U}^{(J)}_{b,b+j}\Biggr)^{1/n_J}} \,\frac{\widetilde{U}^{(I)}_{0}}{\widetilde{U}^{(J)}_{0}} \,,
\end{split}
\end{equation}
Note that the adjoint variables~${U}^{(I)}_{ab}$ are equivalent to the adjoint tilded variables~$\widetilde{U}^{(I)}_{ab}:=\widetilde{U}^{(I)}_{a,b}\,$.

The contour prescription of~\cite{Lee:2022vig} indicates that all physical poles selected by~$\Gamma_{gauge}$ should be located at
\begin{equation}\label{eq:DegPole}
{U}^{(I)}_{a}\,=\,0\,, \qquad a=1\,,\,\ldots\,,\, n_I\,, \qquad  I\,=\,1\,,\,2\,,\,3\,. 
\end{equation}
with generic adjoint and bi-fundamental ratios~${U}^{(I)}_{ab}\,$ and~${U}^{(I,J)}_{ab}\,$. In the tilded variables this means that all physical poles should be located at
\begin{equation}\label{eq:DegPole2}
\widetilde{U}^{(I)}_0\,=\, 0 \,, \qquad I\,=\,1\,,\,2\,,\,3\,,
\end{equation}
with generic ratios~$\widetilde{U}^{(I,J)}_{ab}\,$ and~$\widetilde{U}^{(I,J)}_{ab}\,$.
This means that in the new variables the contour of integration can be divided in two components, a co-dimension 3 loop that we denote below as~$\widetilde{\Gamma}_{gauge}$ and a 3-dimensional infinitesimal loop picking up the residue at~$\widetilde{U}^{(I)}_0\,=\, 0$
\begin{equation}
\,\oint_{\Gamma_{gauge}} \text{d}\mu_1 \text{d}\mu_2 \text{d}\mu_3\,\to\, \oint_{\widetilde{\Gamma}_{gauge}}\prod_{I=1}^3\,\Biggl(\prod_{a=1}^{n_I-1}\frac{\text{d$\widetilde{U}^{(I)}$}_{a,a+1}
}{2\pi\i \widetilde{U}{(I)}_{a,a+1}}\Biggr)\,\cdot \,\oint_{\widetilde{U}^{(I)}_0=0}\,\prod_{I=1}^3\frac{\text{d$\widetilde{U}^{(I)}_{0}$}}{2\pi\i \widetilde{U}^{(I)}_0}\,\, (\text{Vandermonde Det's}).
\end{equation}
At this point we can proceed to evaluate the 3-dimensional integral over the diagonal modes~ $\widetilde{U}^{(I)}_{0}\,$. However, this is not the most convenient way to proceed, because the pole~\eqref{eq:DegPole2} is degenerate. In the original variables this degeneracy is reflected in the vanishing of all the positions~${U}^{(I)}_{a}\,$. In the new variables the complication is translated into~$0/0$'s indefiniteness in the naive residue evaluation. The latter indefiniteness arises after evaluating the bi-fundamental positions~$U^{(I J)}_{a b}$ defined in the third line of equation~\eqref{eq:InitialVarsTildedVars}, at the position of the pole~$\widetilde{U}^{(I)}_{0}\,=\,0\,$. This technical complication makes ill-defined the naive residue evaluation, due to the contribution coming from the fundamental strings stretching among different stacks of branes~$\mathcal{I}_{2d}\,$.

To simplify this residue computation it is convenient to deform the integration measure by substituting
\begin{equation}
\prod_{I=1}^3\frac{\text{d$\widetilde{U}^{(I)}$}_{0}}{2\pi\i \widetilde{U}^{(I)}_0}\,\to\,\prod_{I=1}^3\frac{\text{d$\widetilde{U}^{(I)}$}_{0}}{2\pi\i \Bigl(\widetilde{U}^{(I)}_0\,-\,\mu\Bigr)}\, ,
\end{equation}
where~$\mu$ should be thought of as a parameter that will be taken to zero after evaluating the non-degenerate residues. After this modification the degenerate poles transform into non-degenerate ones
\begin{equation}
\widetilde{U}^{(I)}_{0}\,=\,0\,\quad  \to \quad \widetilde{U}^{(I)}_{0}\,=\,\mu\,.
\end{equation}
For~$\mu\neq0$ the physical poles do not condense to the very same position~$U^{(I)}_a=0\,$ and one can proceed to evaluate~\footnote{ This deformation is an example of the resolutions used in~\cite{Lee:2022vig} to evaluate residues.}
\begin{equation}
\begin{split}
\oint_{\widetilde{U}^{(I)}_0=0}\,\prod_{I=1}^3\frac{\text{d$\widetilde{U}^{(I)}$}_{0}}{2\pi\i \Bigl(\widetilde{U}^{(I)}_0\,-\,\mu\Bigr)} \,\cdot\,\mathcal{I}^{4 d}_{n_1,n_2,n_3}\, \mathcal{I}^{2 d}_{n_1,n_2,n_3}&\,=\, \Biggr(\mathcal{I}^{4 d}_{n_1,n_2,n_3}\, \mathcal{I}^{2 d}_{n_1,n_2,n_3}\Biggr)\Biggl|_{\widetilde{U}^{(I)}_0=\mu}\\ &\,=\,
\mathcal{I}^{4 d}_{n_1,n_2,n_3}\, \Biggl(\mathcal{I}^{2 d}_{n_1,n_2,n_3}\Biggr)\Biggl|_{\widetilde{U}^{(I)}_0=\mu} \\&\,=:\,\mathcal{I}^{4 d}_{n_1,n_2,n_3}\,\widetilde{\mathcal{I}}^{2 d}_{n_1,n_2,n_3}\,.
\end{split}
\end{equation}
For later convenience we note that
\begin{equation}\label{eq:Indices}
\widetilde{\mathcal{I}}^{2 d}_{n_1,n_2,n_3}\,=\,\Biggl(\mathcal{I}^{2 d}_{n_1,n_2,n_3}\Biggr)\Biggl|_{\widetilde{U}^{(I)}_0=1}\,.
\end{equation}
At last, we can write
\begin{equation}\label{eq:IntegralResolved}
\mathcal{I}_{n_1,n_2,n_3} \,=\, \oint_{\widetilde{\Gamma}_{gauge}}\prod_{I=1}^3\,\Biggl(\prod_{a=1}^{n_I-1}\frac{\text{d$\widetilde{U}^{(I)}$}_{a,a+1}
}{2\pi\i \widetilde{U}{(I)}_{a,a+1}}\Biggr)\,\mathcal{I}^{4 d}_{n_1,n_2,n_3}\,\widetilde{\mathcal{I}}^{2 d}_{n_1,n_2,n_3}\,,
\end{equation}
where we have not written down the limit~$\mu\to0$ in the right-hand side because the integrand~$\mathcal{I}^{4 d}_{n_1,n_2,n_3}\,\widetilde{\mathcal{I}}^{2 d}_{n_1,n_2,n_3}$ does not depend on~$\mu\,$.

~{In what follows we assume either that there is no other remaining degenerate residue in the affine integration variables~$\widetilde{U}^{(I)}_{a,a+1}\,$, or that, if there is any one such, then it has been  resolved~\cite{Lee:2022vig,Beccaria:2023zjw}. } Anyways, the poles that dominate the expansion~$\Delta_a\to 0$ at fixed ratios among~$\Delta$'s, which are the ones we will be concerned with, are non-degenerate in the affine variables~$\widetilde{U}^{(I)}_{a,a+1}$ and thus they do not require any further resolution. This will be explained below.

\subsection*{The residues at~$\Delta_a\to 0$} The integral~\eqref{eq:IntegralResolved} can be written as
\begin{equation}\label{eq:ResidueGGApp}
\mathcal{I}_{n_1,n_2,n_3}\, =\, \sum_{\alpha} \text{Res}\Bigl[\ldots \mathcal{I}^{4 d}_{n_1,n_2,n_3} \widetilde{\mathcal{I}}^{2 d}_{n_1,n_2,n_3};\,U\,=\,U_\alpha\Bigr]\,.
\end{equation}
where~$\alpha$ runs over whichever are the poles selected by the choice of contour~$\widetilde{\Gamma}_{gauge}\,$. 

In these expressions we have removed the indices~$I$ and~$a,a+1$, and the products over~$I=1,2,3$ and~$a=1,\ldots, n_I-1$, to ease presentation. 
For generic values of~$n_1$,~$n_2$ and~$n_3\,$
\begin{equation}\label{eq:ConditionNonNegative}
\sum_{\alpha} \text{Res}\Bigl[\ldots \mathcal{I}^{4 d}_{n_1,n_2,n_3} \widetilde{\mathcal{I}}^{2 d}_{n_1,n_2,n_3};\,\widetilde{U}=\widetilde{U}_\alpha\Bigr] \,\neq\, 0\,.
\end{equation}
The results in subsection~\ref{sec:LargeBlackHole} imply the following asymptotic condition for residues~\footnote{To derive this relation below it is important not to truncate the infinite products in the residues and to work with their plethystic exponential representations.}~\footnote{We recall that the symbol~ $\underset{\Delta_a  \,\to\, 0\atop\text{with ratios fixed}}{\to}$ means that the quotient between the left and right-hand side expressions tends to~$1$ in the corresponding limit.}

\begin{equation}\label{eq:ASymptExpGG}
\text{Res}\Bigl[\ldots \mathcal{I}^{4 d}_{n_1,n_2,n_3}\widetilde{\mathcal{I}}^{2 d}_{n_1,n_2,n_3};\,\widetilde{U}=\widetilde{U}_\alpha\Bigr] \underset{\Delta_a  \,\to\, 0\atop\text{with ratios fixed}}{\to} \widetilde{\text{Res}}_\alpha[\underline{x},\widetilde{U}_\alpha]\,e^{-\widetilde{s}(\underline{x}, \widetilde{U}_\alpha)}\,.
\end{equation}
In this equation the function~$\widetilde{s}(\underline{x},\widetilde{U})$ equals the {localized effective action} reported in equation~\eqref{eq:CoarseGrainedEffActionGiantG0},
\begin{equation}
\widetilde{s}(\underline{x},\widetilde{U})\,:=\,\widetilde{s}^{(n_1,n_2,n_3)}(\underline{x},\frac{\log U}{2\pi\i})\,,
\end{equation}
when the latter is expressed as a function of the new affine variables~$U\to \widetilde{U}\,$, and restricted to the~$(n_1+n_2+n_3-3)$-dimensional section
\begin{equation}\label{eq:Restriction}
\widetilde{U}^{(I)}_{0} 
\,:=\,1\,.
\end{equation}
This action~$\widetilde{s}(\underline{x},\widetilde{U})$ defines the exponential singularity of the integrand of~$\mathcal{I}_{n_1,n_2,n_3}\,$ in the expansion~$\Delta_a\to0$ at fixed ratios. The asymptotic relation~\eqref{eq:ASymptExpGG} and the explicit form of the function~$\widetilde{\text{Res}}_\alpha[\underline{x},\widetilde{U}]$ to be presented below in~\eqref{eq:RemainderSubleading}, follow from the fact that the polynomial~$\prod_{I=1}^{3}\prod_{a=1}^{n_I-1}(\widetilde{U}^{(I)}_{a,a+1}-\widetilde{U}^{(I)}_{\alpha;a,a+1})$ that needs to be multiplied to the integrand in order to extract its residue at~$\widetilde{U}=\widetilde{U}_{\alpha}\,$, does not affect the leading exponential growth of the integrand in the limit~$\Delta_a\to 0\,$ at fixed ratios of~$\Delta_a$'s. 

The function~$\widetilde{\text{Res}}_\alpha[\underline{x},\widetilde{U}]$ is a subleading contribution defined as
\begin{equation}\label{eq:RemainderSubleading}
\widetilde{\text{Res}}_\alpha[\underline{x},\widetilde{U}] := e^{-s^{(0)}(\underline{x},\widetilde{U}+0^+)+\sum_{I,a}\log (\widetilde{U}^{(I)}_{a,a+1}-\widetilde{U}^{(I)}_{\alpha;a,a+1}+0^+)}\, ,
\end{equation}
where the~$0^+$ is an auxiliary regulator whose only function is to keep finite the two terms in the exponent of~
\eqref{eq:RemainderSubleading} at~$\widetilde{U}=\widetilde{U}_\alpha$ (for the combination of the two quantities, this regulator plays no role because the logarithmic term is cancelled by the first term).

The function
\begin{equation}
e^{-s^{(0)}(\underline{x},{ \widetilde{U}})}\underset{\Delta_a  \,\to\, 0\atop\text{with ratios fixed}}{\leftarrow} \,\Bigl(\ldots \mathcal{I}^{4 d}_{n_1,n_2,n_3} \widetilde{\mathcal{I}}^{2 d}_{n_1,n_2,n_3}\Bigr)\times  e^{\widetilde{s}(\underline{x},{ \widetilde{U}})}\, ,
\end{equation}
is the subleading contribution that we have discarded in the evaluation of~$e^{-\widetilde{s}^{(n_1,n_2,n_3)}(\underline{x}, \frac{\log U}{2\pi\i})}\,$. Namely, the ambiguous contributions of type~$F$ that were defined around~\eqref{eq:FTypeContribution}.

As it was explained in the main body of the paper in a certain region of chemical potentials~$\underline{x}$ the magnitude of the exponential growth of the factor~$|e^{-\widetilde{s}^{(n_1,n_2,n_3)}(\ldots)}|$ is maximized by the configuration~$U^{(I)}_{a}\,=\,1\,$. More generally, we explained how~$U^{(I)}_{a}\,=\,1$ is a stationary point of~$e^{-\widetilde{s}^{(n_1,n_2,n_3)}(\ldots)}\,$. In virtue of this last statement and of~\eqref{eq:CoarseGrainedEffActionGiantG0} with the restriction~\eqref{eq:Restriction}, it follows that the configuration~$\widetilde{U}^{(I)}_{a,a+1}\,=\,1$ maximizes the exponential growth of the leading factor~$|e^{-\widetilde{s}(\ldots)}\,|$ in certain regions of chemical potentials~$\underline{x}\,$, and more generally, that it is a stationary point of~$e^{-\widetilde{s}(\ldots)}\,$.  This means that in the small-chemical potential expansion above-quoted, the sum over residues 
\begin{equation}\label{eq:ResidueOriginal}
\sum_{\alpha} \widetilde{\text{Res}}_\alpha[\underline{x},\widetilde{U}_\alpha]\,e^{-\widetilde{s}(\underline{x},{ \widetilde{U}_\alpha})}\,,
\end{equation}
is dominated by poles~$\{\beta\}\,\subset\,\{\alpha\}$ that obey the asymptotic condition
\begin{equation}\label{eq:LimitUtildesFinal}
\widetilde{U}^{(I)}_{\beta;a,a+1}(x)\,\underset{\Delta_a  \,\to\, 0\atop\text{with ratios fixed}}{\to}\, 1\,,
\end{equation}
if and only if:
\begin{itemize}
\item $\widetilde{\Gamma}_{gauge}$ encloses some of them and the sum over their residues is non-vanishing. 
\end{itemize}
For the contour prescription proposed in~\cite{Imamura:2021ytr} there are infinitely many such poles.
In the integrand~$\ldots \mathcal{I}^{4 d}_{n_1,n_2,n_3} \widetilde{\mathcal{I}}^{2 d}_{n_1,n_2,n_3} $ these poles always come in pairs~\footnote{At least at large charges, these pairs mutually cancel each other, as it will be shown below.} (denoted as \emph{positive} and \emph{negative} poles). For example, assuming~$n_I>1$ there are simple poles defined by selecting~$n_I-1$ pairs~$(a,b)$ for each~$I=1,2,3$ such that (for~$I\neq J\neq K$ and generic~$w_{I,J,K}$~\footnote{It is sufficient, not necessary, to assume~$w_I$ to be different from any product of rational powers of~$w_J$.})
\begin{equation}\label{eq:PolesExamples}
{\widetilde{U}_{a b}^{(I)}}\,=\,\prod_{j=a}^{b-1} \widetilde{U}^{(I)}_{j,j+1}\,=\,{U}_{a b}^{(I)}\,=\, \frac{w_I}{(w_J)^{c_1(a,b)}}\quad\text{or}\quad (w_J w_K)^{c_2(a,b)}\,, 
\end{equation}
for any two choices of integers~$c_{1}(a,b)\geq0$ and~$c_{2}(a,b)>0\,$.
These poles come from the elliptic gamma functions~\cite{Felder2000} in the vector contributions~\eqref{eq:4dAdjointContribution}. The first family comes from the poles of the first factor in the numerator of~\eqref{eq:4dAdjointContribution}. The second family comes from the zeroes of the denominator of~\eqref{eq:4dAdjointContribution}.
They can be organized in two groups that map into each other under a~$\mathbb{Z}_2$ operation. One could denote such two subsets as \emph{positive} and \emph{negative}. This separation in two, which is non unique, comes from the fact that for every pole~$\widetilde{U}=\widetilde{U}_{\alpha}$ there is a pole located at the inverse position~$\widetilde{U}=\widetilde{U}^{-1}_{\alpha}\,$. This bijection implies the existence of many~$\mathbb{Z}_2$ operations, out of which one can pick up one, and declare that it maps~\emph{half of} the number of poles coming from vector multiplets (\emph{positive}) into the other half (\emph{negative}). For the indices studied here there are~$\infty\,$ many such poles.~\footnote{If the giant graviton expansion is complete and not asymptotic, then one must expect, and we will assume so, that the corresponding infinite sum over poles will be convergent in some continuous domain of rapidities.} As it will be shown below, in order to have a non-trivial answer at large charges,~$\widetilde{\Gamma}_{gauge}$ must necessarily pick up an unbalanced number of~\emph{positive} and~\emph{negative poles} in order for the corresponding integral not to vanish trivially at large charges.

In the concrete example of~$\mathcal{I}_{0,0,2}$ it is easy to identify poles in the first family in~\eqref{eq:PolesExamples} for the choices~$c_1=0 \text{ and } 1$ as:
\begin{equation}\label{eq:PolesExamples2}
\widetilde{U}^{(3)}_{12} \,=\, w_3\,,\,\frac{w_3}{w_1}\,,\,\frac{w_3}{w_2}\,.
\end{equation}
Using both, the identifications and the constraint below
\begin{equation}
\widetilde{U}^{(3)}_{12}\to z^{-1}_{\text{there}}\,,\qquad  w_a \to q u_{a\text{there}}\,,\qquad u_{1\text{there}}u_{2\text{there}}u_{3\text{there}}\,=\,1\,,
\end{equation}
the three positive poles~\eqref{eq:PolesExamples2} map into the three positive poles corresponding to the tachyonic and zero mode terms~$f_1$, $f_2$ and~$f_3$ depicted in figure 2 of~\cite{Imamura:2021ytr}.

As recalled in the latter example, the pole-selection prescriptions of~\cite{Imamura:2021ytr} and~\cite{Lee:2022vig}, pick up an unbalanced number of positive and negative poles of type~$\beta\,$ which happen to come solely from vector multiplets (the positions of the poles coming from the chiral multiplets reduce to some power of~$p_1$ in the scaling~$\Delta_a\to0$. Please refer to~\eqref{eq:2dBiadjoint index})~\footnote{The poles for this bi-fundamental contribution come from the zeroes of the Jacobi theta functions~\cite{Felder2000} in the denominator.}. 

Let us proceed to explain why the sum over residues of type~$\beta$ selected by contours~$\widetilde{\Gamma}_{gauge}$ breaking the~$\mathbb{Z}_2$ symmetries among the latter, does not vanish for generic values of chemical potentials. For the poles of type~$\beta$, we can always use the Taylor expansion around~$\underline{\Delta}=0$
\begin{equation}\label{eq:ExpansionU0}
\widetilde{U}_\beta(x)\,=\, 1+ \sum_{i_{1},i_2,i_3 \,\in\, \mathbb{Z}^*_+ \atop \underline{i}\neq \underline{0}}\,c_{\underline{i},\beta} \Delta_1^{i_1}\Delta_2^{i_2}\Delta_3^{i_3}\,,
\end{equation}
where the coefficients~$c_{\underline{i},\beta}:=c_{i_1,i_2,i_3,\beta}$ are c-numbers. In particular for every~$\beta$,
\begin{equation}
c_{\underline{1},\beta}\,:=\,\bigl\{c_{1,0,0,\beta}\,,\,c_{0,1,0,\beta}\,,\,c_{0,0,1,\beta}\bigr
\} \,\neq\, \underline{0}
\end{equation}
and generically for~$\beta\neq \beta^\prime$~\footnote{We identify poles~$\beta$ and~$\beta^\prime$ that are identical after a permutation of their gauge indices~$a_I=1,\ldots n_I-1\,$.}
\begin{equation}\label{eq:LinearCombinationsNo}
c_{\underline{1},\beta}\,\neq \,c_{\underline{1},\beta^\prime}\,. 
\end{equation}
Moreover, within this family of poles~$\{\beta\}$ the remainder function
\begin{equation}
\widetilde{\text{Res}}_\beta[\underline{x},\widetilde{U}_\beta]  \,\underset{\Delta_a  \,\to\, 0\atop\text{with ratios fixed}}{\to} \, e^{-\widetilde{s}_0(\underline{x},1+{c_{\underline{1},\beta}\cdot\underline{\Delta}})+\log[{c_{\underline{1},\beta}\cdot\underline{\Delta
}}]}=: \, e^{-\widetilde{s}^{(log)}_{0,\beta}(\underline{x})}
\end{equation}
reduces to the exponential of a function of~$\underline{x}\,$,~$-\widetilde{s}^{(log)}_{0,\beta}(\underline{x})\,$, whose dependence on~$\beta$ can be constrained in a simple way. We will do so in the following subsection.
After substituting~\eqref{eq:ExpansionU0} in the leading contribution to~\eqref{eq:ResidueOriginal} coming from the residues of type~$\beta$, and keeping in the exponent the terms that do not vanish trivially as~$\Delta_a\to0\,$, one obtains
\begin{equation}
\,e^{-\widetilde{s}(\underline{x},1)\,}\,\sum_{\beta} \,e^{-\,\widetilde{s}^{(log)}_{0,\beta}
(\underline{x})-\widetilde{s}_\beta(\underline{x})}\,,
\end{equation} 
where -- after reinstating the indices~$I$ and~$a$  --- it follows that
\begin{equation}
\begin{split}
\widetilde{s}_\beta (\underline{x})&\,\underset{\Delta_a  \,\to\, 0\atop\text{with ratios fixed}}{\leftarrow}\,\sum_{I=1}^3\sum_{a=1}^{n_I-1
}\partial_{U^{(I)}_{a,a+1}}\Biggl( \widetilde{s}\biggl(\underline{x},{ \widetilde{U}}\biggr) \Biggr)\Bigg|_{\widetilde{U}=1} \,\times\,(c^{(I);a,a+1}_{\underline{1},\beta}\cdot\underline{\Delta})\,\\
 &\,\underset{\Delta_a  \,\to\, 0\atop\text{with ratios fixed}}{\leftrightarrow}\,0\, ,
\end{split}
\end{equation}
vanishes trivially because~$\widetilde{U}=1$ is a saddle point of the action~$\widetilde{s}\,$.
Thus, we conclude that in the limit~$\Delta_a\to0$ with ratios fixed, the total -- and leading -- residue contribution to~$\mathcal{I}_{n_1,n_2,n_3}$ takes the asymptotic form
\begin{equation}\label{eq:EqIntermApp}
\,e^{-\widetilde{s}(\underline{x},0)\,}\,\sum_{\beta} \,e^{-\,\widetilde{s}^{(log)}_{0,\beta}
(\underline{x})} \underset{\Delta_a\to0\atop\text{with ratios fixed}}{\leftarrow}\, \mathcal{I}_{n_1,n_2,n_3}\,.
\end{equation} 
As we will show below, the sum over~$\beta$ can be a series only if~$\widetilde{\Gamma}_{gauge}$ selects an infinite number of unpaired positive or negative poles.   The equation~\eqref{eq:LinearCombinationsNo} implies that for a generic choice of a 
contour~$\widetilde{\Gamma}_{gauge}$ selecting an unbalanced number of positive or negative poles of type~$\beta\,$, the exponential factors~$\{e^{-\widetilde{s}^{(log)}_{0,\beta
(\underline{x})}}\}$, whose quotient will be reported in equation~\eqref{eq:DefinitionLargeChargeGrav} below, are linear independent  functions of~$\underline{x}$ and thus, for generic values of~$\underline{x}$
\begin{equation}\label{eq:FinalAppAsymp}
\sum_{\beta} \,e^{-\widetilde{s}^{(log)}_{0,\beta}(\underline{x})}\, \neq \,0\,.
\end{equation}
In such a case, in virtue of~\eqref{eq:EqIntermApp}, one concludes that, provided the sum over~$\beta$'s is either finite or a convergent series, 
\begin{equation}\label{eq:AsympAppA}
\mathcal{I}_{n_1,n_2,n_3} \,\simExp\, e^{-\widetilde{s}(\underline{x},1)-\widetilde{s}^{(log)}_{0,\beta_0}(\underline{x})}\,,\qquad \widetilde{s}(\underline{x},1)+\widetilde{s}^{(log)}_{0,\beta_0}(\underline{x})\,\underset{\Lambda\to\infty}{\sim}\, \widetilde{s}^{\Xi, (n_1,n_2,n_3)}
(\underline{x},0)\,.
\end{equation}
where the ambiguity in the choice of~$\beta_0$ is shielded in the ambiguity in the relations~$\sim\,$.
\subsection*{Constraining the relative contribution of poles}
Let us come back to the function (\emph{reduced residue})
\begin{equation}\label{eq:ResApp}
\widetilde{\text{Res}}_\beta[\underline{x},U] \,= \,  e^{-s^{(0)}(\underline{x},\widetilde{U})+\log (\widetilde{U}-\widetilde{U}_{\beta})}\,.
\end{equation}
The goal is to constrain the dependence on~$\beta$ of the quantity 
\begin{equation}
\widetilde{\text{Res}}_\beta[\underline{x},\widetilde{U}_\beta+0^+]\,<\,\infty,
\end{equation}
in the asymptotic expansion near its singularity~$\widetilde{U}_\beta\,\to\,1\,$. It is convenient to compute such asymptotic expansion in two-steps, starting from the function~\eqref{eq:ResApp}
\begin{equation}
\widetilde{U}_\beta\to \widetilde{U}\,, \quad \widetilde{U}\,\to\, 1\,.
\end{equation}
For example, in a two-step expansion defined by the quadratic differential variations
\begin{equation}
\widetilde{U} = 1+\delta \widetilde{U} \,,\, \widetilde{U}_\beta = \widetilde{U}(1 +\delta \widetilde{U}_\beta)\, ,
\end{equation}
where~$\delta \widetilde{U}_\beta \ll \delta \widetilde{U}$ the exponent of~\eqref{eq:ResApp} takes the form
\begin{equation}
\label{eq:Beta1}
-s^{(0)}(\underline{x},U)+\log (\widetilde{U}-\widetilde{U}_{\beta}) \,{\sim}\, -s^{(0)}(\underline{x},1+\delta \widetilde{U})+\log(\delta \widetilde{U})+ \log \delta \widetilde{U}_\beta\,.
\end{equation}
Using the expansion~\eqref{eq:Beta1} for two different poles~$\beta=\beta_1$ and~$\beta=\beta_2\,$ we conclude, after exponentiation, that in a limit~$\delta{\widetilde{U}}_{\beta_{1,2}}\,\to\,0$ the quotient among the reduced residues of roots of type~$\beta$ approaches a universal expression,
\begin{equation}
\frac{\widetilde{\text{Res}}_{\beta_1}[\underline{x},\widetilde{U}_{\beta_1} = 1+ \delta \widetilde{U}_{\beta_1}]
}{\widetilde{\text{Res}}_{\beta_2}[\underline{x},\widetilde{U}_{\beta_2}= 1+\delta \widetilde{U}_{\beta_2}]
}\,\to\, \frac{\delta \widetilde{U}_{\beta_1}}{\delta \widetilde{U}_{\beta_2}}\,.
\end{equation}
This expression and~\eqref{eq:ExpansionU0} imply the following relation
\begin{equation}\label{eq:DefinitionLargeChargeGrav}
\frac{\widetilde{\text{Res}}_{\beta_1}[\underline{x},\widetilde{U}_{\beta_1}(\underline{x})]
}{\widetilde{\text{Res}}_{\beta_2}[\underline{x},\widetilde{U}_{\beta_2}(\underline{x})]
}\,\underset{\Delta_a\to 0\atop \text{at fixed ratio}}{\to}\,\frac{\sum_{I=1}^3\sum_{a=1}^{n_I-1}c^{(I);a,a+1}_{\underline{1},\beta_1}
\cdot \underline{\Delta}}{\sum_{I=1}^3 \sum_{a=1}^{n_I-1}\,c^{(I);a,a+1}_{\underline{1},\beta_2}
\cdot \underline{\Delta}}\,=:\,\frac{e^{-\widetilde{s}^{(log)}_{0,\beta_1}(\underline{x})}}{e^{-\widetilde{s}^{(log)}_{0,\beta_2}(\underline{x})}}\,.
\end{equation}
This relation is telling us that the relative contribution of poles of type~$\beta$ is defined, unambiguously, by their corresponding coefficients~$c_{\underline{1},\beta}\,$. This is very useful, because the latter coefficients can be computed easily, and consequently using~\eqref{eq:DefinitionLargeChargeGrav} one can straightforwardly predict what poles in the integrand would cancel among each other should~$\widetilde{\Gamma}_{gauge}$ pick them all.

For example, from~\eqref{eq:DefinitionLargeChargeGrav} one concludes that if the positions~$\beta_1$ and~$\beta_2$ are inverse to each other, then
\begin{equation}
\frac{e^{-\widetilde{s}^{(log)}_{0,\beta_1}(\underline{x})}}{e^{-\widetilde{s}^{(log)}_{0,\beta_2}(\underline{x})}}\,=
\,-1\,.
\end{equation}
which means that both contributions would cancel each other in the sum
\begin{equation}
e^{-\widetilde{s}(\underline{x},0)\,}\,\sum_{\beta} \,e^{-\,\widetilde{s}^{(log)}_{0,\beta}
(\underline{x})}\,.
\end{equation}
This implies that, should~$\widetilde{\Gamma}_{gauge}$ not break the~$\mathbb{Z}_2$ symmetries for poles of type~$\beta$,
then the contributions of the latter would vanish at large charges. On the contrary for a~$\widetilde{\Gamma}_{gauge}$ that breaks the~$\mathbb{Z}_2$ symmetries for poles of type~$\beta$ the analytic analysis above presented predicts that the answer will not vanish. 

For choices of~$\widetilde{\Gamma}_{gauge}$ that pick up an infinite number of unpaired positive and negative poles of type~$\beta$ it may be possible that the sum 
\begin{equation}
\sum_{\beta} \,e^{-\,\widetilde{s}^{(log)}_{0,\beta}
(\underline{x})}\,,
\end{equation}
could not be resumed into a finite function. Equation~\eqref{eq:DefinitionLargeChargeGrav} can be used to understand this point better. Just to give an idea, assume $\Delta_1=\Delta_2=\Delta_3$ and take~$\beta=\beta_{min}$ to denote the pole(s) with the minimum value~$n_{min}$ of
\begin{equation}
|n|=|n(\beta
)|:=|\sum_{I=1}^3\sum_{a=1}^{n_I-1} c^{(I);a,a+1}_{1,0,0,\beta}+c^{(I);a,a+1}_{1,0,0,\beta}+c^{(I);a,a+1}_{1,0,0,\beta}|\,.
\end{equation}
Then we can write
\begin{equation}\label{eq:SumNPoles}
\begin{split}
\sum_{\beta} \,e^{-\,\widetilde{s}^{(log)}_{0,\beta}
(\underline{x})}&\,=\,e^{-\,\widetilde{s}^{(log)}_{0,\beta_{min}}
(\underline{x})}\,\frac{1}{n_{min}}\,\sum_{\beta}{n(\beta)}\\&\,=\,\,e^{-\,\widetilde{s}^{(log)}_{0,\beta_{min}}
(\underline{x})}\,\frac{1}{n_{min}}\,\sum_{n\in\mathbb{Z}\atop n\geq n_{min}}{deg(n)\,n}\, ,
\end{split}
\end{equation}
where the integer number~$deg(n)$ receives contributions from every~$\beta$ selected by~$\widetilde{\Gamma}_{gauge}\,$ with~$n(\beta)=n\,$: precisely,~$+1$ contributions from positive poles and~$-1$ contributions from negative poles. Obviously, only if~$deg(n)=0$ for~$n> L$ where~$L$ is a positive integer, then the sum in the right hand side of~\eqref{eq:SumNPoles} becomes finite. Assuming~$\widetilde{\Gamma}_{gauge}$ does select an infinite number of unpaired positive and negative poles of type-~$\beta\,$, we interpret the infinity above as signature that the infinite sum over residues can not be blindly commuted with the expansion~$\Delta_a\to0$ at fixed ratios. At the level of computing asymptotic expansions though, it is enough to truncate the convergent sum over poles~$\beta$ to a large sum, say with only~$L
\gg1\,$ elements, those with the minimum values of~$n(\beta)$ out of the infinitely many selected by the contour. In the presence of this intermediate cut-off~$L$ the asymptotic relation~\eqref{eq:AsympAppA} follows from the fact that the dependence on~$L$ is shielded in the subleading ambiguity of the relations~$\sim\,$.~\footnote{In other words, in the regions of chemical potentials~$\Delta_a$'s where an infinite sum over poles of type~$\beta$ converges,~$\widetilde{U}^{(I)}_{ab}=0\,$ happens to be an accumulation point for such type of poles, i.e., in those regions of~$\Delta_{a}'s$ the larger~$|n(\beta)|$ the closer~$\beta$ is to~$\widetilde{U}^{(I)}_{ab}=0\,$. This is the reason why a series over poles of type~$\beta$ can not be commuted with the limit~$\Delta_a \to 0$ with ratios fixed: for a fast enough limit of poles towards~$\widetilde{U}^{(I)}_{ab}=0\,$ it is not always true that the posterior limit~$\Delta_a \to 0$ implies the condition~\eqref{eq:LimitUtildesFinal}. That said, for any finite sum over poles of type~$\beta$ the latter issue is not present. }

\section{Large charge entropy from averages over free Fermi systems}\label{sec:ExactRep}

\paragraph{Brief summary of results in this appendix}
In~\cite{Murthy:2022ien} the author proposed an exact giant graviton-like expansion for a large family of matrix integrals that include the $\frac{1}{16}$-BPS index as a particular example. Schematically, this expansion looks like
\begin{equation}
\mathcal{I}(\tf) \,=\,\sum_{n} \,\int dt\,\mathcal{I}_{n,\zeta}(\tf)\,,
\end{equation}
where
\begin{equation}
\mathcal{I}_{n,\zeta}(\tf,\zeta) = \sum_{Q} \,a_{n,t}(Q) e^{2\pi\i \tf Q}\,.
\end{equation}
~$\zeta=e^{-{t}}$ is an auxiliary integration variable, whose string theory interpretation is unclear to us, and which we find evidence that --at least at large charges -- it may be related to the linear combination of giant graviton numbers~$c_{1,\pm}\cdot\underline{n}$ in the representation of~\cite{Imamura:2021ytr}. On the other hand,~$n$ is a non-negative natural number that reminisces, as well, one of the three numbers of giant gravitons in the expansions of~\cite{Imamura:2021ytr} and~\cite{Gaiotto:2021xce}. From now on, when referring to the representation of~\cite{Murthy:2022ien}, $n$ will be called the giant graviton-like number.

At large enough charges,  the microcanonical index grows slower than the giant graviton-like contributions~$\int dt\, a_{n,t}(Q)$~\cite{Liu:2022olj}
\begin{equation}
\frac{|a(Q)|}{|\int d\zeta\, a_{n,\zeta}(Q)|} \,\underset{Q\to \infty}{\sim}\, 0\,.
\end{equation}
 This means that at large~$Q$'s a
large number of cancellations happen after evaluating
\begin{equation}
\sum_{n}\int dt a_{n,t}(Q)\,.
\end{equation}
Indeed, we will check that these cancellations can be understood as a transition in between two pairs of complex conjugated saddle-point configurations of
\begin{equation}\label{eq:EntropyFunctional}
\log \mathcal{I}_{n,\zeta}(\tf) + 2\pi \i \tf Q \, ,
\end{equation}
at large~$Q\,$. 

The large-charge localization Lemma of subsection~\eqref{sec:LCLocalization} implies that the integral over~$t$ must localize  -- at large-R charges and fixed~$J$ and~$n\,$ -- around essential singularities of the integrand~$\mathcal{I}_{n,\zeta}(\tf)\,$. Indeed, we find that the two relevant exponential singularities are located around~$\zeta\,=\,\pm 1$, respectively. If we denote the asymptotic expansion of~$\mathcal{I}_{n,\zeta}(\tf)$ around them as
\begin{equation}\label{eq:subsSaddleMinus}
\mathcal{I}_{n,\zeta}(\tf)\,\to\,\mathcal{I}_{n,\pm1\,+\,t}(\tf)\,,
\end{equation}
then the saddle points obtained after extremizing the answer obtained after the substitution of the choice~$``-"$ in~\eqref{eq:subsSaddleMinus} on~\eqref{eq:EntropyFunctional}, are the ones  determining~$a_n(Q)$ at large~$Q$ and fixed~$n\,$. On the other hand the saddle points obtained after extremizing the substitution of the choice of sign~$+$ in~\eqref{eq:subsSaddleMinus} on~\eqref{eq:EntropyFunctional} dominate the counting after the sum over~$n$ is evaluated and exponentially large cancellations happen. The details of this analysis will be summarized in section~\ref{sec:ExactRep}. 

In summary, we will check that at large charges~$Q\to\infty$ (for all~$N$) the following asymptotic formulae hold
\begin{equation}\label{eq:GGSummaryAssympt2}
\sum_{{n}} \int dt\,a_{{n},t}(Q)\,{\sim}\,a^{loc}_{n,{t}_+^\star(Q)}(Q)\,+\,a^{loc}_{n,{t}_-^\star(Q)}(Q)\,\sim\, a(Q)\, \sim\,  e^{\left(\sqrt{3}\right) 3^{1/3}\pi  \,{c
J^{2/3}}{ N^{2/3}}}\, ,
\end{equation}
(and a more general version of it as well making contact with the complete moduli space of dual black hole solutions) where the complex conjugated contributions
\begin{equation}
a^{loc}_{{n},t^\star_{\pm}}(Q)\, ,
\end{equation}
come from the saddle points of the localized effective action~$-\log \mathcal{I}_{n,\zeta}(\tau)$ around the singular region~$\zeta\to1$. In this case the two complex conjugated saddle point values are
\begin{equation}\label{eq:CondensateValue}
{t^\star}_\pm \,=\,\widetilde{c}_\pm\, \frac{J^{2/3}}{N^{1/3}}\,,
\end{equation}
where again, the~$\widetilde{c}_\pm$ are order~$1$ complex contributions that happen to match the above-quoted~$c_{2\pm}$ in equation~\eqref{eq:CondensateValue0} --up to a normalization factor--. 

The asymptotic relations~\eqref{eq:GGSummaryAssympt2} tell us how the exponential growth of $\frac{1}{16}$-BPS states in the boundary gauge theory is recovered from the giant graviton-like representation of~\cite{Murthy:2022ien}. The relevant computations are summarized below. Curiously, the similarity among~\eqref{eq:CondensateValue0} and~\eqref{eq:CondensateValue} suggests that there may be a relation between the sum of the~$n$ auxiliary integration variables of type~$t=-\log \zeta$ (in representation~\eqref{eq:GGFormula}) and a single linear combination of giant graviton numbers~$\underline{n}$ in the proposal of~\cite{Imamura:2021ytr} (in representation~\eqref{eq:GiantGravImamura}).

\subsection*{The cancellation mechanism}

Let us explain how the cancellation mechanism among giant-graviton-like contributions happens in the exact expansion~\eqref{eq:GGFormula}. In this expansion the microcanonical index of giant graviton-like contribution,~$\sum_{n} \mathcal{J}_n\,$, can be written as:
\begin{equation}\label{eq:GGMicroIndexDefinition}
\widetilde{d}_{M}(\underline{\widetilde{Q}}^\prime)\,=\, \int_{\Gamma} d\underline{x}\,\sum_{n\,=\,0}^{\lfloor \text{max}\Qq_{1,2,3}/N\rfloor}\,\int \frac{d \underline{u}}{ n!}\int \frac{d\underline{t}}{n!}\, e^{-S^{(n)}_{M}(\underline{x};\underline{z},\underline{\zeta})-\i \underline{x}\cdot \underline{\widetilde{Q}}^\prime}\,,
\end{equation}
where~$t_i=-{\log(\zeta_i)}\in [0,2\pi\i)$ and
\begin{equation}\label{eq:EffActSam}
\begin{split}
-S^{(n)}_{M}(\underline{x};\underline{z},\underline{\zeta})&\,:=\,\sum_{l=1}^{\infty}\,\frac{\left(1-w_1^l\right) \left(1-w_2^l\right) \left(1-w_3^l\right)-\left(1-p^l\right)
   \left(1-q^l\right)}{l \left(1-w_1^l\right) \left(1-w_2^l\right) \left(1-w_3^l\right)}\\&\qquad\,\times\,\sum_{i\neq j=1}^{n} \left(\frac{z_i}{z_j}\right)^l\left(1-\zeta _i^l\right) \left(1-\zeta _j^{-l}\right)\,-\,\mathcal{T}^{(n)}(\underline{x};\underline{z},\underline{\zeta}).
   \end{split}
\end{equation}
We collect the determinant and the zero-mode contributions in the quantity
\begin{equation}
\begin{split}
-\mathcal{T}^{(n)}(\underline{x};\underline{z},\underline{\zeta})&\,=\,\sum_{l=1}^{\infty}\,\frac{\left(1-w_1^l\right) \left(1-w_2^l\right) \left(1-w_3^l\right)-\left(1-p^l\right)
   \left(1-q^l\right)}{l \left(1-w_1^l\right) \left(1-w_2^l\right) \left(1-w_3^l\right)}\sum_{i=1}^{n}\left(1-\zeta_i^l\right) \left(1-\zeta _i^{-l}\right)\\
   &+\log (\text{Det}[z,\zeta ])-\sum_{i=1}^{n} \log \left(\frac{(1-\zeta _i)^2}{\zeta_i}\right)+N\sum_{i=1}^n\text{log}\left(\zeta _i\right)- \pi\i  n\,.
   \end{split}
\end{equation}
The identity
\begin{equation}
\begin{split}
-\sum_{i=1}^{n}\sum_{l=1}^{\infty}\,&\frac{1}{l }\left(\zeta_i^l+\zeta _i^{-l}\right)-\sum _i \log \left(\frac{1-\zeta _i}{\zeta
   _i}\right)- \pi\i  n\,\, = 0\,\text{mod}\,(2\pi\i)\, ,
   \end{split}
\end{equation}
simplifies~$\mathcal{T}^{(n)}$ as follows
\begin{equation}
\begin{split}
-\mathcal{T}^{(n)}(\underline{x};\underline{z},\underline{\zeta})&\,=\,\sum_{i=1}^{n}\sum_{l=1}^{\infty}\,\frac{-\left(1-p^l\right)
   \left(1-q^l\right)}{l \left(1-w_1^l\right) \left(1-w_2^l\right) \left(1-w_3^l\right)}\left(1-\zeta_i^l\right) \left(1-\zeta _i^{-l}\right)\,+\,N\text{Log}\left(\zeta _i\right)\\
   &\,+\, \sum_{l=1}^{\infty}\,\frac{\left(1-w_1^l\right) \left(1-w_2^l\right) \left(1-w_3^l\right)}{l \left(1-w_1^l\right) \left(1-w_2^l\right) \left(1-w_3^l\right)}\,(2 n)\,+\,\log (\text{Det}[z,\zeta ])\,.
   \end{split}
\end{equation}
To apply our large-charge localization Lemma, we must compute the asymptotic expansion of the effective action around  the relevant power-like singularity(ies). There are many singularities, but we will show that the two relevant ones~$(\pm)$ are located at~$\Delta_a\to0$ and~$\zeta_i\to\pm 1\,$.~\footnote{In particular, from now on we will only pay attention to the leading asymptotic behaviour, thus will not pay attention to the $F$-type contributions (See the definitions given around~\eqref{eq:FTypeContribution}) coming from the~$\log \text{Det}[z,\zeta]$ term.} We find and check (in the following section), that the singularity locus at~$\zeta_i=-1$ is the one relevant to compute asymptotic growth of states at fixed giant graviton-like number~$n$. 

As in the cases before, these previous singularities serve as attractors to saddle-points. The localization of the effective action around~$\zeta_i=\pm 1$ determines different saddle-point contributions to the total integral~\eqref{eq:GGMicroIndexDefinition}. In this subsection, we will focus on the vicinity of the singularity locus (or equivalently, on the saddle-points obtained after localization) at~$\zeta_i=1$, which is the one making explicit contact with the index at large charges.

If we substitute
\begin{equation}
\label{eq:DeltasSRep}
\Delta_a\, \to\,  
\epsilon \Delta_{a}\,,
\end{equation}
and
\begin{equation}\label{eq:ZetaSRep}
\zeta_i \to e^{-\epsilon t_i} 
\,,\qquad z_i\,\to\, e^{-2 \pi \i \, u_i} \, ,
\end{equation}
in the effective action~\eqref{eq:EffActSam} and expand it~\footnote{Really, we first make the substitution in the denominator, then expand the result and keep leading contributions. Then, finally, we re-sum over the variable~$l$ and obtain a sum over polylogarithms at diverse level. Then we substitute~\eqref{eq:DeltasSRep} and~\eqref{eq:ZetaSRep} and expand the answer around~$\epsilon=0$ up to the desired order. In this way we are able to avoid finding undesired infinities due to mistreatment of logarithmic divergencies (See the discussion in pargraph~\ref{paragraph:ZeroModes}). } around~$\epsilon=\frac{1}{\Lambda}=0\,$ then the first term in the right-hand side of~\eqref{eq:EffActSam} reduces to\,
\begin{equation}
\begin{split}
\sum_{i\neq j=1}^n\,&\frac{\pi ^2  \left(-\overline{B}_2\left(- u_{{ij}}-\frac{i \omega _1}{2 \pi
   }\right)+2 \overline{B}_2\left(-
    u_{{ij}}\right)-\overline{B}_2\left(- u_{{ji}}
   +\frac{i
   \omega _1}{2 \pi }\right)\right)t_i t_j}{\Delta _1 \Delta _2 \Delta _3} \\ & \,-\,\sum_{i\neq j=1}^n\,\frac{\omega _1\left(\Delta _1+\Delta _2+\Delta _3\right)  t_i t_j}{2 \Delta _1 \Delta
   _2 \Delta _3}\,+\,O(\epsilon^2)\,.
   \end{split}
\end{equation}
Evaluating this at the asymptotic form of the~$n!$ inequivalent saddle-points for gauge-rapidities~$u_{ij}\,\to\,u_{ij}^\star=0$, and expanding the~$\mathcal{T}^{(n)}$ we obtain
\begin{equation}
\begin{split}
-S^{(n)}_{M}+\mathcal{T}^{(n)}_{M}&\,=\,\sum_{i\neq j=1}^n\frac{1}{\epsilon}\,T(\underline{x}) \, t_i t_j 
\,+\,O(\epsilon^2)\,,
\\
\mathcal{T}^{(n)}_{M}&\,=\,\frac{1}{\epsilon}\,T(\underline{x}) \sum_{i=1}^n\Bigl( t_i\Bigr)^2 
\,+\,\epsilon\,\sum_{i=1}^n t_i\,N
\,+\,O(\log \epsilon)\,+\,O(\epsilon^2)\,.
\end{split}
\end{equation}
Adding both results we obtain
\begin{equation}
-S^{(n)}_{M}\,=\,\frac{1}{\epsilon}\,T(\underline{x}) \Bigl(\widetilde{t}\Bigr)^2 
\,+\,\epsilon\,\widetilde{t}\,N
\,+\,O(\log \epsilon)\,+\,O(\epsilon^2)\, ,
\end{equation}
where we have changed to variables (with unit Jacobian)
\begin{equation}
\widetilde{t} :={\sum_{i=1}^n t_i} \,,\,\qquad \widetilde{t}_{1,2,\ldots, n-1}\,=\,t_{1,2,\ldots, n-1}\,.
\end{equation}~\footnote{We assume the change of variables is implemented before taking any expansion that breaks the periodicity~$t_i \to {t_i}+\frac{2\pi \i}{\epsilon}\,$. In this way we are safe to consider that the~$n$ new variables are such that~$\frac{\epsilon}{2\pi\i} \widetilde{t}$ and $\frac{\epsilon}{2\pi\i} \widetilde{t}_{1,2,3,\ldots, n-1}$ range over the segment~$[0,1)\,$.}Lastly, we apply the large-charge localization lemma to the integral
\begin{equation}
\int{d\underline{\widetilde{u}}} {d\underline{\widetilde{t}}}\, e^{-S^{(n)}_{M}(\underline{x};\underline{z},\underline{\zeta})} \simExp\int_0^{\frac{2\pi\i}{\epsilon} }{d\widetilde{t}}\, e^{-S^{(n)}_{M}(\underline{x};\underline{z}^\star,\underline{\zeta})} \,\simExp\, e^{-\widetilde{s}^{(n)}_{M}(\underline{x};\widetilde{t}^\star)}\, ,
\end{equation}
where
\begin{equation}
\widetilde{t}^\star\=-\frac{N}{2T(\underline{x})}
\, ,\end{equation}
is the saddle point of
\begin{equation}
\widetilde{s}^{(n)}_{M}(\underline{x};\widetilde{t})\,=\,-T(\underline{x}) \Bigl(\widetilde{t}\Bigr)^2 
\,-\,N\, \widetilde{t}
\,,
\end{equation}
with onshell value
\begin{equation}\label{eq:GGTotalLeadingSaddle}
 e^{-\widetilde{s}^{(n)}_{M}(\underline{x};\widetilde{t}^\star)}\,\simExp\, e^{\,\frac{N^2\Delta _1
   \Delta _2 \Delta _3 }{2\omega _1\,\left({\Delta _1+\Delta _2+\Delta _3}-\omega _1\pm2  \pi\i
   \right) }}\,.
\end{equation}
This result matches the exponential of the entropy function accounting for the asympotic growth of states at large charges and spin, and thus one concludes that
\begin{equation}
\widetilde{d}_{M}(\underline{\widetilde{Q}}^\prime) \simExp  \widetilde{d}^{\prime}(\underline{\widetilde{Q}}^\prime) \qquad (\forall\, N)\,.
\end{equation}
This had to be the case because the representation~\eqref{eq:GGFormula} is an exact representation of the index.
\subsection*{The large charge growth at fixed~$n$}
We finalize by showing that the contribution coming from the (pair of complex conjugated) saddle points of the localized action at~$\zeta_i=-1\,$, dominate the counting of states relative to a single giant graviton-like block~$n\,$, for any finite~$n\,$. And second, by understanding from a macrocanonical perspective how the latter contributions cancel at large charges after summing over~$n\,$, letting the complex conjugated pairs of solutions of the localized action at~$\zeta_i=1$ to dominate the counting of $\frac{1}{16}$-BPS states at large charges.

This time we substitute
\begin{equation}
\Delta_a\, \to\,  
\epsilon \Delta_{a}\,,
\end{equation}
and
\begin{equation}
\zeta_i \to (-1)\,e^{-\epsilon t_i} 
\,,\qquad z_i\,\to\, e^{-2 \pi \i \, u_i}\,, 
\end{equation}
in the effective action
\begin{equation}
-S^{(n)}_{M}(\underline{x};\underline{z},\underline{\zeta})\,.
\end{equation}
Then we expand the answer around~$\epsilon=\frac{1}{\Lambda}=0\,$. Being careful with contributions coming from zero-modes (as detailed in previous analysis) and picking up the leading gauge saddle point~$u^{*}_{ij}=0$ we obtain the following leading contribution
\begin{equation}\label{eq:EntropyFunctionFixedGG}
\frac{4\pi ^4}{3}\frac{{\overline{B}_4}\left[\frac{1}{2}+\frac{\i \omega _1}{2 \pi }\right]-
   {\overline{B}_4}\left[\frac{\i \omega _1}{2 \pi }\right]-\frac{1}{16}}{ \Delta _1 \Delta _2
   \Delta _3 \epsilon ^3}\times n^2\,+\,O\left(\frac{1}{\epsilon ^2}\right)\times n^2\, ,
\end{equation}
where
\begin{equation}
\overline{B}_4(x)\,:=\,B_4(x-\lfloor x \rfloor)\,,\qquad B_4(x)\,:=\,x^4-2x^3+x^2-1/30\,.
\end{equation}
In the large charge region
\begin{equation}\label{eq:LocusGG}
\Qq_1\,=\,\Qq_2\,=\,\Qq_3\,=\,\Qq \,=\, q \,\Lambda^4, \qquad \widetilde{J}^\prime \,=\,0\,.
\end{equation}
~\eqref{eq:EntropyFunctionFixedGG} predicts an entropy growth-rate of~$\sim {\Qq}{}^{3/4}\,$. To show this we follow the approach presented in subsection~\ref{ref:Interference}. To count states in the charge locus~\eqref{eq:LocusGG} we must extremize
\begin{equation}\label{eq:SaddlePTilded}
\frac{4 \pi ^4}{3}\,n^2\,\frac{ {\overline{B}_4}\left[\frac{1}{2}+\frac{\i \omega _1}{2 \pi }\right]-
   {\overline{B}_4}\left[\frac{\i \omega _1}{2 \pi }\right]-\frac{1}{16}}{\Delta_1\Delta_2\Delta_3} +(\Delta_1+\Delta _2+\Delta_3 )\,\Qq\,.
\end{equation}
The relevant saddle point values are
\begin{equation}
\begin{split}
\frac{\i\omega_1}{2\pi}\,=\,\frac{1}{4}\,, \qquad
\Delta_1=\Delta_2=\Delta_3=\frac{\left(\frac{1}{2}\,\mp\,\frac{i}{2}\right) \sqrt[4]{\frac{113}{5}} \pi  }{2^{3/4}
   \sqrt{3} } \,\frac{\sqrt{n}}{\sqrt[4]{Q}}
   \,,
\end{split}
\end{equation}
and the prediction for entropy growth at fixed~$n$ is 
\begin{equation}
|\widetilde{d}^{(\text{at fixed}~n)}_{M}|[\underline{\widetilde{Q}}^\prime] \,\simExp\, |e^{\frac{(1-i) \sqrt[4]{\frac{226}{5}} \pi  }{\sqrt{3}}\sqrt{n} \Qq{}^{3/4}}\,+\,c.c.|\,\simExp\,e^{\frac{\sqrt[4]{\frac{226}{5}} \pi  }{\sqrt{3}}\sqrt{n} \Qq{}^{3/4}  }\,.
\end{equation}
Notice first that this does not depend on~$N$ and second, that at finite~$N$ it grows faster than the $\frac{1}{16}$-BPS microcanonical index~$|\widetilde{d}_{M}|\,$ which grows exponentially fast in~$O(N^{2/3}\Qq{}^{2/3})\,$.

\begin{figure}\centering
\includegraphics[width=12cm]{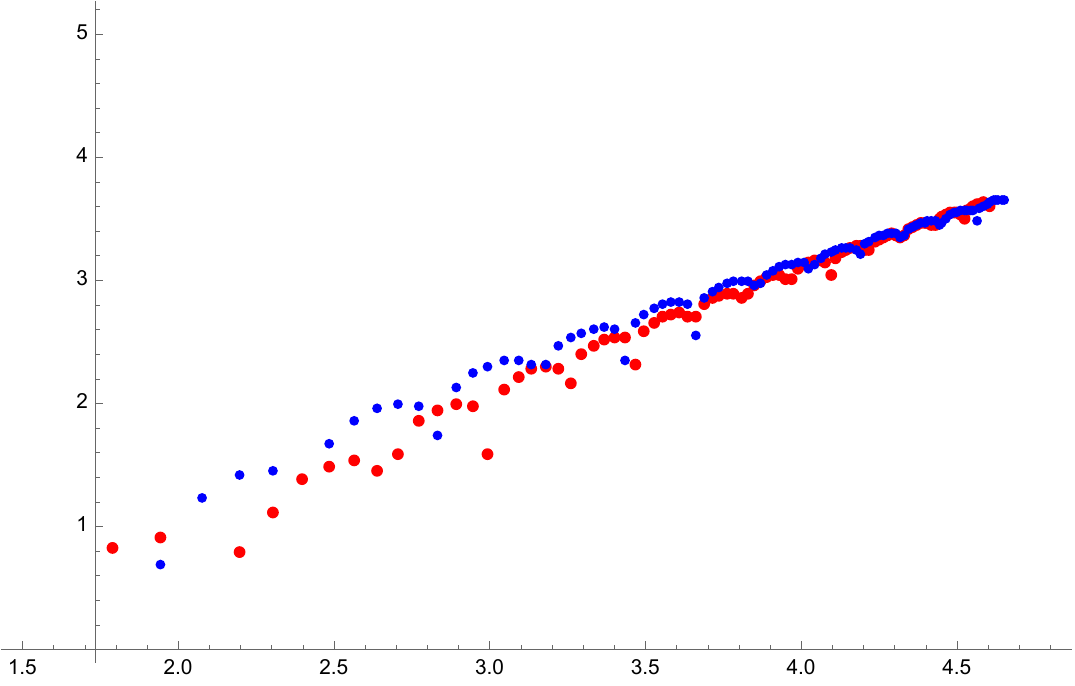} 
\caption{The vertical axis represents~$\log\Bigl(\log|d^{(n=1)}_M[Q]|\Bigr)\,$ and the horizontal axis represents~$\log Q\,$, with $Q$ ranging from~$6$ to~$80\,$. 
The vertical coordinates of the blue points come from evaluating the asymptotic formula
~$\log(\frac{1}{3}X
+\log |2 \cos X|
)\,$ 
where $X=\sqrt[4]{\frac{226}{5}} \pi\,\sqrt{3} {(Q/6)}^{3/4}$.
The vertical positions of the red points come from 
taking~$\log (\log|\cdot|)$ of the coefficients in~\eqref{eq:QSeriesGGM}. }
\label{fig:GGIndexComparison}
\end{figure}

Let us define
\begin{equation}
q=e^{-\frac{\Delta_1}{2}}=e^{-\frac{\Delta_2}{2}}=e^{-\frac{\Delta_3}{2}}=e^{-\frac{\omega_1}{3}}\, ,
\end{equation}
and proceed to compute the~$q$-series~$\mathcal{J}_{n=1}=\mathcal{J}_{n=1}(q)$ by computing residues of the integral~\eqref{eq:JN} (up to order~$Q=80$)
\begin{equation}\label{eq:QSeriesGGM}
{\fontsize{10}{12}\selectfont \begin{split}
\mathcal{J}_{n=1}(q)\,=\,&\sum_{Q} d^{(n=1)}_M[Q]\, q^{Q} \\ =& -10 q^6 + 12 q^7 - 9 q^8 + 21 q^{10} - 54 q^{11} + 83 q^{12} - 102 q^{13} + 72 q^{14} + 128 q^{15} \\
& -585 q^{16} + 1122 q^{17} - 1513 q^{18} + 1380 q^{19} + 138 q^{20} - 3900 q^{21} + 9996 q^{22} \\
& -17376 q^{23} + 22568 q^{24} - 18114 q^{25} - 6030 q^{26} + 58474 q^{27} - 142020 q^{28} + 244116 q^{29} \\
& -320713 q^{30} + 287250 q^{31} - 25656 q^{32} - 592766 q^{33} + 1645122 q^{34} - 3038934 q^{35} \\
& + 4370499 q^{36} - 4792836 q^{37} + 2942865 q^{38} + 2915380 q^{39} - 14343372 q^{40} \\
& + 31698240 q^{41} - 52605856 q^{42} + 70039506 q^{43} - 70602105 q^{44} + 34228542 q^{45} + 63154131 q^{46} \\
& -242185620 q^{47} + 506010016 q^{48} - 819250914 q^{49} + 1082818902 q^{50} - 1111506156 q^{51} \\
& + 627383301 q^{52} + 710585424 q^{53} - 3216045014 q^{54} + 7001989140 q^{55} - 11715308649 q^{56} \\
& + 16199071728 q^{57} - 18156237900 q^{58} + 13963219146 q^{59}\\& + 1135248962 q^{60} - 32145290706 q^{61} 
 + 82429426092 q^{62} - 150565817086 q^{63}\\& + 226011251286 q^{64} - 284147263932 q^{65} + 282109482979 q^{66} 
 -157874585688 q^{67}\\& - 163795361382 q^{68} + 754154356216 q^{69} - 1647734709546 q^{70} + 2791649406978 q^{71}\\& -3979126322771 q^{72} + 4777631630670 q^{73} - 4473905312412 q^{74} + 2073206793162 q^{75} \\&+ 3594625665549 q^{76}-13599471353058 q^{77} + 28405997629926 q^{78} - 47107802836014 q^{79} \\& + 66434292154434 q^{80}\,+\,\cdots\,.
\end{split}}
\end{equation}
We note that~$Q\neq \Qq$ (see the discussion in section~\ref{sec:ChoicesConstraint}), however they are related in the asymptotic limit~\eqref{eq:LocusGG} as follows
\begin{equation}
Q \underset{\Lambda\to \infty}{\sim} 6 \Qq\,.
\end{equation}
Using this relation we can compare 
\begin{equation}
\begin{split}
|d^{(n=1)}_M[Q]|&\,\simExp\,|\widetilde{d}^{(\text{at fixed}~n=1)}_{M}[{\widetilde{Q}}^\prime]|  \\&\,\simExp\, \,e^{\frac{\sqrt[4]{\frac{226}{5}} \pi  }{\sqrt{3}} \Qq{}^{3/4}  }|2\cos{\frac{\sqrt[4]{\frac{226}{5}} \pi  }{\sqrt{3}} \Qq{}^{3/4}  }|\,.
\end{split}
\end{equation}
The result is presented in figure~\ref{fig:GGIndexComparison}.

Notice that the giant graviton index at fixed~$n$ grows faster than the total giant graviton index in the limit~\eqref{eq:LocusGG}. How are these cancellations explained in the present approach?
Let us define the variable~$\delta n$
\begin{equation}
\delta n := \frac{N}{\Lambda^4}\,n\, ,
\end{equation}
which ranges over a continuum domain in the limit as~$\Lambda\to \infty$. Then we can trade the sum over~$n$ by an integral over a \underline{finite} segment of length~$\widetilde{q}^\prime=O(1)$ as $\Lambda\to \infty$ that can be evaluated by saddle-point approximation (as it is Gaussian):
\begin{equation}
\begin{split}
\,\sum_{n\,=\,0}^{\lfloor \Qq/N\rfloor}\, e^{\Bigl(\frac{4\pi ^4}{3}\frac{{\overline{B}_4}\left[\frac{1}{2}+\frac{\i \omega _1}{2 \pi }\right]-
   {\overline{B}_4}\left[\frac{\i \omega _1}{2 \pi }\right]-\frac{1}{16}}{ \Delta _1 \Delta _2
   \Delta _3 \epsilon ^3}\Bigr)\, n^2 }&\,\simExp\, \int^{\widetilde{q}^\prime}_0 d[\delta n] e^{\Bigl(\frac{4\pi ^4}{3}\frac{{\overline{B}_4}\left[\frac{1}{2}+\frac{\i \omega _1}{2 \pi }\right]-
   {\overline{B}_4}\left[\frac{\i \omega _1}{2 \pi }\right]-\frac{1}{16}}{ \Delta _1 \Delta _2
   \Delta _3 \epsilon ^3}\Bigr)\,\frac{\Lambda^8}{N^2}\, \delta n^2 }\\ & \simExp\,1\,.
   \end{split}
\end{equation}
This mechanism explains how these contributions do not compete with the ones coming from the singularity locus at~$\zeta_i=1$ (encoded in~\eqref{eq:GGTotalLeadingSaddle}) in determinining the total microcanonical giant graviton index~\eqref{eq:GGMicroIndexDefinition} at large charges.


\begin{thebibliography}{10}

\bibitem{Kinney:2005ej}
J.~Kinney, J.~M. Maldacena, S.~Minwalla, and S.~Raju, {\it {An Index for 4
  dimensional super conformal theories}},  {\em Commun. Math. Phys.} {\bf 275}
  (2007) 209--254, [\href{http://arxiv.org/abs/hep-th/0510251}{{\tt
  hep-th/0510251}}].

\bibitem{Romelsberger:2005eg}
C.~Romelsberger, {\it {Counting chiral primaries in N = 1, d=4 superconformal
  field theories}},  {\em Nucl. Phys.} {\bf B747} (2006) 329--353,
  [\href{http://arxiv.org/abs/hep-th/0510060}{{\tt hep-th/0510060}}].

\bibitem{Imamura:2021ytr}
Y.~Imamura, {\it {Finite-N superconformal index via the AdS/CFT
  correspondence}},  {\em PTEP} {\bf 2021} (2021), no.~12 123B05,
  [\href{http://arxiv.org/abs/2108.12090}{{\tt arXiv:2108.12090}}].

\bibitem{Arai:2019xmp}
R.~Arai and Y.~Imamura, {\it {Finite $N$ Corrections to the Superconformal
  Index of S-fold Theories}},  {\em PTEP} {\bf 2019} (2019), no.~8 083B04,
  [\href{http://arxiv.org/abs/1904.09776}{{\tt arXiv:1904.09776}}].

\bibitem{Gaiotto:2021xce}
D.~Gaiotto and J.~H. Lee, {\it {The Giant Graviton Expansion}},
  \href{http://arxiv.org/abs/2109.02545}{{\tt arXiv:2109.02545}}.

\bibitem{Lee:2022vig}
J.~H. Lee, {\it {Exact stringy microstates from gauge theories}},  {\em JHEP}
  {\bf 11} (2022) 137, [\href{http://arxiv.org/abs/2204.09286}{{\tt
  arXiv:2204.09286}}].

\bibitem{Murthy:2020rbd}
S.~Murthy, {\it {The growth of the $\frac{1}{16}$-BPS index in 4d
  $\mathcal{N}=4$ SYM}},  \href{http://arxiv.org/abs/2005.10843}{{\tt
  arXiv:2005.10843}}.

\bibitem{Agarwal:2020zwm}
P.~Agarwal, S.~Choi, J.~Kim, S.~Kim, and J.~Nahmgoong, {\it {AdS black holes
  and finite N indices}},  \href{http://arxiv.org/abs/2005.11240}{{\tt
  arXiv:2005.11240}}.

\bibitem{Cabo-Bizet:2018ehj}
A.~Cabo-Bizet, D.~Cassani, D.~Martelli, and S.~Murthy, {\it {Microscopic origin
  of the Bekenstein-Hawking entropy of supersymmetric AdS$_{5}$ black holes}},
  {\em JHEP} {\bf 10} (2019) 062, [\href{http://arxiv.org/abs/1810.11442}{{\tt
  arXiv:1810.11442}}].

\bibitem{Choi:2018hmj}
S.~Choi, J.~Kim, S.~Kim, and J.~Nahmgoong, {\it {Large AdS black holes from
  QFT}},  \href{http://arxiv.org/abs/1810.12067}{{\tt arXiv:1810.12067}}.

\bibitem{Benini:2018ywd}
F.~Benini and E.~Milan, {\it {Black Holes in 4D $\mathcal{N}$=4
  Super-Yang-Mills Field Theory}},  {\em Phys. Rev. X} {\bf 10} (2020), no.~2
  021037, [\href{http://arxiv.org/abs/1812.09613}{{\tt arXiv:1812.09613}}].

\bibitem{Murthy:2022ien}
S.~Murthy, {\it {Unitary matrix models, free fermions, and the giant graviton
  expansion}},  {\em Pure Appl. Math. Quart.} {\bf 19} (2023), no.~1 299--340,
  [\href{http://arxiv.org/abs/2202.06897}{{\tt arXiv:2202.06897}}].

\bibitem{Liu:2022olj}
J.~T. Liu and N.~J. Rajappa, {\it {Finite N indices and the giant graviton
  expansion}},  {\em JHEP} {\bf 04} (2023) 078,
  [\href{http://arxiv.org/abs/2212.05408}{{\tt arXiv:2212.05408}}].

\bibitem{Eniceicu:2023uvd}
D.~S. Eniceicu, {\it {Comments on the Giant-Graviton Expansion of the
  Superconformal Index}},  \href{http://arxiv.org/abs/2302.04887}{{\tt
  arXiv:2302.04887}}.

\bibitem{Berenstein:2022srd}
D.~Berenstein and S.~Wang, {\it {BPS coherent states and localization}},  {\em
  JHEP} {\bf 08} (2022) 164, [\href{http://arxiv.org/abs/2203.15820}{{\tt
  arXiv:2203.15820}}].

\bibitem{Lin:2022gbu}
H.~Lin, {\it {Coherent state operators, giant gravitons, and gauge-gravity
  correspondence}},  {\em Annals Phys.} {\bf 451} (2023) 169248,
  [\href{http://arxiv.org/abs/2212.14002}{{\tt arXiv:2212.14002}}].

\bibitem{Beccaria:2023zjw}
M.~Beccaria and A.~Cabo-Bizet, {\it {On the brane expansion of the Schur
  index}},  {\em JHEP} {\bf 08} (2023) 073,
  [\href{http://arxiv.org/abs/2305.17730}{{\tt arXiv:2305.17730}}].

\bibitem{Honda:2019cio}
M.~Honda, {\it {Quantum Black Hole Entropy from 4d Supersymmetric Cardy
  formula}},  {\em Phys. Rev.} {\bf D100} (2019), no.~2 026008,
  [\href{http://arxiv.org/abs/1901.08091}{{\tt arXiv:1901.08091}}].

\bibitem{ArabiArdehali:2019tdm}
A.~Arabi~Ardehali, {\it {Cardy-like asymptotics of the 4d $ \mathcal{N}=4 $
  index and AdS$_{5}$ blackholes}},  {\em JHEP} {\bf 06} (2019) 134,
  [\href{http://arxiv.org/abs/1902.06619}{{\tt arXiv:1902.06619}}].

\bibitem{Kim:2019yrz}
J.~Kim, S.~Kim, and J.~Song, {\it {A 4d $ \mathcal{N} $ = 1 Cardy Formula}},
  {\em JHEP} {\bf 01} (2021) 025, [\href{http://arxiv.org/abs/1904.03455}{{\tt
  arXiv:1904.03455}}].

\bibitem{Cabo-Bizet:2019osg}
A.~Cabo-Bizet, D.~Cassani, D.~Martelli, and S.~Murthy, {\it {The asymptotic
  growth of states of the 4d $ \mathcal{N}=1 $ superconformal index}},  {\em
  JHEP} {\bf 08} (2019) 120, [\href{http://arxiv.org/abs/1904.05865}{{\tt
  arXiv:1904.05865}}].

\bibitem{Cassani:2021fyv}
D.~Cassani and Z.~Komargodski, {\it {EFT and the SUSY Index on the 2nd Sheet}},
   {\em SciPost Phys.} {\bf 11} (2021) 004,
  [\href{http://arxiv.org/abs/2104.01464}{{\tt arXiv:2104.01464}}].

\bibitem{Choi:2022ovw}
S.~Choi, S.~Kim, E.~Lee, and J.~Lee, {\it {From giant gravitons to black
  holes}},  \href{http://arxiv.org/abs/2207.05172}{{\tt arXiv:2207.05172}}.

\bibitem{Gutowski:2004ez}
J.~B. Gutowski and H.~S. Reall, {\it {Supersymmetric AdS$_5$ black holes}},
  {\em JHEP} {\bf 02} (2004) 006,
  [\href{http://arxiv.org/abs/hep-th/0401042}{{\tt hep-th/0401042}}].

\bibitem{Gutowski:2004yv}
J.~B. Gutowski and H.~S. Reall, {\it {General supersymmetric AdS(5) black
  holes}},  {\em JHEP} {\bf 04} (2004) 048,
  [\href{http://arxiv.org/abs/hep-th/0401129}{{\tt hep-th/0401129}}].

\bibitem{Cvetic:2004ny}
M.~Cvetic, H.~Lu, and C.~Pope, {\it {Charged rotating black holes in five
  dimensional U(1)**3 gauged N=2 supergravity}},  {\em Phys. Rev. D} {\bf 70}
  (2004) 081502, [\href{http://arxiv.org/abs/hep-th/0407058}{{\tt
  hep-th/0407058}}].

\bibitem{Cabo-Bizet:2020ewf}
A.~Cabo-Bizet, {\it {From multi-gravitons to Black holes: The role of complex
  saddles}},  \href{http://arxiv.org/abs/2012.04815}{{\tt arXiv:2012.04815}}.

\bibitem{Cabo-Bizet:2019eaf}
A.~Cabo-Bizet and S.~Murthy, {\it {Supersymmetric phases of 4d $ \mathcal{N} $
  = 4 SYM at large $N$}},  {\em JHEP} {\bf 09} (2020) 184,
  [\href{http://arxiv.org/abs/1909.09597}{{\tt arXiv:1909.09597}}].

\bibitem{Cabo-Bizet:2020nkr}
A.~Cabo-Bizet, D.~Cassani, D.~Martelli, and S.~Murthy, {\it {The large-$N$
  limit of the 4d $ \mathcal{N} $ = 1 superconformal index}},  {\em JHEP} {\bf
  11} (2020) 150, [\href{http://arxiv.org/abs/2005.10654}{{\tt
  arXiv:2005.10654}}].

\bibitem{Aharony:2021zkr}
O.~Aharony, F.~Benini, O.~Mamroud, and E.~Milan, {\it {A gravity interpretation
  for the Bethe Ansatz expansion of the $\mathcal{N}=4$ SYM index}},  {\em
  Phys. Rev. D} {\bf 104} (2021) 086026,
  [\href{http://arxiv.org/abs/2104.13932}{{\tt arXiv:2104.13932}}].

\bibitem{Mamroud:2022msu}
O.~Mamroud, {\it {The SUSY Index Beyond the Cardy Limit}},
  \href{http://arxiv.org/abs/2212.11925}{{\tt arXiv:2212.11925}}.

\bibitem{Chen:2023lzq}
Y.~Chen, M.~Heydeman, Y.~Wang, and M.~Zhang, {\it {Probing Supersymmetric Black
  Holes with Surface Defects}},  \href{http://arxiv.org/abs/2306.05463}{{\tt
  arXiv:2306.05463}}.

\bibitem{Hosseini:2017mds}
S.~M. Hosseini, K.~Hristov, and A.~Zaffaroni, {\it {An extremization principle
  for the entropy of rotating BPS black holes in AdS$_{5}$}},  {\em JHEP} {\bf
  07} (2017) 106, [\href{http://arxiv.org/abs/1705.05383}{{\tt
  arXiv:1705.05383}}].

\bibitem{Berenstein:2002jq}
D.~E. Berenstein, J.~M. Maldacena, and H.~S. Nastase, {\it {Strings in flat
  space and pp waves from N=4 superYang-Mills}},  {\em JHEP} {\bf 04} (2002)
  013, [\href{http://arxiv.org/abs/hep-th/0202021}{{\tt hep-th/0202021}}].

\bibitem{Alday:2007mf}
L.~F. Alday and J.~M. Maldacena, {\it {Comments on operators with large spin}},
   {\em JHEP} {\bf 11} (2007) 019, [\href{http://arxiv.org/abs/0708.0672}{{\tt
  arXiv:0708.0672}}].

\bibitem{Basso:2006nk}
B.~Basso and G.~P. Korchemsky, {\it {Anomalous dimensions of high-spin
  operators beyond the leading order}},  {\em Nucl. Phys. B} {\bf 775} (2007)
  1--30, [\href{http://arxiv.org/abs/hep-th/0612247}{{\tt hep-th/0612247}}].

\bibitem{Komargodski:2012ek}
Z.~Komargodski and A.~Zhiboedov, {\it {Convexity and Liberation at Large
  Spin}},  {\em JHEP} {\bf 11} (2013) 140,
  [\href{http://arxiv.org/abs/1212.4103}{{\tt arXiv:1212.4103}}].

\bibitem{Fitzpatrick:2012yx}
A.~L. Fitzpatrick, J.~Kaplan, D.~Poland, and D.~Simmons-Duffin, {\it {The
  Analytic Bootstrap and AdS Superhorizon Locality}},  {\em JHEP} {\bf 12}
  (2013) 004, [\href{http://arxiv.org/abs/1212.3616}{{\tt arXiv:1212.3616}}].

\bibitem{Alvarez_Gaume_2021}
L.~Alvarez-Gaume, D.~Orlando, and S.~Reffert, {\it Selected topics in the large
  quantum number expansion},  {\em Physics Reports} {\bf 933} (oct, 2021)
  1--66.

\bibitem{GonzalezLezcano:2020yeb}
A.~González~Lezcano, J.~Hong, J.~T. Liu, and L.~A. Pando~Zayas, {\it
  {Sub-leading Structures in Superconformal Indices: Subdominant Saddles and
  Logarithmic Contributions}},  \href{http://arxiv.org/abs/2007.12604}{{\tt
  arXiv:2007.12604}}.

\bibitem{Goldstein:2020yvj}
K.~Goldstein, V.~Jejjala, Y.~Lei, S.~van Leuven, and W.~Li, {\it {Residues,
  modularity, and the Cardy limit of the 4d $\mathcal{N}=4$ superconformal
  index}},  \href{http://arxiv.org/abs/2011.06605}{{\tt arXiv:2011.06605}}.

\bibitem{Amariti:2020jyx}
A.~Amariti, M.~Fazzi, and A.~Segati, {\it {The SCI of $ \mathcal{N} $ = 4
  USp(2N$_{c}$) and SO(N$_{c}$) SYM as a matrix integral}},  {\em JHEP} {\bf
  06} (2021) 132, [\href{http://arxiv.org/abs/2012.15208}{{\tt
  arXiv:2012.15208}}].

\bibitem{Amariti:2021ubd}
A.~Amariti, M.~Fazzi, and A.~Segati, {\it {Expanding on the Cardy-like limit of
  the SCI of 4d $ \mathcal{N} $ = 1 ABCD SCFTs}},  {\em JHEP} {\bf 07} (2021)
  141, [\href{http://arxiv.org/abs/2103.15853}{{\tt arXiv:2103.15853}}].

\bibitem{ArabiArdehali:2021nsx}
A.~Arabi~Ardehali and S.~Murthy, {\it {The 4d superconformal index near roots
  of unity and 3d Chern-Simons theory}},
  \href{http://arxiv.org/abs/2104.02051}{{\tt arXiv:2104.02051}}.

\bibitem{Jejjala:2021hlt}
V.~Jejjala, Y.~Lei, S.~van Leuven, and W.~Li, {\it {$SL(3,\mathbb{Z})$
  Modularity and New Cardy Limits of the $\mathcal{N}=4$ Superconformal
  Index}},  \href{http://arxiv.org/abs/2104.07030}{{\tt arXiv:2104.07030}}.

\bibitem{Ardehali:2021irq}
A.~A. Ardehali and J.~Hong, {\it {Decomposition of BPS Moduli Spaces and
  Asymptotics of Supersymmetric Partition Functions}},
  \href{http://arxiv.org/abs/2110.01538}{{\tt arXiv:2110.01538}}.

\bibitem{Cabo-Bizet:2021plf}
A.~Cabo-Bizet, {\it {On the 4d superconformal index near roots of unity: bulk
  and localized contributions}},  {\em JHEP} {\bf 02} (2023) 134,
  [\href{http://arxiv.org/abs/2111.14941}{{\tt arXiv:2111.14941}}].

\bibitem{Cabo-Bizet:2021jar}
A.~Cabo-Bizet, {\it {Quantum phases of 4d SU(N) $ \mathcal{N} $ = 4 SYM}},
  {\em JHEP} {\bf 10} (2022) 052, [\href{http://arxiv.org/abs/2111.14942}{{\tt
  arXiv:2111.14942}}].

\bibitem{Jejjala:2022lrm}
V.~Jejjala, Y.~Lei, S.~van Leuven, and W.~Li, {\it {Modular factorization of
  superconformal indices}},  \href{http://arxiv.org/abs/2210.17551}{{\tt
  arXiv:2210.17551}}.

\bibitem{Amariti:2023rci}
A.~Amariti and A.~Zanetti, {\it {S-duality in the Cardy-like limit of the
  superconformal index}},  \href{http://arxiv.org/abs/2307.08391}{{\tt
  arXiv:2307.08391}}.

\bibitem{Choi:2019zpz}
S.~Choi, C.~Hwang, and S.~Kim, {\it {Quantum vortices, M2-branes and black
  holes}},  \href{http://arxiv.org/abs/1908.02470}{{\tt arXiv:1908.02470}}.

\bibitem{Nian:2019pxj}
J.~Nian and L.~A. Pando~Zayas, {\it {Microscopic entropy of rotating
  electrically charged AdS$_{4}$ black holes from field theory localization}},
  {\em JHEP} {\bf 03} (2020) 081, [\href{http://arxiv.org/abs/1909.07943}{{\tt
  arXiv:1909.07943}}].

\bibitem{GonzalezLezcano:2022hcf}
A.~Gonz\'alez~Lezcano, M.~Jerdee, and L.~A. Pando~Zayas, {\it {Cardy expansion
  of 3d superconformal indices and corrections to the dual black hole
  entropy}},  {\em JHEP} {\bf 01} (2023) 044,
  [\href{http://arxiv.org/abs/2210.12065}{{\tt arXiv:2210.12065}}].

\bibitem{BenettiGenolini:2023rkq}
P.~Benetti~Genolini, A.~Cabo-Bizet, and S.~Murthy, {\it {Supersymmetric phases
  of AdS$_4$/CFT$_3$}},  \href{http://arxiv.org/abs/2301.00763}{{\tt
  arXiv:2301.00763}}.

\bibitem{Amariti:2023ygn}
A.~Amariti, J.~Nian, L.~A. Pando~Zayas, and A.~Segati, {\it {Universal
  Cardy-Like Behavior of 3D A-Twisted Partition Functions}},
  \href{http://arxiv.org/abs/2306.05462}{{\tt arXiv:2306.05462}}.

\bibitem{Bobev:2022bjm}
N.~Bobev, V.~Dimitrov, V.~Reys, and A.~Vekemans, {\it {Higher derivative
  corrections and AdS5 black holes}},  {\em Phys. Rev. D} {\bf 106} (2022),
  no.~12 L121903, [\href{http://arxiv.org/abs/2207.10671}{{\tt
  arXiv:2207.10671}}].

\bibitem{Cassani:2022lrk}
D.~Cassani, A.~Ruip\'erez, and E.~Turetta, {\it {Corrections to AdS$_{5}$ black
  hole thermodynamics from higher-derivative supergravity}},  {\em JHEP} {\bf
  11} (2022) 059, [\href{http://arxiv.org/abs/2208.01007}{{\tt
  arXiv:2208.01007}}].

\bibitem{Cassani:2023vsa}
D.~Cassani, A.~Ruip\'erez, and E.~Turetta, {\it {Boundary terms and conserved
  charges in higher-derivative gauged supergravity}},
  \href{http://arxiv.org/abs/2304.06101}{{\tt arXiv:2304.06101}}.

\bibitem{Berkooz:2006wc}
M.~Berkooz, D.~Reichmann, and J.~Simon, {\it {A Fermi Surface Model for Large
  Supersymmetric AdS(5) Black Holes}},  {\em JHEP} {\bf 01} (2007) 048,
  [\href{http://arxiv.org/abs/hep-th/0604023}{{\tt hep-th/0604023}}].

\bibitem{Berkooz:2008gc}
M.~Berkooz and D.~Reichmann, {\it {Weakly Renormalized Near 1/16 SUSY Fermi
  Liquid Operators in N=4 SYM}},  {\em JHEP} {\bf 10} (2008) 084,
  [\href{http://arxiv.org/abs/0807.0559}{{\tt arXiv:0807.0559}}].

\bibitem{Chang:2023zqk}
C.-M. Chang, L.~Feng, Y.-H. Lin, and Y.-X. Tao, {\it {Decoding stringy
  near-supersymmetric black holes}},
  \href{http://arxiv.org/abs/2306.04673}{{\tt arXiv:2306.04673}}.

\bibitem{Budzik:2023vtr}
K.~Budzik, H.~Murali, and P.~Vieira, {\it {Following Black Hole States}},
  \href{http://arxiv.org/abs/2306.04693}{{\tt arXiv:2306.04693}}.

\bibitem{Caetano:2023zwe}
J.~a. Caetano, S.~Komatsu, and Y.~Wang, {\it {Large Charge 't Hooft Limit of
  $\mathcal{N}=4$ Super-Yang-Mills}},
  \href{http://arxiv.org/abs/2306.00929}{{\tt arXiv:2306.00929}}.

\bibitem{Boruch:2022tno}
J.~Boruch, M.~T. Heydeman, L.~V. Iliesiu, and G.~J. Turiaci, {\it {BPS and
  near-BPS black holes in $AdS_5$ and their spectrum in $\mathcal{N}=4$ SYM}},
  \href{http://arxiv.org/abs/2203.01331}{{\tt arXiv:2203.01331}}.

\bibitem{Turiaci:2023jfa}
G.~J. Turiaci and E.~Witten, {\it {$\mathcal{N}=2$ JT Supergravity and Matrix
  Models}},  \href{http://arxiv.org/abs/2305.19438}{{\tt arXiv:2305.19438}}.

\bibitem{Chang:2013fba}
C.-M. Chang and X.~Yin, {\it {1/16 BPS states in $\mathcal N=$ 4
  super-Yang-Mills theory}},  {\em Phys. Rev.} {\bf D88} (2013), no.~10 106005,
  [\href{http://arxiv.org/abs/1305.6314}{{\tt arXiv:1305.6314}}].

\bibitem{Dolan:2008qi}
F.~A. Dolan and H.~Osborn, {\it {Applications of the Superconformal Index for
  Protected Operators and q-Hypergeometric Identities to N=1 Dual Theories}},
  {\em Nucl. Phys. B} {\bf 818} (2009) 137--178,
  [\href{http://arxiv.org/abs/0801.4947}{{\tt arXiv:0801.4947}}].

\bibitem{Benini:2018mlo}
F.~Benini and E.~Milan, {\it {A Bethe Ansatz type formula for the
  superconformal index}},  {\em Commun. Math. Phys.} {\bf 376} (2020), no.~2
  1413--1440, [\href{http://arxiv.org/abs/1811.04107}{{\tt arXiv:1811.04107}}].

\bibitem{Narukawa}
A.~{Narukawa}, {\it {The modular properties and the integral representations of
  the multiple elliptic gamma functions}},
  \href{http://arxiv.org/abs/math/0306164}{{\tt math/0306164}}.

\bibitem{Felder2000}
G.~Felder and A.~Varchenko {\em Advances in Mathematics} {\bf 156} (2000),
  no.~1 44--76, [\href{http://arxiv.org/abs/9907061}{{\tt 9907061}}].

\end{thebibliography}

\providecommand{\href}[2]{#2}\begingroup\raggedright\endgroup

\bibliographystyle{JHEP}

\end{document}